\documentclass[traditabstract]{aa}
\usepackage{amsmath}
\usepackage{graphicx}
\usepackage{wrapfig}
\usepackage{lscape}
\usepackage{captions}
\usepackage{arydshln}
\usepackage{txfonts}
\usepackage{natbib}
\bibpunct{(}{)}{;}{a}{}{,}
\usepackage{ulem}

\newcommand{\halpha}{H$\alpha$}

\newcommand{\lk}{$\,L_{\rm K}$}

\newcommand{\mi}{$\mu$m}

\newcommand{\ginj}{g_{\rm inj}}
\newcommand{\gammabreak}{\gamma_{\rm b}}
\newcommand{\Ebreak}{E_{\rm b}}

\newcommand{\nubreak}{\nu_{\rm b}}
\newcommand{\nuc}{\nu_{\rm c}}

\newcommand{\vc}{V_{\rm c}}

\newcommand{\sth}{S_{th}}

\newcommand{\sthzero}{S_{th,0}}
\newcommand{\snthzero}{S_{\rm nth,0}}
\newcommand{\stot}{S_{\rm tot}}
\newcommand{\anth}{\alpha_{\rm nth}}
\newcommand{\fth}{f_{\rm th}}

\newcommand{\im}[1]{\includegraphics[width=.47\textwidth]{#1}}
\newcommand{\image}[1]{\includegraphics[width=.21\textwidth]{#1}}

\newcommand{\Msun}{{M$_{\odot}$}}

\def\fmag{\hbox{$.\!\!^{\rm m}$}} 

\def\degr{\hbox{$^\circ$}}

\def\arcsec{\hbox{$^{\prime\prime}$}}
\def\utw{\smash{\rlap{\lower5pt\hbox{$\sim$}}}}
\def\udtw{\smash{\rlap{\lower6pt\hbox{$\approx$}}}}

\begin{document}

\title{Radio synchrotron spectra of star-forming galaxies}
\titlerunning{Synchrotron spectra of galaxies}
\subtitle{}
\author{U. Klein\inst{1}
    \and
	U. Lisenfeld\inst{2,3}
    \and
	S. Verley\inst{2,3}
	}
\institute{AIfA, Universit\"{a}t Bonn, Auf dem H\"ugel 71, 53121 Bonn, Germany\\
\email{[uklein]@astro.uni-bonn.de}
\and Departamento de F\'isica Te\'orica y del Cosmos, Universidad de Granada, Spain\\
\email{[ute,simon]@ugr.es}
\and Instituto Universitario Carlos I de F\'isica Te\'orica y Computacional, Facultad 
de Ciencias, 18071 Granada, Spain}

\date{Received month ??, ????; accepted month ??, ????}

\abstract{

The radio continuum spectra of 14 star-forming galaxies are investigated by 
fitting nonthermal (synchrotron) and thermal (free-free) radiation laws. The 
underlying radio continuum measurements cover a frequency range of $\sim$325~MHz 
to 24.5~GHz (32~GHz in case of M\,82). It turns out that most of these synchrotron 
spectra are not simple power-laws, but are best represented by a low-frequency 
spectrum with a mean slope $\alpha_{nth} = 0.59 \pm 0.20$ ($S_{\nu} \propto 
\nu^{-\alpha}$), and by a break or an exponential decline in the frequency 
range of 1~--~12~GHz. Simple power-laws or mildly curved synchrotron 
spectra lead to unrealistically low  thermal flux densities, and/or to 
strong deviations from the expected optically thin free-free spectra with 
slope $\alpha_{th} = 0.10$ in the fits. The break or cutoff energies are 
in the range of 1.5 - 7~GeV. We briefly discuss the possible origin of such 
a cutoff or break. If the low-frequency spectra obtained here reflect the 
injection spectrum of cosmic-ray electrons, they comply with the mean spectral 
index of Galactic supernova remnants. A comparison of the fitted thermal flux 
densities with the (foreground-corrected) H$\alpha$ fluxes yields the extinction, 
which increases with metallicity. The fraction of thermal emission is higher than 
believed hitherto, especially at high frequencies, and is highest in the dwarf 
galaxies of our sample, which we interpret in terms of a lack of containment in 
these low-mass systems, or a time effect caused by a very young starburst.

}


\maketitle

\section{Introduction}
\label{sect:intro} 

Star-forming galaxies emit both thermal (free-free) and nonthermal (synchrotron) 
radiation in the radio regime. It is known since a few decades that the synchrotron 
component dominates at frequencies up to about 10~GHz \citep{klein81, gioia82} and 
that the overall slope of the observed spectra has a mean value of about 0.70 to 0.75, 
with little dispersion (about 0.10). And for decades, the perception has been that the 
spectra are the superposition of two power-laws,  
\begin{equation}
S_{tot}(\nu) = S_{th}(\nu_0)  \left(\frac{\nu}{\nu_0}\right)^{-0.1}
      	     + S_{nth}(\nu_0
) \left(\frac{\nu}{\nu_0}\right)^{-\anth} \; , 
\label{stot}
\end{equation}
where $\alpha_{nth}$ is the spectral index of the synchrotron radiation spectrum. 
The superposition of thermal and synchrotron radiation thus  produces a 
flattening of the total radio spectrum towards high frequencies, with the 
asymptotic value of $-0.1$. This, however, can hardly be seen, owing to the onset 
of thermal radiation from dust, which becomes significant above about 40~GHz. 

In the past decade, numerous studies have been dedicated to characterize the
shape of the radio spectrum. The two major questions have been: What is the 
contribution  of the thermal emission (thermal fraction, $\fth = \sth/\stot$)?
What is the synchrotron spectral index, $\anth$, and how does it change with frequency?
Past studies have given different answers, owing to the differences in the covered
frequency range and the sample selection. \citet{gioia82} found a mean spectral index 
of the total radio continuum emission of $\alpha_{tot} \approx 0.7$, based on a sample 
of 56 galaxies, which they had observed at 408~MHz, 4.75 and 10.7~GHz. They concluded 
that the spectral index of the synchrotron radiation is $\alpha_{syn} \approx 0.8$, 
with the synchrotron emission dominating below 10~GHz. They estimated the fraction 
of thermal radiation to be $f_{\rm th,10~GHz} < 40\%$. \citet{1988PhDT.......136K} found 
a mean value of $\anth = 0.88 \pm 0.06$, and a fraction of thermal emission of 
$\la 40$\% at 10~GHz. \citet{duric88}, however, claimed a large variation of the thermal 
fraction in a sample of 41 spiral galaxies. \citet{niklas97} found a thermal fraction 
$f_{\rm th,1~GHz} = 8\% \pm 1\%$, based upon observations of a sample of 74 Shapley-Ames 
galaxies at 10.7~GHz. \citet{williams10} obtained radio continuum spectra of galaxies 
with continuous frequency coverage between 1 and 7~GHz. They found curved synchrotron 
spectra for the starburst galaxies NGC\,253, M\,82, and Arp\,220. \citet{marvil14} 
presented radio continuum spectra of 250 bright galaxies, reporting a curvature of 
the spectra between 74~MHz and 4.85~GHz, with $\anth$ changing from 0.45 to 
0.69. These latter two papers strongly indicate that the synchrotron spectra of 
star-forming galaxies cannot be simple power-laws, but probably decline with 
increasing frequency in the cm-wavelength range. Recently, \citet{tabatabaei17} 
presented measurements of 61 galaxies 1.4, 4.8, 8.5, and 10.5~GHz. Their analysis 
delivered a total spectral index of $\alpha_{\rm tot} = 0.79 \pm 0.15$, a mean 
synchrotron spectral index of $\anth= 0.97 \pm 0.16$, and a fraction of thermal 
radiation of $f_{\rm th,1~GHz} = 10\% \pm 9\%$. 

There are several reasons that may have led to the variety of the derived 
synchrotron spectra and thermal fractions summarized above, some of which are 
contradicting. First, the quality of early ($\la$ 1990) continuum measurements 
of galaxies has been inferior to that of observations conducted thereafter. 
Second, the sample sizes used in the analyses were rather different, and the 
frequency coverage of the measurements culled from the literature too small. 
Third, the careful data selection described in Sect.\ref{sect:data} may not 
have been made in some of the works.

The spectrum of the diffuse synchrotron emission is shaped by the processes that 
characterize the propagation of relativistic electrons, in particular the type of 
propagation (diffusion or convection), of energy losses (synchrotron, inverse-Compton, 
adiabatic losses or bremsstrahlung), and the confinement to the galaxy. It is generally 
accepted that the acceleration takes place in the shocks of supernova remnants (SNRs), 
which have an average spectral index of -0.5 \citet{green14}. From there, the 
relativistic particles propagate into the interstellar medium (ISM) via diffusion 
and/or convection, and lose energy. The dominant energy losses at high frequencies 
are synchrotron and inverse-Compton losses, which steepen the synchrotron spectrum.
At low frequencies thermal absorption would flatten the spectra and may even produce 
a turn-over at the lowest frequencies \citep[see the papers and discussion by] [] 
{israel90, hummel91}. However, for the frequency range considered in the present paper, 
we are not concerned with this latter process, as it would require extreme emission 
measures in order to become relevant for the frequency range considered here.

In this paper, we present an analysis of accurately determined radio continuum 
spectra of 14 star-forming galaxies, comprising massive grand-design, closely 
interacting, and low-mass (dwarf) galaxies. The spectra cover a frequency range 
$\sim$325~MHz~--~24.5~GHz (32~GHz in case of M\,82), with the 24.5~GHz 
measurements performed with the Effelsberg 100-m telescope in the mid 80's, 
and mostly unpublished to date. All other flux densities were carefully 
selected from the literature. For the first time, the data allow a reliable 
separation of the thermal and nonthermal components to be performed, hence an 
analysis of the resulting synchrotron spectra, and a firm extrapolation towards 
lower frequencies. This kind of work is also considered valuable towards 
establishing and interpreting low-frequency spectra of galaxies obtained 
with LOFAR observations \citep[see, e.g.,][]{mulcahy14}. However, for this 
paper we decided not to use measurements at frequencies below 325~MHz, since 
astrophysical processes that shape the spectra there would increase the number 
of free parameters unnecessarily. 

In Sect.\,\ref{sect:gals} we present the galaxy 
sample and their selection criteria. The data selection made for the 
present investigation is described in Sect.~\ref{sect:data}. In 
Sect.\,\ref{sect:spec_fits} the fitting procedure of the spectra is 
described, and in Sect.\,\ref{sect:results} the results are presented. 
These are discussed in Sect.\,\ref{sect:disc}, while Sect.\,\ref{sect:sum} 
contains a summary and lists our conclusions.

\section{The Galaxies}
\label{sect:gals}

The galaxy sample used for our analysis was determined by the inclusion of 
high-frequency radio continuum data (see Sect.~\ref {subsect:old_obs}), which 
are avaliable for a small number of galaxies mapped with the Effelsberg 100-m 
telescope years ago. This results in a mixture of rather different galaxy types, 
spanning the whole mass range of star-forming galaxies and incorporating closely 
interacting pairs and ongoing mergers. These galaxies  have very high star 
formation rates per unit surface, hence all of them are rather radio-bright. 
In what follows we shall briefly introduce the galaxies and their pertinent 
properties, subdividing them into several categories. All subsequent tables 
of this paper follow this categorization using dashed horizontal lines. 

{\bf Dwarf galaxies}: 
There are five low-mass galaxies in our sample. II~Zw~40 and II~Zw~70 are 
classical blue compact dwarf galaxies (BCGDs). BCDGs are also called  \ion{H}{ii}
galaxies because they are dominated by giant \ion{H}{ii} regions occupying much of 
their total volumes. Their emission-line spectra indicate that they are 
metal-poor, while they are gas-rich and are experiencing intense bursts of 
star formation. II~Zw~40 was observed in the radio continuum by \citet{klein84a}, 
\citet{klein91}, and by \citet{deeg93}. II~Zw~70 is a small and distant BCDG so 
that existing radio continuum measurements by \citep{klein84a, skillman88, 
deeg93} only provided integrated flux densities. The high-frequency radiation 
of these two galaxies is purely thermal. 

IC\,10, NGC\,1569, and NGC\,4449 are nearby starbursting dwarf galaxies. They
have properties that are rather similar to those of BCDGs. They are gas-rich, 
too, but their metallicities are not so low. 

Because of its proximity to the Galactic plane, IC\,10 is a tricky case 
for radio continuum observations. In particular at low frequencies, at 
which interferometric observations are indispensable, imaging suffers from 
contamination by spurious sidelobes from radio continuum structures in the 
Galactic plane. Another worry is thermal absorption through the plane at 
the lowest frequencies. Useful measurements for our purpose have been 
carried out  \citet{klein83a}, \citet{klein86}, \citet{chyzy03}, and 
by \citet{2016ApJ...819...39C}. 

NGC\,1569 is considered as a template of a low-mass galaxy with an evolved 
starburst. \citet{israel88} were the first to notice that the synchrotron 
spectrum is not a simple power-law, but `has a high-frequency cutoff 
at $8 \pm 1$~GHz'. This rapid decline is not obvious in the total radio 
spectrum, due to a high fraction of thermal radio emission, which was 
derived by \citet{israel88} from the \halpha\ emission. Numerous radio 
continuum studies at a multitude of frequencies, which aimed at an 
understanding of cosmic-ray propagation into the halo regime of this 
dwarf galaxy \citep{klein86, lisenfeld04, 2010ApJ...712..536K, 
purkayastha14} prove its intense star formation. It is among the 
radio-brightest in our sample and possesses a low-frequency radio halo. 

Another such nearby template of a starburst dwarf galaxy, NGC\,4449, has 
been observed in the radio continuum by \citet[][24.5~GHz]{klein86}, 
\citet[][4.9, 8.5~GHz]{2000A&A...355..128C}, \citet[][2.7, 4.9~GHz]{chyzy11}, 
\citet[][150, 325, 610~MHz]{2014MNRAS.443..860S}, and by 
\citet[][350~MHz]{purkayastha14}. This galaxy, too, is characterized by a 
high star formation rate (SFR), which was most likely triggered by the close 
passage of another - yet lower-mass - galaxy. NGC\,4449 also possesses a 
low-frequency radio halo.

{\bf Interacting galaxies}:
Upon close inspection, almost all galaxies are gravitationally interacting 
with another to some extent. In the present sample, NGC\,4490/85, NGC\,5194/95 
(M\,51), and NGC\,4631 (with nearby dwarf elliptical NGC\,4627 and the more 
massive disk galaxy NGC\,4652) have nearby and obviously interacting companion 
galaxies. These interactions are the likely cause of the intense ongoing star 
formation in these galaxies, with the close companions stirring up the gas and 
giving rise to gas compression and subsequent star formation out of the molecular 
gas. 

Radio continuum studies of NGC\,4490/85, a closely interacting pair of galaxies, 
were reported by \citet[][1.4~GHz]{viallefond80}, \citet[][1.49, 4.86, 8.44, 
15.2~GHz]{1999MNRAS.307..481C}, and by \citet[][24.5~GHz]{klein83a}. 
\citet{nikiel16} published a multi-frequency radio continuum study 
of this system using archival VLA data and new GMRT observations at 
610~MHz. The highest frequency at which NGC\,4490/85 has been studied 
was 25~GHz \citep{klein83b}. 

M\,51 has been the target of numerous radio continuum studies, which in the 
early days aimed at measuring the total radio continuum spectrum and studying 
cosmic-ray propagation \citep{klein84b}, and later on focused on investigating 
the structure of the large-scale magnetic field in it 
\citep[e.g.][]{1992A&A...263...30N, 2011MNRAS.412.2396F, 2015ApJ...800...92M}. 
The rich set of flux-density measurements available for this galaxy required 
a careful selection, with only the most reliable data in terms of confusion 
and missing short spacings culled by us. \citet{klein84b} inferred a simple 
power-law for the synchrotron spectrum of M\,51. 

After the discovery of an extended radio continuum halo at 610 and 1412~MHz around 
NGC\,4631 by \citet{ekers77} there has been a large number of observations at higher 
frequencies \citep{1990A&A...236...33H, 1991A&A...248...23H, 1999A&A...345..778G, 
2012AJ....144...44I}. These mostly aimed at studying cosmic-ray transport out of 
the disk and at investigating the morphology of the magnetic field of this galaxy. 
The high star-formation rate is probably caused by the gravitational interaction 
with the neighbouring galaxies NGC\,4627 and NGC\,4652, rendering this galaxy a 
rather radio-bright one.

{\bf Merging galaxies}:
In case of ongoing mergers, star formation is extreme, owing to the collision 
of molecular clouds in the merging centres of the two galaxies. 

NGC\,4038/39, the `Antenna Galaxy', represents one of the classical nearby ongoing 
mergers of two galaxies. \citet{2004A&A...417..541C} and \cite{2017MNRAS.464.1003B} 
performed thorough radio continuum studies, with the emphasis on the linear 
polarization and on the resulting morphology of the large-scale magnetic field in  
NGC\,4038/39. In separating the thermal from the nonthermal radio emission, they 
inferred a simple power-law for the total synchrotron spectrum of this system. 

NGC\,6052, which became known as a so-called `clumpy irregular galaxy' under 
its label Mkn\,297 \citep{1979aers.conf..205H}, is revealed as a close merger 
of two galaxies oriented perpendicularly by HST images \citep{1996AJ....112..416H}.
It has a high radio brightness. High-resolution radio continuum observations of 
this system were reported by \citet{deeg93}.

{\bf Starburst galaxies}:
NGC\,2146 and NGC\,3034 (M\,82) are classical nearby starburst galaxies, again 
with the intense star-forming activity resulting from gravitational interaction 
with a nearby galaxy or from a past merger event. A comprehensive radio continuum 
study of NGC\,2146 was reported by \citet{1996MNRAS.281..301L}, who investigated 
the cosmic-ray transport in this galaxy. NGC\,3034 (M82) has been the target of a 
large number of radio continuum studies across a large frequency range. 
\citet{adebahr13} presented observations at $\lambda\lambda$3, 6, 22, and 92 cm, 
using the VLA and the WSRT. \citet{2015A&A...574A.114V} made dedicated measurements 
with LOFAR to image the radio continuum of M\,82 at 118 and 154~MHz. However, such 
low-frequency measurements are not relevant for our present study, since at 
frequencies below about 1~GHz various effects may shape the synchrotron spectra 
in such a way as to produce deviations from a simple power-law 
(see Sect.~\ref{subsect:old_obs}). \citet{klein88} published observations of 
M\,82 at 32~GHz, the highest frequency at which thermal radiation from dust is 
still insignificant. 

NGC\,3079 and NGC\,3310 are starburst which can be considered as transition phases 
towards AGN (active galactic nucleus) activity. NGC\,3079 exhibits a focused wind 
emerging from its centre and giving rise to a `figure-8' bow-shock structure seen 
in the nonthermal radio continuum \citep{1983ApJ...273L..11D, 1988ApJ...326..574D}. 
NGC\,3310 has recently swallowed one of its dwarf companion galaxies \citep[see, 
e.g.,][]{2001A&A...376...59K}, this having led to the enormous starburst. 
High-resolution radio continuum observations were performed by 
\citet{1986ApJ...304...82D}, with the aim to study the distribution of cosmic rays 
in this galaxy. 

\section{The data}
\label{sect:data}

\subsection{24.5~GHz measurements}
\label{subsect:old_obs}

The observations at 24.5~GHz had been performed in perfect weather during test 
measurements in February 1982. The K-band maser receiver system with a bandwidth 
100~MHz had been installed in the primary focus of the Effelsberg 100-m telescope. 
The half power beam width (HPBW) was measured to be 38\arcsec\ $\pm$ 1\arcsec. 
The frontend was equipped with two horns, giving an angular beam separation of 
114\arcsec\ $\pm$ 2\arcsec\ on the sky. The aperture efficiency of the 100-m 
telescope was 19\% at the time of the observations. The pointing and focus of 
the telescope were checked using the point source 3C\,286, with small maps 
centered on this source providing also the flux calibration. The flux density 
scale is that of \citet{baars77}. For each galaxy, six coverages were taken 
between 50\degr\ and 80\degr\ elevation (corresponding to system temperatures 
of 79~K and 49~K, respectively, on the sky), using a scan separation of 10\arcsec. 
These were stacked to yield final maps with an root mean square (rms) noise of 
2.9~mJy/beam area. The observing procedure and reduction technique is essentially 
the same as described by \citet{emerson79}. The sample selection was governed by 
the radio brightness and sizes of the target galaxies, dictated by the rather high 
frequency and the comparatively small bandwidth of the maser receiver. Since those 
were the first (and only) measurements of this kind, it was attempted to include 
star-forming galaxies with a variety of properties (low- and high-mass galaxies, 
interacting systems).

\subsection{Literature search}
\label{subsect:old_obs}

In what follows we describe the compilation of the data used below 24.5~GHz. 
All published flux densities used here were checked for the calibration scale 
and, wherever necessary, were scaled to the common scale of \citet{baars77}. 
At the highest frequencies (24.5 and 32~GHz), the Baars scale is $\sim 1 \%$ 
higher than that properly extended to higher frequencies by \citet{2013ApJS..204...19P}.
This small diifference does not affect our analysis. 

A careful selection was made to ensure that the most reliable data would be 
used. We discarded interferometric measurements that might underestimate the 
flux density as a result of missing short spacings (mostly relevant above about 
1.4~GHz), but also single-dish measurements that might overestimate the flux 
density owing to source confusion (mostly relevant below about 5~GHz). If 
several measurements are available at the same or a neighbouring frequency, 
we calculated the error-weighted average at the corresponding mean frequency. 
The resulting data compilation is presented in tabular form in 
App.\,\ref{app:data}. For each galaxy, the tables give the mean frequency in 
GHz (Col.\,1), the flux density (Col.\,2) and its error (Col.\,3) in mJy, 
and the reference to the flux density measurement (Col.\,4). In a few cases, 
we have determined flux densities at 325~MHz and 1.4~GHz using the Westerbork 
Northern Sky Survey (WENSS) \citep{rengelink97} and the NRAO VLA Sky Survey (NVSS) 
\citep{condon98}, respectively. These are galaxies with small angular sizes 
so that flux loss by interfreometric measurements is not an issue. Finally, some 
flux densities were recently obtained in the course of other observing programmes 
(VLA, Effelsberg). 

We decided to utilize only flux densities above 0.3~GHz, since at lower frequencies 
a number of astrophysical effects such as thermal absorption may shape the synchrotron 
spectra in such a way as to produce deviations from a simple power-law. It would require 
more free parameters, rendering the fit results less reliable when taking these processes 
into account. The synchrotron spectra obtained here may serve though to extrapolate 
them to lower frequencies and thus provide a firm `leverage' for low-frequency studies. 

\subsection{Ancillary data}
\label{subsect:ancillary_data}

Apart from the radio data, we collected a set of ancillary data from the literature 
for the galaxies that allowed us to quantify their star formation rate (SFR), 
\halpha\ flux, stellar mass, metallicity, and optical sizes. The data, together 
with the distances, are listed in Table~\ref{tab:ancillary_data}.

As a measure of the stellar mass we used the infrared $K_{\rm S}$ luminosity, 
${\it L_{\rm K}}$, which we calculated from the total (extrapolated) 
$K_{\rm S}$ flux, $f_{\rm K}$, as ${\it L_{\rm K}}$ = $\nu f_{\rm K}(\nu) 
\, 4\pi D^2$ (where $D$ is the distance and $\nu$ is the frequency of the 
K-band\footnote{Note that - unfortunately - there exits definitions for the 
K-band in both, optical and radio astronomy, both of which have to be used 
in the present paper and must not be confused.}, 1.38$\times$10$^{14}$ Hz). 
The fluxes in the $K_{\rm S}$ (2.17 $\mu$m) band were taken from the 2MASS 
Extended Source Catalog \citep{jarrett00}, the 2MASS Large Galaxy Atlas 
\citep{jarrett03}, and from the 2Mass Extended Objects Final Release.

We calculated the star formation rate (SFR) as a combination of 24~\mi\ 
and \halpha\ fluxes, following \citet{kennicutt09}:

\begin{equation}
{\mathrm SFR} = 5.5 \times 10^{-42} \, \frac{L({\rm H\alpha}) + L(24\, \mu 
\mathrm m)}{\rm erg\,s^{-1}} \,\,\, \rm M_\odot \mathrm yr^{-1}.
\label{eq:sfr}
\end{equation}

For the 24~\mi\ flux we used, whenever possible, the flux measured with MIPS 
on the {\it Spitzer} satellite. Only in those cases for which no {\it Spitzer} 
data was available (IC\,10 and NGC\,4038), data from the IRAS satellite at 
25~\mi\ were used, neglecting the small central wavelength difference. By 
combining mid-infrared and \halpha\ fluxes we take into account both 
unobscured and dust-enshrouded star formation and circumvent the 
uncertain extinction correction of the \halpha\ flux.
 
\begin{table*}
\caption{Ancillary data for the galaxy sample}
\begin{tabular}{lrrccrcccc}
\hline
Galaxy & $D$~~~~ & $d$\tablefootmark{1}~~~ & 12+log(O/H) & Ref.\tablefootmark{2}   & log(\lk)\tablefootmark{3}  &  log($L_{ 24\mu m}$)\tablefootmark{4}  & $F$(\halpha)\tablefootmark{5} & Ref.\tablefootmark{6}  & log(SFR)\tablefootmark{7}   \\
    &[Mpc] & [kpc] & & & [$L_{\odot,k}$]~~~  &  [erg s$^{-1}$]~~~&  [erg s$^{-1}$ cm$^{-2}$] & &  [\Msun yr$^{-1}$]  \\
\hline
     IC~10   &   0.9   &   1.6   &   8.30   &   1   &   8.79   &  40.61   &  $-9.63$   &   2   &  $-0.93$  \\
  II~Zw~40   &  10.3   &   1.3   &   8.12   &   2   &   8.42   &  42.41   & $-10.81$   &   1   &   0.02  \\
  II~Zw~70   &  23.1   &   4.8   &   7.86   &   3   &   8.94   &  41.83   & $-12.08$   &   1   &  $-0.56$  \\
  NGC~1569   &   2.9   &   3.3   &   8.16   &   4   &   9.05   &  42.03   & $-10.05$   &   2   &  $-0.21$  \\
  NGC~4449   &   3.7   &   5.0   &   8.23   &   5   &   9.57   &  41.83   & $-10.64$   &   2   &  $-0.55 $ \\
\hdashline
  NGC~4490   &   8.1   &  15.8   &   8.39   &   9   &  10.21   &  42.63   & $-10.86$   &   2   &   0.03  \\
  NGC~4631   &   6.3   &  26.4   &   8.75   &   6   &  10.35   &  42.69   & $-10.70$   &   2   &   0.02  \\
  NGC~5194   &   7.6   &  30.2   &   8.86   &   6   &  10.90   &  43.03   & $-10.79$   &   2   &   0.26  \\
\hdashline
  NGC~4038   &  21.3   &  33.4   &   8.74   &   8   &  11.12   &  43.65   & $-10.93$   &   3   &   2.01  \\
  NGC~6052   &  70.4   &  16.8   &   8.85   &  10   &  10.76   &  43.74   & $-11.66$   &   3   &   1.04  \\
\hdashline
  NGC~2146   &  22.4   &  34.5   &   8.68   &   5   &  11.21   &  44.11   & $-11.18$   &   3   &   1.20  \\
  NGC~3034   &   3.8   &  12.1   &   9.12   &   6   &  10.63   &  43.85   & $-10.11$   &   2   &   0.93  \\
  NGC~3079   &  19.1   &  45.3   &   8.89   &   7   &  10.99   &  43.22   & $-11.42$   &   3   &   0.35  \\
  NGC~3310   &  18.1   &   9.4   &   8.75   &   7   &  10.40   &  43.39   & $-10.95$   &   3   &   0.64  \\
\hline
\label{tab:ancillary_data}
\end{tabular}

\tablefoottext{1} {Optical diameter in kpc, calculated from $d_{25}$ in  LEDA.}

\tablefoottext{2} {References for the oxygen abundance: (1) \citet{magrini09}, (2) \citet{thuan05}, 
(3) \citet{kehrig08}, (4) \citet{kobulnicky97}, (5) \citet{engelbracht08}, (6) \citet{moustakas10}, 
(7) \citet{robertson13}, (8) \citet{bastian09}, (9) \citet{pilyugin07}, (10)  \citet{sage93}.
}

\tablefoottext{3} {Decimal logarithm of the luminosity in the K-band in units of 
the solar luminosity in the $K_{\rm S}$-band ($L_{K,\odot} = 5.0735 \times 
10^{32}$ erg s$^{-1}$).}

\tablefoottext{4} {Decimal logarithm of the luminosity at 24~$\mu$m, derived from 
{\it Spitzer} MIPS fluxes as $L_{ 24~\mu m} = \nu f_{ 24~\mu m} \times 4 \pi D^2$}. 
The fluxes were obtained from \citet[][II~Zw~40, NGC~1569, NGC~2146, NGC~3310, 
NGC~4449, NGC~3079]{engelbracht08}, \citet[][NGC~3034, NGC~4490, NGC~4631, 
NGC~5194]{dale09}, \citet[][NGC~6052]{brown14}, or directly from the {\it 
Spitzer} archive (https://irsa.ipac.caltech.edu) (II~Zw~70). For two galaxies 
(IC~10 and NGC~4038), no {\it Spitzer} MIPS data were available and we used 
IRAS 25~\mi\ instead, neglecting the slight difference in the central wavelength.

\tablefoottext{5} {Decimal logarithm of the \halpha\ flux, corrected for  
\ion{N}{ii} contribution and for Galactic foreground extinction.}
 
\tablefoottext{6} {References for the \halpha\ fluxes. (1) \citet{gildepaz03} (2) 
\citet{kennicutt08}, (3) \citet{moustakas06}.}

\tablefoottext{7} {Decimal logarithm of the SFR, calculated with Eqn.~\ref{eq:sfr}.}
\end{table*}

\section{Spectral fitting}
\label{sect:spec_fits}

\subsection{Radio spectra}
\label{subsect:radio_spec}

Following Eqn.~\ref{stot} we fitted the radio data with the sum of thermal and 
nonthermal radio emission.

In order to predict the synchrotron emission of a model galaxy, the first step is
to calculate the relativistic electron particle density, $N(E)$, i.e. the number 
density of relativistic electrons at energy $E$ within the energy interval $dE$. 
The synchrotron emission can then be calculated by convolving this distribution 
with the synchrotron emission of a single electron:

\begin{equation}
S_{\rm nth}(\nu) \propto \int_1^{\infty} \left( \frac{\nu}{\nu_c}\right ) ^{0.3} 
e^{- \nu/\nu_{\rm c}} \, N(\gamma) \, d\gamma, 
\label{sconv}
\end{equation}
where $\gamma = E/(m_e c^2)$ is the Lorentz factor, $m_e$ the electron rest mass
and 
\begin{equation}
\nuc = \frac{3}{4\pi} \frac{e \, B_\bot}{m_e \, c} \gamma^2
\label{nu_crit}
\end{equation}
is the critical frequency, $e$ is the electron charge and $B_\bot$ the strength of 
the magnetic field component perpendicular to the direction of the relativistic 
electron's velocity. The critical frequency is close to the peak of the electron 
spectrum and represents, within a factor of a few, the frequency where the relativistic 
electrons emit most of their energy \citep[see, e.g.,][]{klein15}. The shape of the 
synchrotron spectrum emitted by a single electron is such that for frequencies above 
$\nuc$ the emission decreases exponentially, whereas for lower frequencies there is
power-law, $\nu^{0.3}$. 

If the electron energy distribution is a power-law, $N(E)~\propto~E^{-g}$, we can 
calculate the synchrotron emission to a good accuracy by assuming that an electron
with $E$ emits the entire synchrotron radiation at the critical frequency  
\citep[see][]{klein15}, 

\begin{equation}
S_{\rm nth}(\nu) = N[E(\nu_c)] \, \frac{dE}{dt}\bigg|_\mathrm{syn} \cdot 
\frac{dE}{d\nu} \propto B_\bot^{\frac{g+1}{2}} \nu^{-\frac{g-1}{2}} .
\label{eq:syn_single_el}
\end{equation}

In the following we are going to explain the four different models that we considered 
for the synchrotron spectrum. 
We chose these models with the goal to cover the range from a spectrum with constant slope 
to a maximally curved synchrotron spectrum within simplified yet realistic scenarios. \\

{\bf Spectrum with a constant slope:}
In the first case we adopt a synchrotron spectrum with  a constant slope over
the entire frequency range, 

\begin{equation}
S_{\rm nth}(\nu) = S_{\rm nth,0} \left(\frac{\nu}{\nu_0}\right)^{-\alpha_{nth}}, 
\label{snth_const}
\end{equation}
where we take $\nu_0 = 1$ GHz in this paper. For the other three cases, we consider 
simple  models, spatially integrated over the entire galaxy, which allows us to predict 
the expected synchrotron emission in different scenarios. \\

{\bf Spectrum curved due to energy losses:}
Here, we consider a steady-state, closed-box model. Since we are only interested in 
the total $N(E)$, integrated over the entire galaxy, we do not need to take into account
propagation of the relativistic particles, as long as it is energy-independent. 
In this approximation, the corresponding equation 
for the relativistic electron particle density is:

\begin{equation}
    {\partial\over \partial E} \biggr[b(E) \, N(E)\biggr] =
\bigg(\frac{E}{m_{\rm e} c^2}\bigg)^{-\ginj} \, q_{SN} \, \nu_{SN}.
\label{eq:rel_el_ss}
\end{equation}

This equation takes into account acceleration in supernova remnants (SNRs) with a 
source spectrum as $\gamma^{-\ginj} \, q_{SN} \, \nu_{SN}$ (where $\nu_{SN}$ is the 
supernova rate, $q_{SN}$ is the number of relativistic electrons produced per 
supernova and per unit energy interval, and $\ginj$ is the injection spectral 
index), and the radiative energy losses of the relativistic electrons $dE/dt=b(E)$. 
The injection spectral index, $\ginj$ is predicted by shock acceleration theory to 
be 2.1 \citep{drury94,berezhko97}. Eqn.~\ref{eq:rel_el_ss}  can be solved by 
integration and gives 

\begin{equation}
N(E) = \bigg(\frac{E}{m_{\rm e} c^2}\bigg)^{-\ginj+1} \frac{q_{SN}\nu_{SN}m_{\rm e} c^2}{b(E)(\ginj-1)}. \, 
\end{equation}

Thus, the spectral shape of $N(E)$ and of the resulting synchrotron emission depend 
on the energy dependence of the energy losses. The most relevant energy losses for CR 
electrons in the GHz range are inverse-Compton and synchrotron losses, 

\begin{equation}
b(E)_\mathrm{syn+iC} = \frac{\mathrm{d}E}{\mathrm{d}t} \bigg|_\mathrm{syn+iC} = C_\mathrm{syn+iC} \, E^2, 
\label{de_dt_syn_ic}
\end{equation}
with $C_\mathrm{syn+iC}  \propto (U_\mathrm{rad} + U_\mathrm{B}$),  $U_\mathrm{rad}$ being 
the energy density of the radiation field below the Klein-Nishina limit, and $U_\mathrm{B}$ 
the energy density of the magnetic field. We furthermore consider bremsstrahlung and adiabatic 
losses, which depend linearly on $E$ and can thus become relevant at lower energies/frequencies.

\begin{equation}
b(E)_\mathrm{ad+brems} =  \frac{\mathrm{d}E}{\mathrm{d}t} \bigg|_\mathrm{ad+brems} = C_\mathrm{ad+brems}  \, E.
\label{de_dt_brems_advec}
\end{equation}

Here $C_\mathrm{ad+brems} $ is a constant that depends on the gas density and advection 
velocity gradient. Taking both kinds of energy losses into account results in an energy 
distribution of the relativistic electrons that exhibits a (very shallow) change of slope 
centered at the break energy, $E_{\mathrm b} = C_\mathrm{ad+brems}/C_\mathrm{syn+iC} $, 

\begin{equation}
N(E)  \propto \frac{E^{-\ginj}} {E/E_{\mathrm b}+1},
\label{eq:ne_closed_box}
\end{equation}

The corresponding curved synchrotron spectrum reads 

\begin{equation}
S_{\rm nth}(\nu) = S_{\rm nth,0} \, \frac{\left(\frac{\nu}{\nu_0}\right)^{-\anth}}  
{\left(\frac{\nu}{\nubreak}\right)^{0.5}+1}, 
\label{snth_curved}
\end{equation}
where $\nubreak = \nu(\gammabreak)$ is the break frequency at which the change of slope 
takes place, and $\anth = \frac{\ginj-1}{2}$ is the low-frequency synchrotron spectral 
index. \\ 

{\bf Spectrum with a break:} 
We can produce a much more pronounced break in an open-box model, in which the relativistic 
electrons  can escape from a finite halo with vertical extent $z_{\rm halo}$ \citep[see][for 
a detailed discussion]{lisenfeld04}. The sharpest break occurs when the relativistic electrons 
propagate by convection because the scalelength (i.e. the distance over which relativistic
electrons can propagate before losing their energy) of diffusion, being a stochastic process, 
has a much weaker dependence on energy and therefore produces a shallower break. We only take 
into account inverse-Compton and synchrotron losses, according to Eqn.~\ref{de_dt_syn_ic}. The 
propagation equation for the relativistic electrons accelerated in SNRs in the galactic plane 
at $z = 0$ and moving with constant convection speed $\vc$ in the z-direction perpendicular to 
the disk is in this case 

\begin{equation}
{\partial N(E,z)\over \partial z} \, \vc -
{\partial\over \partial E} \biggr[ C_\mathrm{syn+iC} E^2 \, N(E,z)\biggr] =
\delta(z) \,  \bigg(\frac{E}{m_{\rm e}c^2}\bigg)^{-\ginj}\, q_{\rm SN} \, \nu_{\rm SN}, 
\end{equation}
with $\delta(z)$ being the one-dimensional $\delta$-function. The solution of this equation 
is given by 

\begin{equation}
N(E,z) = \left\{ \begin{array}{ll}
\frac{q_{\rm SN} \, \nu_{\rm SN}}{2 \, \vc} \,  \bigg(\frac{E}{m_{\rm e}c^2}\bigg)^{-\ginj} \,
\biggl(1-\frac{z}{z_{\rm max}}\biggr)^{\ginj-2} & z < z_{\rm max} \\
0 & z \ge z_{\rm max}  \\
\end{array} \right.
\label{eq:ne_open_vs}
\end{equation}
where  $z_{\rm max}=\vc/E \, C_\mathrm{syn+iC} $ is the maximum distance that a relativistic 
electron can travel before having lost its energy to below $E$. The total relativistic electron 
density in the galaxy is calculated by integrating Eqn.\ref{eq:ne_open_vs} from $z=0$ to $z = 
\min(z_{\rm max},z_{\rm halo})$. We then assume that the relativistic electrons emit all their 
energy at $\nuc$ and obtain for the synchrotron emission, 

\begin{equation}
S_{\rm nth}(\nu) = \left\{ \begin{array}{ll}
S_{\rm nth,0} 
\left(\frac{\nu}{\nu_0}\right)^{-\anth-0.5}   \bigg[ 1-\Bigg( 1-\sqrt{\frac{\nu}{\nubreak}} 
\Bigg)^{\ginj-1}\bigg] & \nu \le \nubreak \\
S_{\rm nth,0} \left(\frac{\nu}{\nu_0}\right)^{-\anth-0.5} & \nu > \nubreak \\
\end{array} \right.
\label{snth_break}
\end{equation}

The resulting spectrum has a break at~$\nubreak = \nuc(\Ebreak)$, with break energy 
$\Ebreak = \vc/(z_{\rm halo}\,C_\mathrm{syn+iC})$.
The spectral index changes from $\anth + 0.5$ at $\nu > \nubreak$ to
$\anth$  for $\nu \ll \nubreak$, i.e. it changes by 0.5.
The break produced  is much more pronounced than the shallow
change of spectral index due to  different energy losses 
(this can be clearly seen in the figures in App.~\ref{app:fits} where the synchrotron spectra
of the different models are shown and can be compared.)

{\bf Spectrum with an exponential cutoff:}
Finally, we consider the case that the distribution of relativistic electrons ends 
abruptly at a certain Lorentz factor $\gamma_{\rm max}$. Assuming continuous pitch-angles, 
thus following the model of \citet{jaffe73}, an exponential cutoff is produced in the 
synchrotron spectrum according to Eqn.~\ref{eq:syn_single_el} whenever there is an 
abrupt cutoff in the energy spectrum of the relativistic electrons \citep{jaffe73, 
kardashev62}. A cutoff in energy can occur in a single-injection scenario because 
the highest-energy electrons lose their energy much faster than lower-energy electrons 
so that their population becomes completely depopulated after a characteristic energy 
loss time. The location of the exponential cutoff in the radiation spectrum depends 
on the relative time scales for acceleration, energy losses, and escape of the particles 
\citep{schlickeiser84}. The synchrotron spectrum in this case is 

\begin{equation}
S_{\rm nth}(\nu) = S_{\rm nth,0}  \left(\frac{\nu}{\nu_0}\right)^{-\anth} e^{-\frac{\nu}{\nubreak}}
\label{snth_cutoff}
\end{equation}

Table~\ref{tab:syn_models} summarizes the four models, together with the free parameters
determined in the spectral fitting.

\begin{table*}[t]
\caption{Shapes of the radio (synchrotron and free-free) emission considered in the fitting 
         process} 
\label{tab:syn_models}
\begin{flushleft}
\begin{tabular}{lll}
\hline\noalign{\smallskip}
Model name & Radio spectrum & free parameters \\
\hline\noalign{\smallskip}
constant &  $\sthzero  \left(\frac{\nu}{\nu_0}\right)^{-0.1} +  \snthzero \left(\frac{\nu}{\nu_0}\right)^{-\anth} $  & $\sthzero$, $\snthzero$, $\anth$ \\\\
\vspace{0.1cm}
curved & $\sthzero  \left(\frac{\nu}{\nu_0}\right)^{-0.1} + \snthzero  \frac{\left(\frac{\nu}{\nu_0}\right)^{-\anth}}  {\left(\frac{\nu}{\nubreak}\right)^{0.5}+1}$   &  $\sthzero$, $\snthzero$, $\nubreak$, $\anth$ \\
\vspace{0.1cm}
break & $\sthzero  \left(\frac{\nu}{\nu_0}\right)^{-0.1} + \snthzero \left(\frac{\nu}{\nu_0}\right)^{-\anth-0.5}   \bigg[ 1-\Bigg( 1-\sqrt{\frac{\nu}{\nubreak}} \Bigg)^{\ginj-1}\bigg] $ ~~~for $\nu \le \nubreak$ & $\sthzero$, $\snthzero$, $\nubreak$, $\anth$ \\
          & $\sthzero  \left(\frac{\nu}{\nu_0}\right)^{-0.1} + S_{\rm nth,0} \left(\frac{\nu}{\nu_0}\right)^{-\anth-0.5} $ ~~~~~~~~~~~~~~~~~~~~~~~~~~~~~~~~~~~~~for  $\nu > \nubreak$  & \\\\
\vspace{0.2cm} 
cutoff & $\sthzero \left(\frac{\nu}{\nu_0}\right)^{-0.1} +  \snthzero  \left(\frac{\nu}{\nu_0}\right)^{-\anth} e^{-\frac{\nu}{\nubreak}}$ &  $\sthzero$, $\snthzero$, $\nubreak$, $\anth$ \\
\hline\noalign{\smallskip}
\end{tabular}
\tablefoot{We take $\nu_0 = 1$ GHz in this paper.}
\end{flushleft}
\end{table*}

\subsection{Fitting method}
\label{sect:fitmethod}

For each galaxy, we fit four models: constant, curved, break, and cutoff. The shapes of 
the model radio spectra are presented in the previous subsection and recapitulated in 
Table~\ref{tab:syn_models}, along with the list of free parameters. The only fixed 
parameter was the slope of the (optically thin) thermal emission, i.e. $\alpha_{th} 
= 0.1$. We require $S_{th}$ to be positive for all four models. In addition, we provide 
the following bounds for some parameters: $S_{nth} \ge 0$ and $1 \le \nu_b \le 50$~GHz 
for the break model and $\nu_b \ge 0$ for the cutoff model.

For each model, the best fit is obtained via the Trust Region Reflective ("trf") 
algorithm for optimization, well suited to efficiently explore the whole space of 
variables for a bound-constrained minimization problem \citep{MR1722103}. In 
App.~\ref{app:fits}, we provide more detailed information  and we show the best 
fits of all four models (constant, curved, break, and cutoff) for an easier 
comparison, together with a table of their best-fit parameters.

\section{Results}
\label{sect:results}

Fig.~\ref{results:spectra} shows the best fits out of the four tested cases (constant, 
curved, break, or cutoff, solid red line). In these plots, we also display the free-free 
(thermal, dotted cyan line) and synchrotron (nonthermal, dashed green line) components 
of the models. The observed radio continuum flux densities minus the fitted nonthermal 
component is represented by cyan stars, while the observed minus the fitted thermal 
component is depicted by green crosses. The optimal values of the parameters are 
listed in the lower left part of the figure, and their errors are given in terms of 
the standard deviation of 1000 generated models (for more detailed information see 
App.~\ref{app:fits}). 

In Table\,\ref{tab:fits} we compile the fit results obtained from the best-fit model, with 
the total flux density at 1~GHz and its error listed in Col.\,2, the fraction of thermal 
emission at 1~GHz and its error in Col.\,3, the nonthermal spectral index and its error 
in Col.\,4, the break frequency in GHz and its error in Col.\,5, the reduced $\chi_\nu^2$ 
in Col.\,6, and the name of the best-fit model in Col.\,7.

\begin{table*}
\caption{Fit results}
\label{tab:fits}
\begin{center}
\begin{tabular}{lrccccr}
\hline
Galaxy & $S_{tot,\ 1\,\rm GHz}$ & $f_{th,\ 1\,\rm GHz}$ & $\alpha_{nth}$ & $\nu_b$ & $\chi^2_\nu$ & best fit \\
      &        [mJy]           &                       &               & [GHz]   &              & \\
\hline
II\,Zw\,40 & $ 32 $ & $ 0.80 $ & $ 0.35 $ & $ - $ & $ - $ & - \\
II\,Zw\,70 & $ 5.2 \pm 0.2 $ & $ 0.62 \pm 0.03 $ & $ 1.15 \pm 0.12 $ & $ - $ & $ 0.33 $ & constant \\
IC\,10 & $ 446 \pm 40 $ & $ 0.21 \pm 0.07 $ & $ 0.58 \pm 0.08 $ & $ - $ & $ 0.18 $ & constant \\
NGC\,1569 & $ 494 \pm 12 $ & $ 0.22 \pm 0.02 $ & $ 0.42 \pm 0.02 $ & $ 12.4 \pm 2.4 $ & $ 0.12 $ & cutoff \\
NGC\,4449 & $ 342 \pm 32 $ & $ 0.27 \pm 0.03 $ & $ 0.40 \pm 0.08 $ & $ 4.7 \pm 0.9 $ & $ 0.17 $ & cutoff \\
\hdashline
NGC\,4490 & $ 1051 \pm 206 $ & $ 0.13 \pm 0.04 $ & $ 0.57 \pm 0.09 $ & $ 6.8 \pm 1.4 $ & $ 0.20 $ & cutoff \\
NGC\,4631 & $ 1637 \pm 75 $ & $ 0.14 \pm 0.01 $ & $ 0.57 \pm 0.03 $ & $ 6.6 \pm 0.4 $ & $ 2.08 $ & cutoff \\
NGC\,5194 & $ 1788 \pm 464 $ & $ 0.11 \pm 0.03 $ & $ 0.67 \pm 0.10 $ & $ 7.2 \pm 0.6 $ & $ 0.62 $ & cutoff \\
\hdashline
NGC\,4038 & $ 683 \pm 160 $ & $ 0.13 \pm 0.04 $ & $ 0.71 \pm 0.09 $ & $ 8.2 \pm 3.8 $ & $ 0.01 $ & break \\
NGC\,6052 & $ 129 \pm 10 $ & $ 0.08 \pm 0.01 $ & $ 0.56 \pm 0.04 $ & $ 2.4 \pm 0.5 $ & $ 0.32 $ & break \\
\hdashline
NGC\,2146 & $ 1359 \pm 18 $ & $ 0.16 \pm 0.01 $ & $ 0.51 \pm 0.02 $ & $ 6.2 \pm 0.5 $ & $ 0.47 $ & cutoff \\
NGC\,3034 & $ 9043 \pm 175 $ & $ 0.14 \pm 0.01 $ & $ 0.40 \pm 0.02 $ & $ 11.2 \pm 1.0 $ & $ 0.45 $ & cutoff \\
NGC\,3079 & $ 1111 \pm 104 $ & $ 0.10 \pm 0.02 $ & $ 0.74 \pm 0.01 $ & $ 9.0 \pm 2.0 $ & $ 1.59 $ & break \\
NGC\,3310 & $ 470 \pm 16 $ & $ 0.19 \pm 0.01 $ & $ 0.60 \pm 0.03 $ & $ 1.2 \pm 0.3 $ & $ 0.40 $ & break \\
\hline
\end{tabular}
\end{center}
\end{table*}

In general, the best fit was selected as the lowest reduced chi-squared, $\chi_{\nu}^2$.
Apart from this, the resulting thermal emission was an asset of whether the fits are 
physically meaningful. Thus, we rejected fits in spite of an acceptable  $\chi_{\nu}^2$ 
if the fitted thermal radio emission was unphysically low. In addition, an important 
test was to look at the resulting thermal flux densities obtained at each frequency 
after subtracting the computed nonthermal flux density from the total (observed). A fit 
makes only sense if the resulting thermal flux densities so obtained obey to the expected 
optically thin spectrum with slope $-0.1$. This is shown by the blue data points in each 
diagram, with the errors taken from those of the measured total flux densities. 

The most striking result of our fits is that most objects need strongly curved (break
or cutoff) synchrotron spectra in order to satisfactorily fit the total radio spectrum.
Apparent exceptions are the three dwarf galaxies IC~10, II~Zw~40, and II~Zw~70. IC~10 
and II~Zw~70 can be well fitted by a constant synchrotron spectrum, while in case of 
II~Zw~40 it is only at the lowest frequencies that nonthermal emission becomes evident. 
We therefore decided to fit a power-law only to its high-frequency data, obtaining an 
optically thin thermal spectrum with slope $\alpha_{th} = 0.10 \pm 0.02$. This spectrum 
is then extrapolated to the lower frequencies at which we retrieve the nonthermal flux 
densities by subtracting the extrapolation from the measured fluxes. In 
Fig.\,\ref{results:spectra} the dashed cyan line shows the fit to the three high-frequency 
data, yielding a perfect optically thin thermal spectrum. At the two low frequencies we 
have subtracted the extrapolated thermal flux densities from those observed, such as to 
yield the nonthermal spectrum. 

We have also produced fits to this spectrum in which $\anth$ is limited to 1.0, 0.9, 
0.8, 0.7, and 0.6. We note that $\anth$ may take values as low as 0.8, and the total 
fit is still compatible with the error bars of the observational data. The thermal 
spectrum is  a little lower in this case and, as expected, $\chi^2_\nu$ is worse. 
Values of 0.7 and 0.6 make the total fit to go below the first data point. We also 
tried to fit with maximum values of 0.5 and 0.4, but the fits did not converge.

Obviously, the BCDGs and IC\,10 are dominated by the thermal radio emission ($\fth \ge 0.2$ 
at 1~GHz) so that we cannot draw any firm conclusions about the shape of the synchrotron 
spectra at all. 

Also for NGC~3079 a straight or curved synchrotron spectrum formally gives the best fit, 
and for NGC~4038, all 4 fits are acceptable. However, the thermal flux density obtained 
for the constant/curved fits are rather low. Compared to the \halpha\ data they would 
require internal extinctions of 2\fmag 5 (for NGC~3079) and 1\fmag 0 (for NGC~4038) 
which seems rather low for these edge-on, highly extinguished galaxies.

For 10 of the 14 galaxies, the best-fit spectrum is that with either a break or a 
cutoff. For 5 galaxies, the $\chi_{\nu}^2$ values are very similar in both cases (within 
a factor of 1.5), and for all 10 galaxies both break and cutoff give acceptable fits. 
{\it In addition, the broad radio range covered by our data in most of these galaxies 
renders both, the steepening of the spectrum starting at $\sim 1 - 5$~GHz, due to 
the strong curvature of the synchrotron emission, and the subsequent flattening due 
to the dominance of the thermal radio emission (above $\sim$ 10 GHz) visible in the 
measured spectra!} The most noticeable cases are NGC~2146, NGC~4449, NGC~4631, and 
NGC~5194, but also in NGC~1569, NGC~3034, and NGC~3310 this trend is visible. The 
fact that we directly observe the steepening and subsequent flattening in the data 
lends support for the reality of the pronounced curvature of the synchrotron spectra 
that our fitting procedure yields.

\begin{center}

\begin{figure*}
\vspace{-0.35cm}
\centerline{
\im{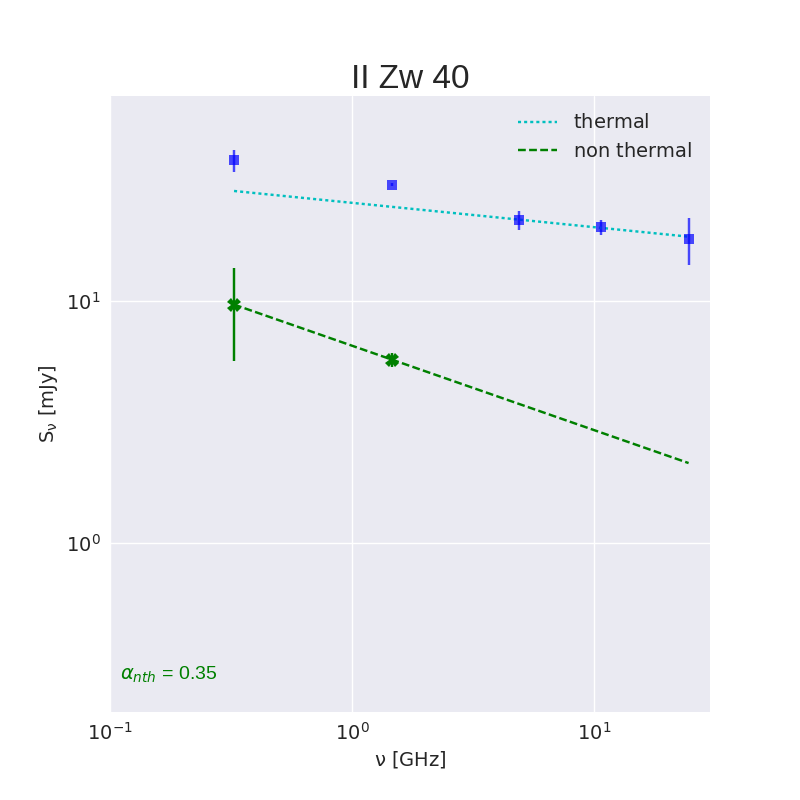}\quad
\im{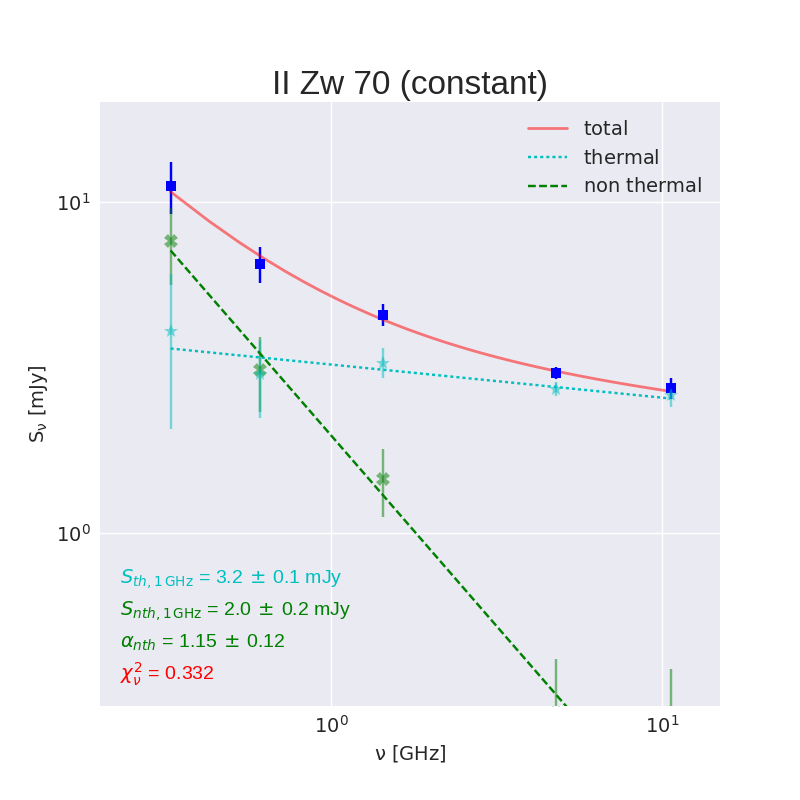}
}
\vspace{-0.40cm}
\centerline{
\im{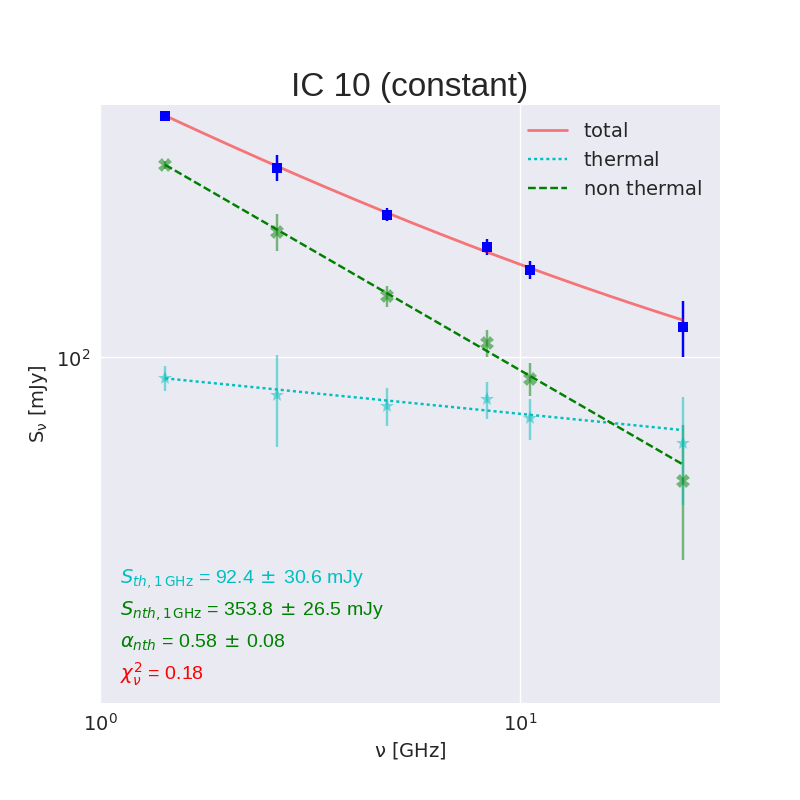}\quad
\im{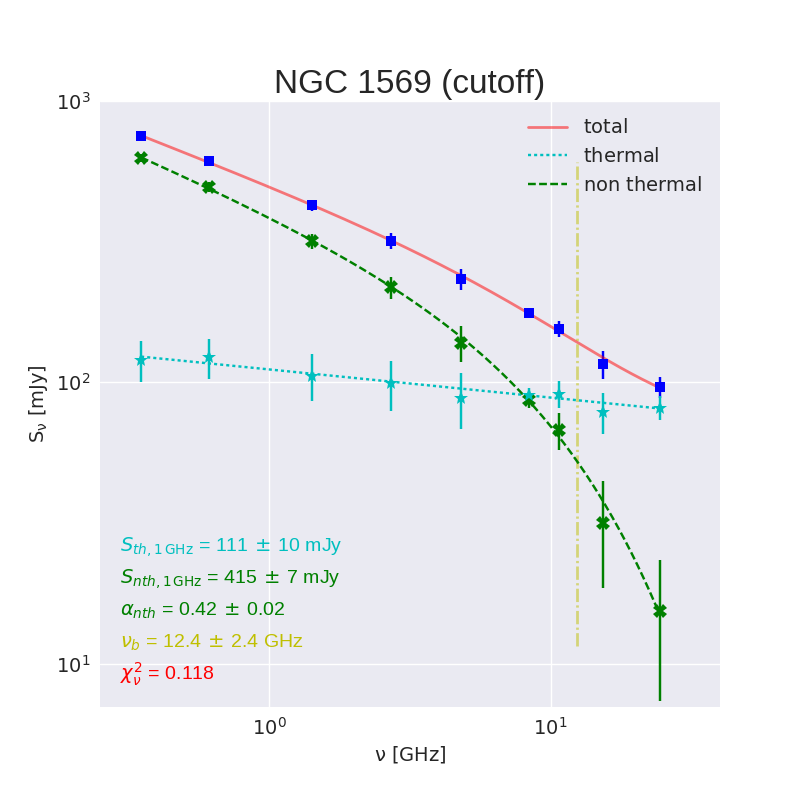}
}
\vspace{-0.45cm}
\centerline{
\im{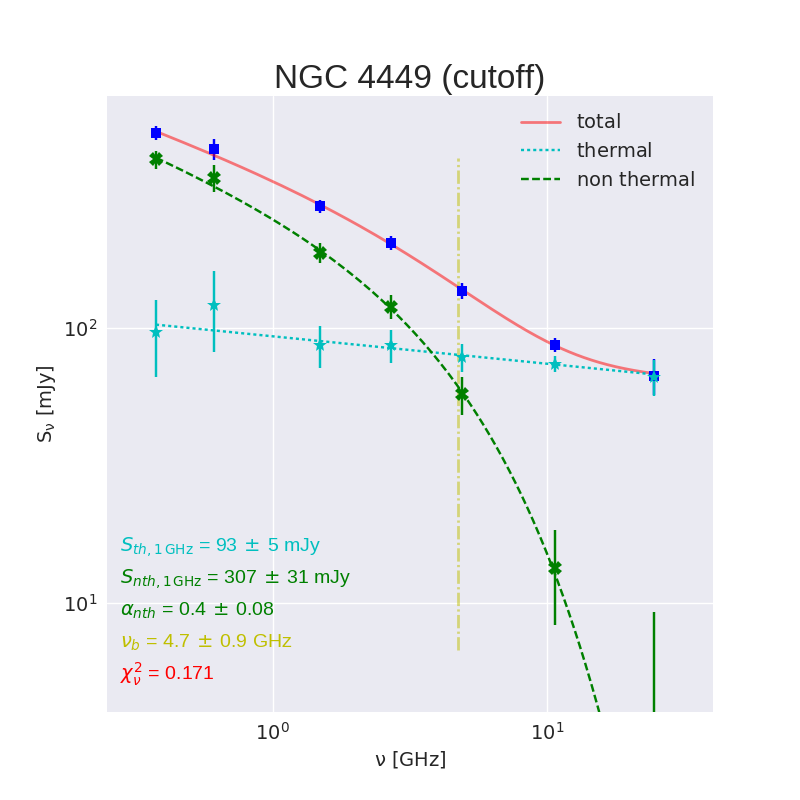}\quad
\im{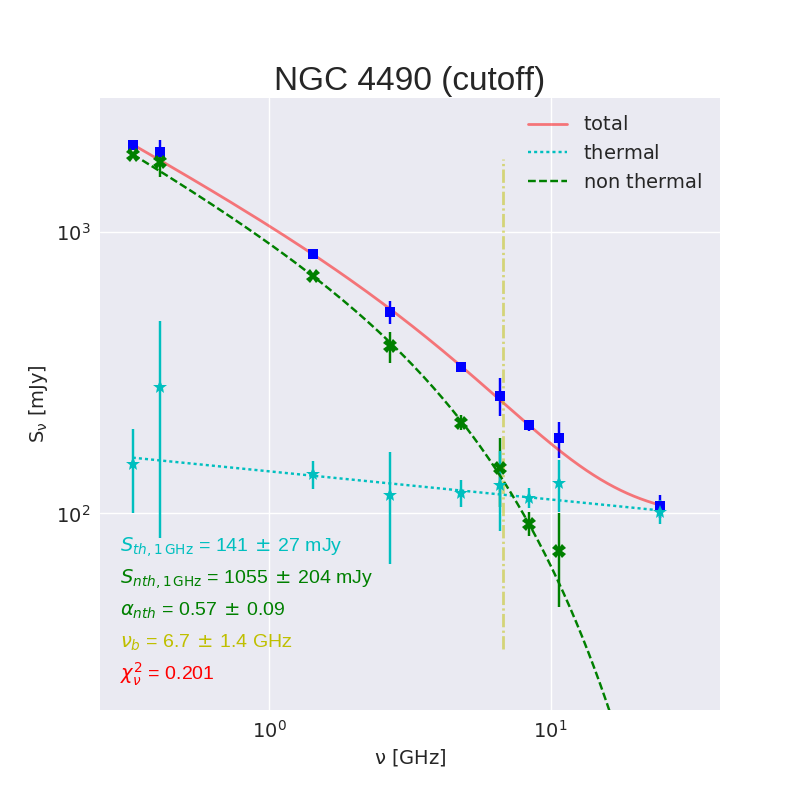}
}
\vspace{-0.2cm}
\caption{Radio continuum spectra of the sample galaxies. Except for II\,Zw\,40 (see 
text for the procedure and plotted lines here), the plots show the following: 
measured flux densities (blue squares), best fit of the radio continuum spectrum 
(solid red line). The best parameters are listed in the lower left part of the 
figure along with the reduced chi-square ($\chi^2_\nu$). The free-free (thermal) 
and synchrotron (nonthermal) components of the models are depicted by the dotted 
cyan line and dashed green line, respectively. The green crosses represent the 
observed nonthermal component (i.e., the modelled thermal component removed from 
the observational data). The cyan stars delineate the observed thermal component 
(i.e., the modelled nonthermal component subtracted from the observational data).
The vertical yellow dash-dotted line marks the break frequency $\nubreak$.}
\label{results:spectra} 
\end{figure*}

\clearpage
\newpage

\begin{figure*}
\vspace{-0.25cm}
\centerline{
\im{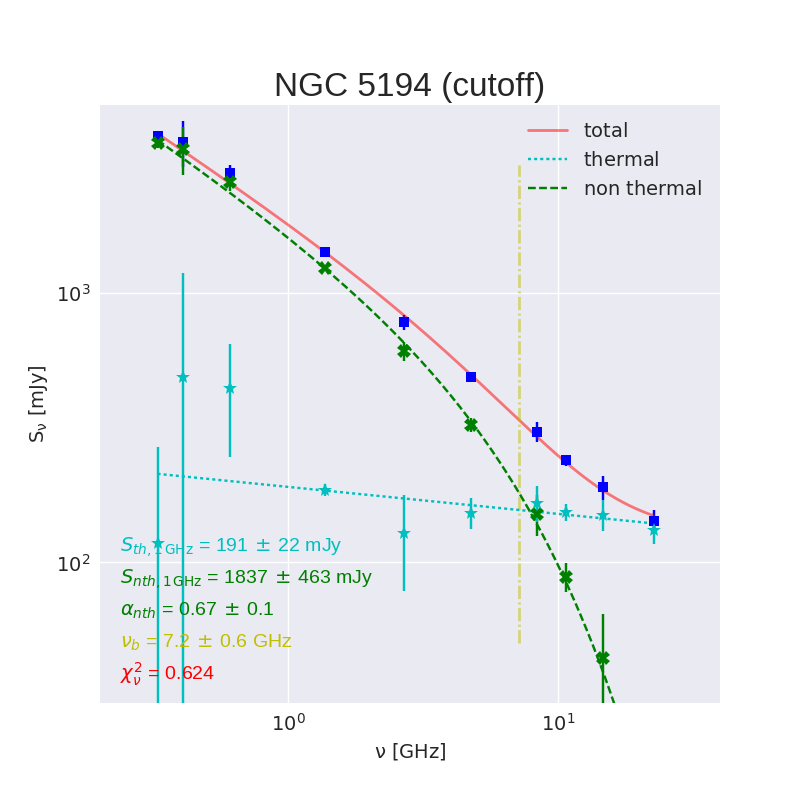}\quad
\im{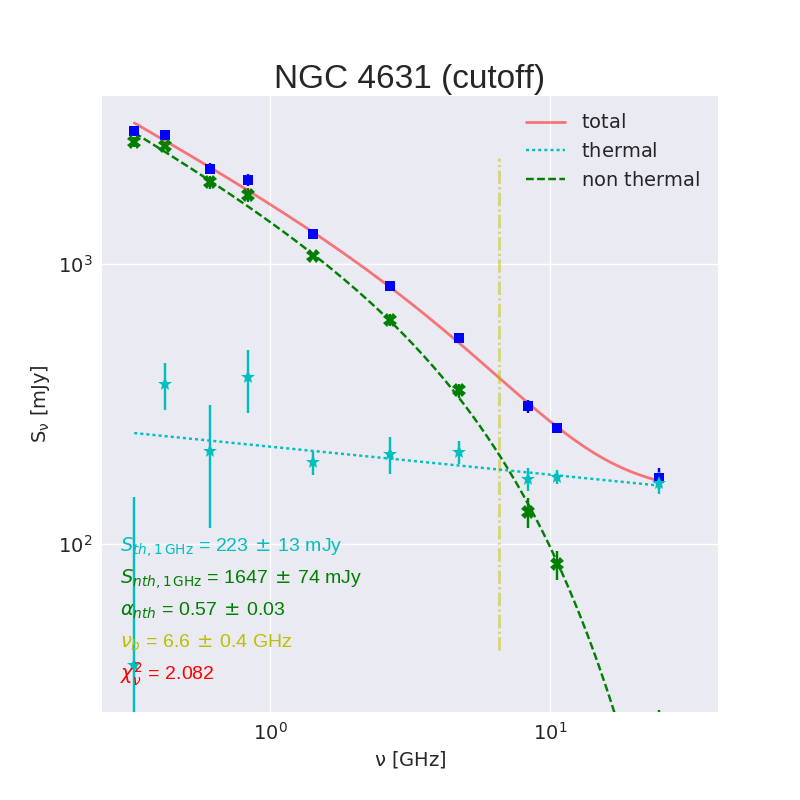}
}
\vspace{-0.4cm}
\centerline{
\im{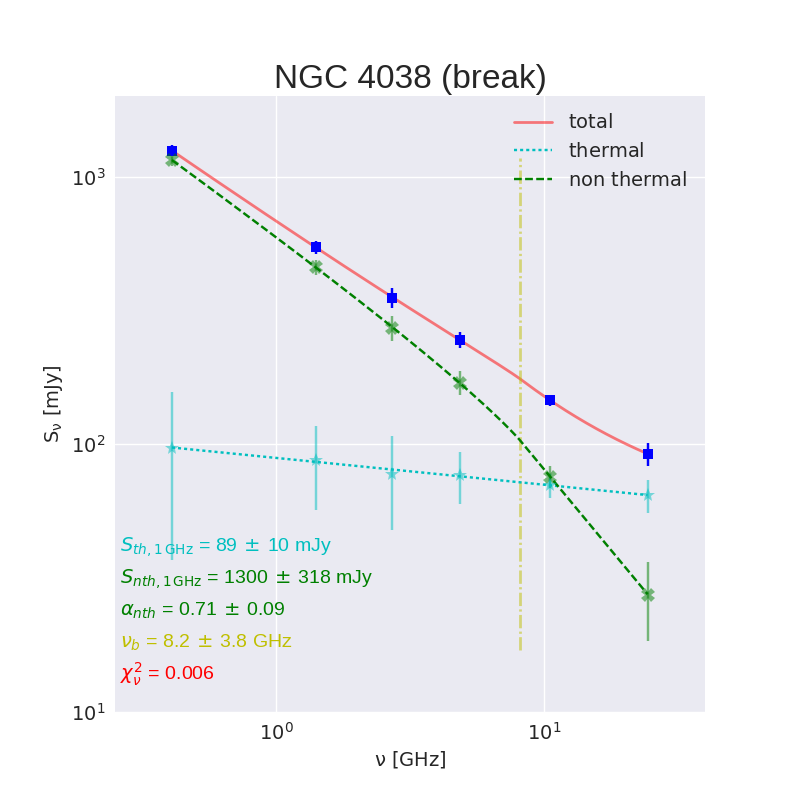}\quad
\im{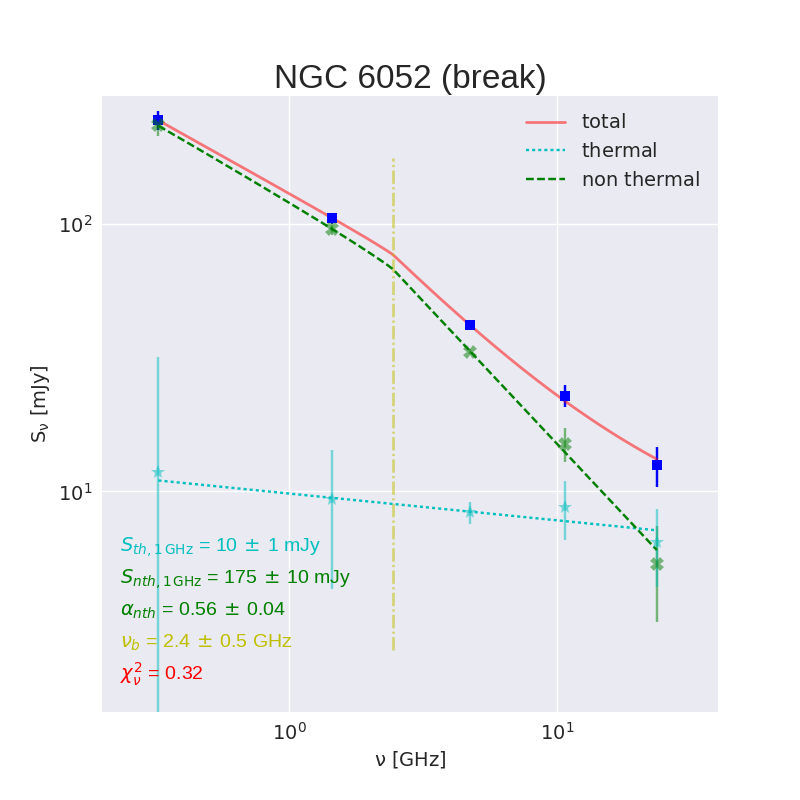}
}
\vspace{-0.4cm}
\centerline{
\im{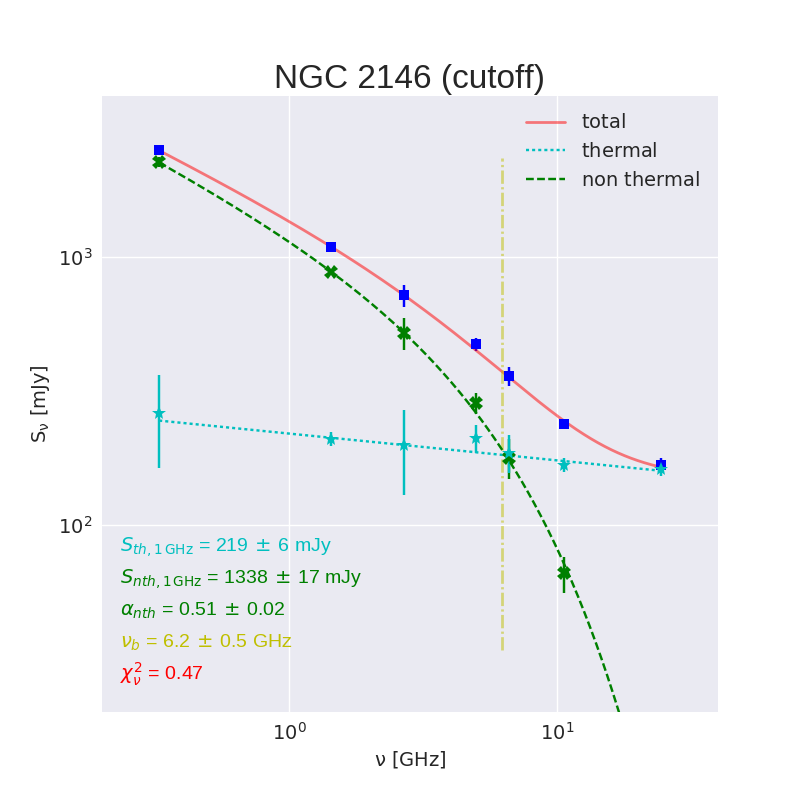}\quad
\im{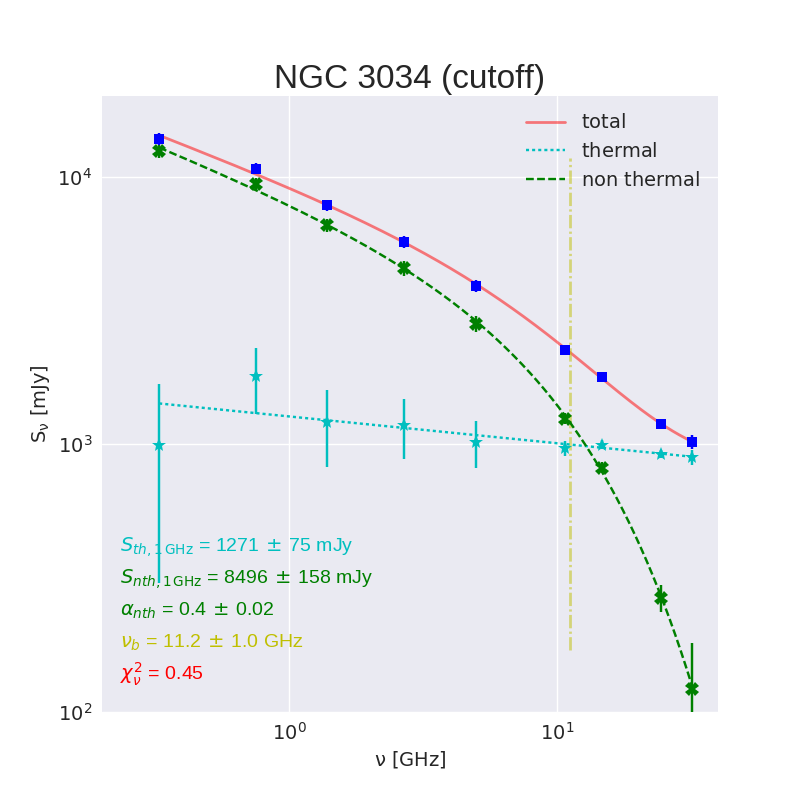}
}
\addtocounter{figure}{-1} 
\caption{(continued)} 
\end{figure*}

\begin{figure*}
\im{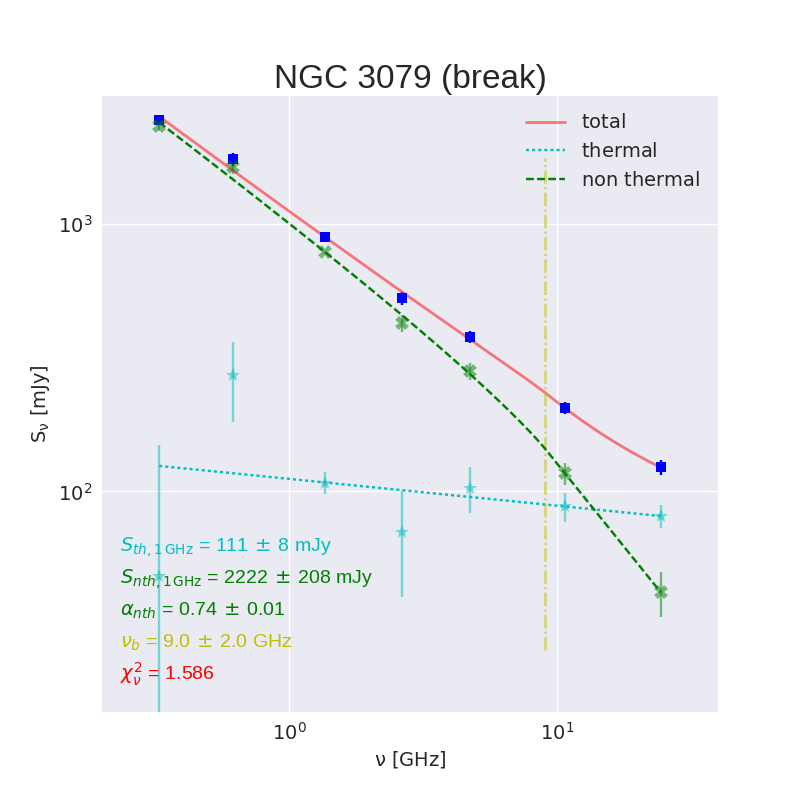}\hfill
\im{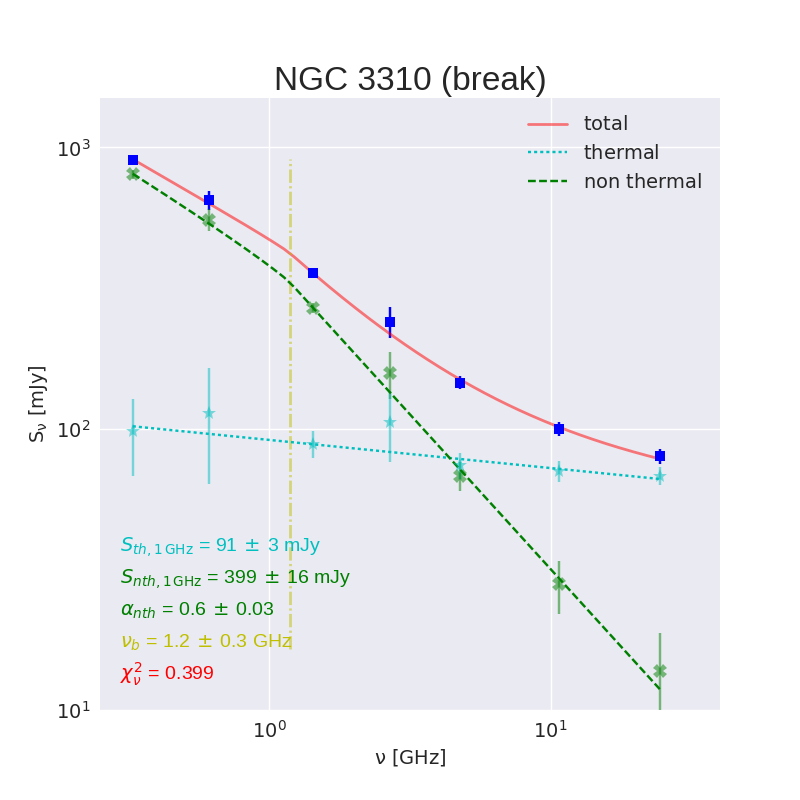}\\
\addtocounter{figure}{-1} 
\caption{(continued)} 
\end{figure*}

\end{center}

\section{Discussion}
\label{sect:disc}

\subsection{Comparision of thermal radio and \halpha\ emission}
\label{subsect:halpha} 

We have used the thermal flux densities resulting from our spectral 
decomposition to compare them with what is predicted from the observed 
\halpha\ fluxes corrected for \ion{N}{ii} emission and Galactic extinction 
(Table\,\ref{tab:ancillary_data}). We used the relation derived by 
\citet{1980stfo.conf...77L}, converted from H$\beta$ to \halpha: 

\begin{equation}
S_{\rm th,H\alpha} = 1.14 \cdot 10^{12}  \left(\frac{\nu}{\rm GHz}\right)^{-0.1} 
\left({\frac{T_{\mathrm e}}{10^4 ~\rm K}}\right)^{0.34} 
\left[\frac{F({\rm H\alpha})}{\rm erg ~s^{-1} ~cm^{-2}}\right] \,\,\, \rm mJy. 
\end{equation}

The extinction is then calculated via 

\begin{equation}
A({\rm H\alpha}) = -2.5 \, \log \left(\frac{S_{\mathrm th,H\alpha}}{S_{\mathrm th,fit}}\right).
\label{eqn:ext}
\end{equation}

\begin{figure}[h]
\includegraphics[width=8cm, angle=0]{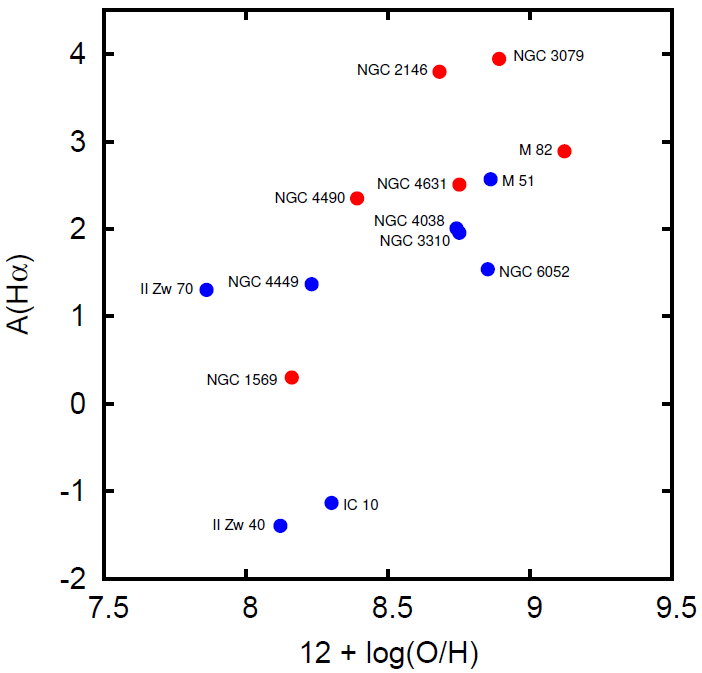}
\caption{\halpha\ extinction vs. metallicity. Highly inclined ($>75$ degrees) 
and edge-on galaxies are marked with red dots}
\label{fig:ext}
\end{figure}

Since extinction is caused by the dust in the galaxy planes, we have plotted 
in Fig.\,\ref{fig:ext} the extinction resulting from Eqn.\,\ref{eqn:ext} vs. 
the metallicity, for which we have collected the quantity $12 + \log(\rm O/H)$ 
from the literature (Table\,\ref{tab:ancillary_data}). As is to be expected, 
there is a trend of increasing extinction with increasing metallicity, with 
the highly inclined spiral galaxies having the largest extinctions. Dwarf 
galaxies are known to possess lower metallicities and thus have a low dust 
content, hence they are found in the lower left part of the diagram. The 
values for IC\,10 and II\,Zw\,40 are even negative, which is most likely 
due to an overestimate of the Galactic extinction which is very high in 
both cases (3\fmag 4 for IC\,10 and 1\fmag 8 for II\,Zw\,40). An 
underestimate of the thermal radio emission is unlikely in these 
two cases because for II\,Zw\,40 the thermal radio emission is 
directly given by the high-frequency data and for IC\,10 the 
result for the thermal fraction yields very similar values 
for all fits.

\subsection{Thermal fraction}
\label{subsect:fth}

\begin{figure}[h!]
\includegraphics[width=8cm, angle=0]{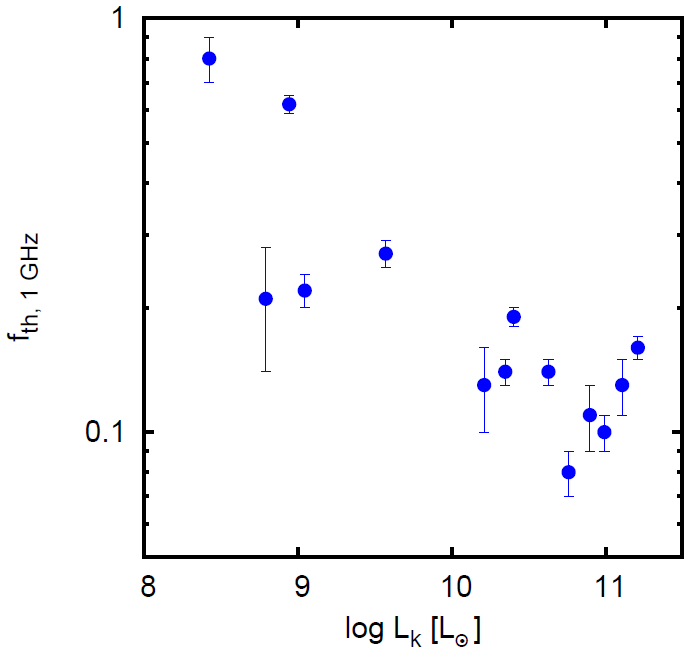}
\caption{Fraction of thermal emission at 1~GHz vs. K-band luminosity, the 
errors resulting from our fits (see Table\,\ref{tab:fits}).}
\label{fig:fth_Lk}
\end{figure}

In Fig.\,\ref{fig:fth_Lk} we show the fraction of thermal emission at 1~GHz 
resulting from our spectral fits, plotted vs. the K-band luminosity of the 
galaxies. Since this luminosity is a measure for the stellar mass, the 
diagram clearly indicates a decreasing relative amount of nonthermal 
radiation as we move to small stellar masses. In fact, the left half 
of this plot contains all the dwarf galaxies in our sample. This result 
corroborates the early conjecture of \citet{klein91} that the lowest-mass 
galaxies are unable to retain the recently produced cosmic rays. Alternatively, 
the high thermal fraction in the most extreme galaxies II~Zw~40 and II~Zw~70 
might be the result of a temporal effect. Both galaxies are currently 
experiencing a strong and very recent starburst, with an age of only 
$3-5$~Myr, derived from the presence of Wolf-Rayet feature in their spectra. 
This means that the newly formed, ionizing stars are contributing to the 
thermal radio emission, but no supernovae may have gone off yet. Hence, 
the observed synchrotron radiation may have been produced during a previous 
epoch of star formation. We also compare the thermal fraction with the SFR 
(from Tab.~\ref{tab:ancillary_data} but found no correlation.

It is important to note here that for some of the galaxies in our sample the 
amount of thermal emission resulting from our analysis is higher than 
thought hitherto, which has consequences for any conjectures based upon the 
thermal radio continuum in star-forming galaxies. This is particularly true for
higher frequencies where the cutoff/break lowers the synchrotron emission
considerably. At 10 GHz, the mean thermal fraction for the galaxies is 0.6 --
considerably more than what was believed so far.

\subsection{Injection spectra}
\label{subsect:nub}

\begin{figure}[h]
\includegraphics[width=8cm, angle=0]{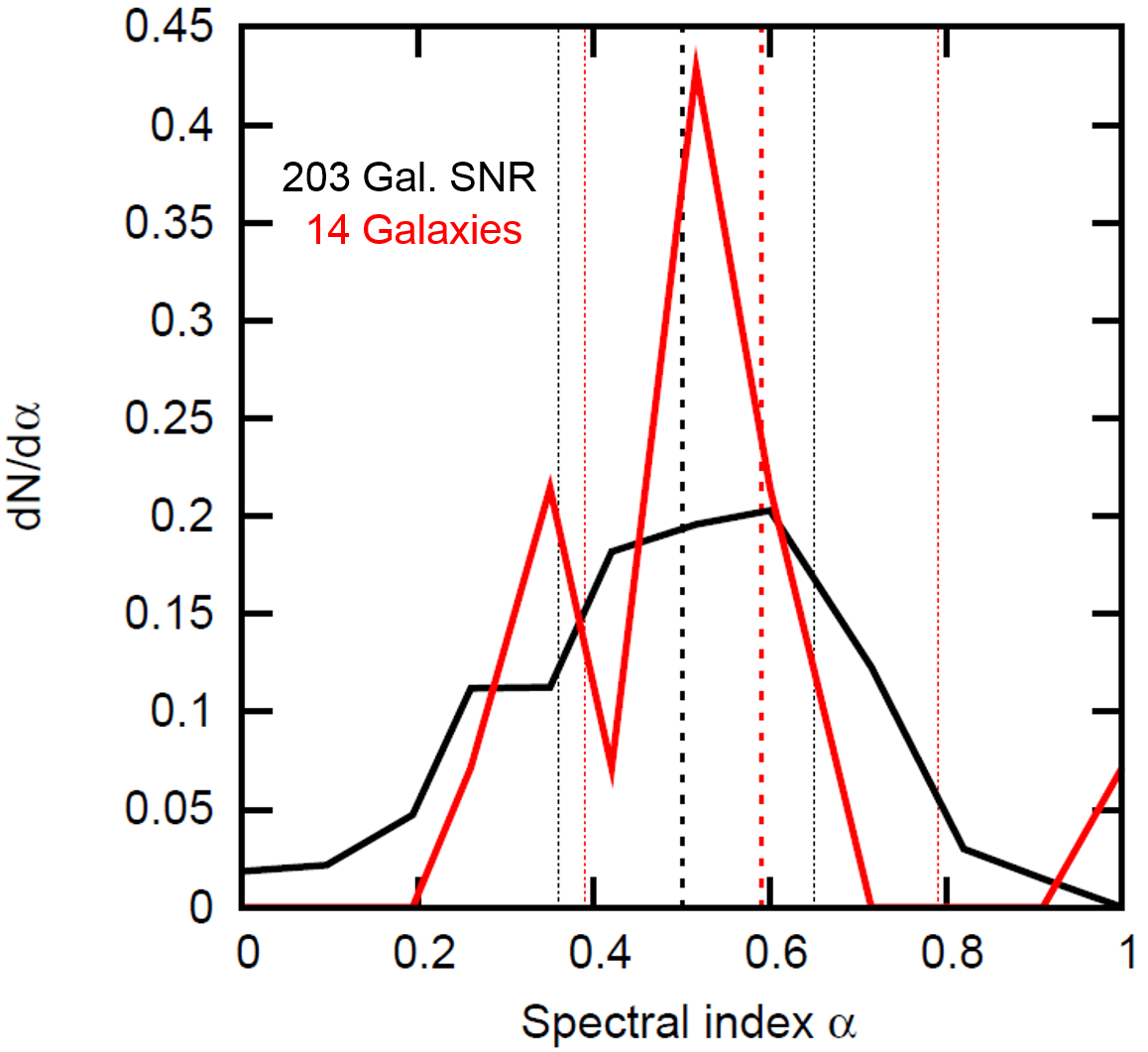}
\caption{Histogram of the spectral indices of the low-frequency synchrotron radiation 
of our sample galaxies (red) and of Galactic supernova remnants (black). The vertical 
dashed lines indicate the mean (thick) and variance (thin) of each distribution.}
\label{fig:snr_hist}
\end{figure}

In Fig.\,\ref{fig:snr_hist} a superposition of the low-frequency spectral indices of 
the synchrotron radiation of our galaxy sample with the spectral indices of Galactic 
supernova remnants (SNR) is shown, the latter data taken from the catalogue of 
\citet{green14}. The two distributions are rather similar, albeit the statistics are 
vastly different (203 vs. 14 objects). The resulting values of the mean and standard 
deviation are $<\alpha_{\rm SNR}> = 0.50$, $\sigma_{\alpha_{\rm SNR}} = 0.33$ and 
$<\alpha_{\rm Gal}> = 0.59$, $\sigma_{\alpha_{\rm Gal}} = 0.20$ for the SNR and galaxies, 
respectively. Both, the mean and the rms are rather similar, possibly suggesting that 
the synchrotron spectra that we measure at low radio frequencies reflect the injection 
spectra of the SNR. 

An outlier is II~Zw~70, which has a low-frequency spectral index of $\alpha_{\rm nth} = 
1.15\pm0.12$. Such a steep spectrum is indicative of the impact of synchrotron and 
inverse-Compton losses which have steepened the injection spectral index by $+0.5$. 
It is unclear why  II~Zw~70 is so different from the rest of the galaxies. The SFR 
per \lk, which can be taken as a measure of the capability of the galaxy to expell 
material into the halo, is similar to that of the other starburst galaxies, such 
as M~82 or NGC~1569 (but more than a factor of 10 lower than for the otherwise 
similar BCDG II~Zw~40).

\begin{table}[h!]
\caption{Radio sizes, magnetic fields, and cutoff/break energies} 
\label{tab:B_E}
\begin{center}
\begin{tabular}{lrrrr}
\hline
Galaxy 	   & Radio Size 	& $B$\tablefootmark{1}  ~~	& $E$ ~~~~    	& Shape 	\\
       	   & [kpc] ~~~		& [$\mu$G]  	& [\rm GeV]  	& 		\\
\hline
II\,Zw\,40 &  	 0.8~~~~~~	&	29~~	&	-~~~~	& -~~~~~ 	\\ 
II\,Zw\,70 & $<$ 0.6~~~~~~	&   $<$ 40~~	&	-~~~~	& constant	\\
IC\,10 	   & 	 1.3~~~~~~	&	14~~	&	-~~~~	& constant	\\
NGC\,1569  & 	 2.5~~~~~~	&	16~~	&	6.8~~	& cutoff	\\
NGC\,4449  & 	 5.0~~~~~~	&	14~~	&	4.6~~	& cutoff	\\
\hdashline
NGC\,4490  &   	  10~~~~~~	&	20~~	&	4.6~~	& cutoff	\\
NGC\,4631  &   	  22~~~~~~	&	13~~	&	5.6~~	& cutoff	\\
NGC\,5194  &   	  15~~~~~~	&	18~~	&	5.0~~	& cutoff	\\
\hdashline
NGC\,4038  & 	  18~~~~~~	&	27~~	&	-~~~~	& break		\\
NGC\,6052  &   	 4.0~~~~~~	&	71~~	&	-~~~~	& break		\\
\hdashline
NGC\,2146  & 	  10~~~~~~	& 	40~~	&	3.1~~	& cutoff	\\
NGC\,3034  & 	 1.7~~~~~~	&	66~~	&	3.3~~	& cutoff	\\
NGC\,3079  & 	  13~~~~~~	&	32~~	&	-~~~~	& break		\\
NGC\,3310  & 	 4.4~~~~~~	&	39~~	&	-~~~~	& break		\\
\hline
\end{tabular}
\end{center}

\tablefoottext{1} {The magnetic field is calculated with the minimum energy assumption.}

\end{table}

\subsection{Spectral breaks and cutoffs}
\label{subsect:breaks_cutoffs}

For the majority of our galaxies, the fitting of their radio spectra shows the 
need of a break of the synchrotron spectra in the range of $1 - 12$~GHz, corresponding 
to particle energies of $1.5 - 7$~GeV, depending on the magnetic-field strengths (see 
Table\,\ref{tab:B_E}). The low-frequency spectrum in all cases except II~Zw~70 has 
a spectral index in the range of what is found for SNRs, suggesting that the 
relativistic electrons emitting in this range have not suffered any significant 
synchrotron and inverse-Compton losses that would steepen their spectrum. The 
steep high-frequency spectrum indicates that synchrotron or inverse-Compton 
losses are important (in the case of a break) or a complete lack of relativistic 
electrons above $\sim 5$ GeV (in the case of a cutoff).

As outlined in Sect.~\ref{subsect:radio_spec}, a sharp break can be explained 
in an open box-model with convection. The break is produced because high-energy 
electrons, emitting at frequencies above the break, suffer substantial synchrotron 
or inverse-Compton losses before they arrive at the edge of the halos, whereas 
low-energy electrons do not. We can estimate the half-lifetime of the relativistic 
electrons emitting the synchrotron radiation at the break freqeuncy, which is 
determined by synchrotron and inverse-Compton losses \citep[see, e.g.,][]{klein15}: 

\begin{equation}
t_{1/2} = 1.25 \cdot 10^{10} \cdot \left [
\left( \frac{B}{\mu \rm G} \right)^2 + \left( \frac{B_{eq}}{\mu \rm G} 
\right)^2 \right]^{-1} \cdot \left( \frac{E}{\rm GeV} \right)^{-1} \,\, 
\rm yr \; .
\label{tau} 
\end{equation} 

Here, $E$ is the energy of the relativistic particles, $B$ is the total 
magnetic-field strength, and $B_{eq}$ is the field strength equivalent 
to the energy density of the local radiation field. The latter can be 
obtained via 

\begin{equation}
B_{eq} =  \sqrt{\frac{32}{c}} \cdot \frac{L_{bol}^{1/2}}{d}\; . 
\label{beq}
\end{equation} 

Here, $L_{bol}$ is the bolometric luminosity of a region of size $d$. 
\citet{kennicutt07} have measured the quantity $\nu \cdot \, L_{\nu}$ 
at 24~$\mu$m (which approximates the luminosity) for a number of \ion{H}{ii} 
regions in M\,51, with typical 
values of $10^{41}$~erg~s$^{-1}$, obtained with aperture sizes of 13\arcsec\ 
(hence, $d = 500$~pc). Plugging this into Eqn.\,(\ref{beq}), we obtain $B_{eq} 
\approx 7 ~\mu$G. The strength of the magnetic field in our sample galaxies 
is between $B = 10 ~\mu$G and $B = 70 ~\mu$G (Table\,\ref{tab:B_E}). This 
implies a range for the particle half-lifetimes of between $7.5 \cdot 
10^5$~yr and $2.5 \cdot 10^7$~yr (Tev and PeV particles enter the GeV 
regime on time scales much shorter than any dynamical time scales under 
conditions considered above). Assuming a vertical halo size of about 250~pc, 
this implies convection speeds of between 10 and 300~km~s$^{-1}$, which is 
reasonable. In order to test this interpretation, it would be useful to carry
out a similar analysis for quiescent galaxies with a low star-formation rate
per area for which we would expect much lower vertical propagation of the
relativistic electrons.

A cutoff in the relativistic electron distribution is more difficult to 
explain. Energy losses can produce a cutoff only in a single-injection 
scenario where radiative ageing depopulates the high-energy part of the 
relativistic electron distribution with time. This situation is unrealistic 
for the integrated radio emission of an entire galaxy. The acceleration 
process of cosmic rays in SNRs is effective up to TeV energies and the 
synchrotron spectra of young SNRs can be observed up to X-ray energies. 
At high energies, the short synchrotron energy loss time in the strong 
magnetic fields ($> 100~\mu$G) in SNR \citep{voelk05} is expected to 
produce a steepening of the electron injection spectrum. This can be 
seen from Eqn.~\ref{tau}, which for a particle energy of, say, 100~GeV, 
in a 100~$\mu$G magnetic field yields a lifetime of $\sim 10^4$~yr. 
This is comparable to the Sedov phase of a SNR ($\tau \sim 3 \cdot 10^4$~yr)
so that at energies above $\sim 100$~GeV synchrotron cooling should be 
relevant and produce a steeper injection spectrum above $\sim 2 \cdot 
10^{13}$~Hz. However, the breaks/cutoffs that we infer are at much lower 
particle energies and do therefore not probe this process. Thus, in the 
framework of the simple models considered here, it is unclear which 
process could produce a cutoff in the synchrotron spectrum. It is,
however, obvious that this must have to do with the relative time 
scales of energy gain, losses and escape of the relativistic electrons,
as discussed by \citet{schlickeiser84}.

\section{Summary and conclusions}
\label{sect:sum}

We have analyzed the radio continuum spectra of 14 star-forming galaxies by 
fitting nonthermal (synchrotron) and thermal (free-free) radiation laws to 
carefully selected measurements, covering a frequency range of $\sim$300~MHz 
to 24.5~GHz (32~GHz in case of M\,82). The 24.5-GHz measurements, mostly 
unpublished to date, are crucial for a more reliable separation of the 
thermal and nonthermal components in this analysis. 

We find that the majority of the synchrotron spectra are not simple power-laws 
as believed hitherto. The curved shape of the synchrotron spectrum is clearly 
visible in many of the total radio spectra, which steepen over a range of 
several GHz and flatten at higher frequencies due to the thermal radio
emission. Our fitting shows that the lowest values of the reduced $\chi_\nu^2$ 
result for synchrotron spectra with a mean slope $\alpha_{nth} = 0.59 \pm 0.20$ 
in the low-frequency regime, and a break or an exponential decline in the frequency 
range of $1- 12$~GHz. There are only one galaxy that shows pure power-laws (the 
dwarf galaxy IC\,10). In the case of the BCDGs II~Zw\,40 and II~Zw\,70 only the 
slope (and not the shape) of the synchrotron spectrum can be worked out, since 
a deviation from a purely thermal spectrum is only evident at the lowest 
frequencies involved. 

For the bulk of the sample galaxies, simple power-laws or mildly curved 
synchrotron spectra lead to unrealistic low thermal flux densities 
(frequently $\sth = 0$), and/or to strong deviations from the expected 
optically thin free-free spectra with slope $\alpha_{th} = 0.10$ in the fits. 
Assuming energy equipartition between relativistic particles and magnetic 
fields, the cutoff and break frequencies translate into energies in the 
range of $1.5 - 7$~GeV. The average spectral index of the low-frequency 
spectra obtained here is comparable to that found for Galactic (shell-type) 
supernova remnants. 

A comparison of the thermal flux densities resulting from our fits with 
the (foreground-corrected) H$\alpha$ fluxes yields the extinction, which 
increases with metallicity. The fraction of thermal emission at 1~GHz is 
higher in some of our galaxies than believed hitherto, and the discrepancy 
increases towards higher frequencies where the mean thermal fraction is 
$\sim 0.6$. It is highest in the dwarf galaxies of our sample, which we 
interpret either in terms of a lack of containment of synchrotron emission 
in these low-mass systems, or alternatively, in the case of II~Zw\,40 and 
II~Zw\,70, as a time effect due to a very young starburst. 

The asymptotic low-frequency synchrotron spectra derived here provide a firm 
‘leverage’ for low-frequency studies, e.g. with LOFAR. In order to significantly 
incraese the number of radio continuum spectra of galaxies allowing analyses as 
presented here, expensive mapping with large single-dish telescopes (Effelsberg, 
GBT, Sardinia Radio Telescope) in the frequency range of 25~--~40~GHz (so-called 
Ka-band in radio astronomy) are indispensible. 

An interpretation of the rapid declines in the synchrotron spectra (break or
exponential cutoff) at GeV particle energies is not obvious, since the 
relativistic particles gain energies up to the TeV range within the supernovae. 
In a simple model, a convective wind carrying away the low-energy relativistic 
electrons could explain a sharp break. The break energy must depend on the relative
time scales of energy gain (acceleration), losses (synchrotron and inverse-Compton)
and escape of the relativistic electrons.

\begin{acknowledgements}

We wish to thank H. Lesch and R. Schlickeiser for helpful discussions. UK 
acknowledges financial support by the German Deutsche Forschungsgemeinschaft, 
DFG project FOR\,1254, and is very grateful for the kind hospitality at the 
Departamento de F\'isica Te\'orica y del Cosmos, Universidad de Granada. UL 
and SV acknowledge support by the research projects AYA2014-53506-P from the 
Spanish Ministerio de Econom\'\i a y Competitividad, from the European Regional 
Development Funds (FEDER) and the Junta de Andaluc\'ia (Spain) grants FQM108.
This research has made use of the NASA/IPAC Extragalactic Database (NED), which 
is operated by the Jet Propulsion Laboratory, California Institute of Technology, 
under contract with the National Aeronautics and Space Administration. We also 
acknowledge the use of the HyperLeda database (http://leda.univ-lyon1.fr). This 
research made use of \textsc{astropy}, a community-developed core \textsc{python} 
({\tt http://www.python.org}) package for Astronomy \citep{2013A&A...558A..33A}; 
\textsc{ipython} \citep{PER-GRA:2007}; \textsc{matplotlib} \citep{Hunter:2007}; 
\textsc{numpy} \citep{citenumpy}; \textsc{scipy} \citep{citescipy}. Finally, we
are very greatful to the referee for her/his comments, which helped to improve 
the manuscript.

\end{acknowledgements}

\bibliographystyle{aa} 
\bibliography{gals} 

\clearpage

\appendix 
\section{Flux densities} 
\label{app:data}

\begin{table*}[h!]
\caption{II\,Zw\,40} 
\label{tab:iizw40}
\begin{flushleft}
\begin{tabular}{rrrl}
\hline\noalign{\smallskip}
Frequency & Flux density & Error & Reference \\
\hline\noalign{\smallskip}
0.325  & 38   & 4   & \citet{deeg93} \\
1.459  & 30.1 & 0.4 & \citet{klein91}, \citet{deeg93} \\ 
4.88   & 21.5 & 1.9 & \citet{jaffe78}, \citet{klein84a}, \citet{klein91} \\ 
10.63  & 20.1 & 1.5 & \citet{skillman88}, \citet{klein84a} \\ 
24.500 & 18   & 4   & \citet{klein84a} \\ 
\hline\noalign{\smallskip}

\end{tabular}
\end{flushleft}
\end{table*}

\begin{table*}[h!]
\caption{II\,Zw\,70} 
\label{tab:iizw70}
\begin{flushleft}
\begin{tabular}{rrrl}
\hline\noalign{\smallskip}
Frequency & Flux density & Error & Reference \\
\hline\noalign{\smallskip}
0.327  & 11.2 & 2.0  & \citet{skillman88} \\
0.609  & 6.5  & 0.8  & \citet{skillman88} \\ 
1.440  & 4.56 & 0.34 & \citet{wynn86}, \citet{balkowski78} \\ 
4.795  & 3.04 & 0.13 & \citet{klein84a}, \citet{wynn86}, \citet{skillman88} \\ 
10.700 & 2.73 & 0.20 & \citet{skillman88}, \citet{klein84a} \\ 
\hline\noalign{\smallskip}
\end{tabular}
\end{flushleft}
\end{table*}

\begin{table*}[t]
\caption{IC\,10} 
\label{tab:ic10}
\begin{flushleft}
\begin{tabular}{rrrl}
\hline\noalign{\smallskip}
Frequency & Flux density & Error & Reference \\
\hline\noalign{\smallskip}
0.327  & 352 & 20 & WENSS, this work \\
1.43   & 377 &  6 & this work \\ 
2.64   & 283 & 20 & \citet{chyzy11} (recomputed) \\
4.82   & 219 &  8 & \citet{klein83a}, \citet{klein86}, \citet{becker91} \\ 
10.575 & 162 &  8 & \citet{klein86}, \citet{chyzy03} \\ 
24.5   & 118 & 18 & \citet{klein86} \\ 
\hline\noalign{\smallskip}
\end{tabular}
\end{flushleft}
\end{table*}

\begin{table*}[t]
\caption{NGC\,1569} 
\label{tab:ngc1569}
\begin{flushleft}
\begin{tabular}{rrrl}
\hline\noalign{\smallskip}
Frequency & Flux density & Error & Reference \\
\hline\noalign{\smallskip}
0.350  & 750 & 20 & Purkayastha (2014); Ph.D. thesis, Univ. Bonn \\ 
0.610  & 610 & 20 & \citet{israel88} \\ 
1.415  & 425 & 20 & \citet{hummel80}, \citet{israel88} \\ 
2.700  & 318 & 20 & \citet{pfleiderer80}, \citet{sulentic76} \\
4.800  & 233 & 20 & \citet{klein86}, \citet{gregory91} \\ 
8.350  & 176 &  5 & this work \\ 
10.700 & 155 & 10 & \citet{klein86} \\ 
15.36  & 116 & 13 & \citet{lisenfeld04} \\ 
24.500 &  96 &  8 & \citet{klein86} \\ 
\hline\noalign{\smallskip}
\end{tabular}
\end{flushleft}
\end{table*}

\begin{table*}[t]
\caption{NGC\,2146} 
\label{tab:ngc2146}
\begin{flushleft}
\begin{tabular}{rrrl}
\hline\noalign{\smallskip}
Frequency & Flux density & Error & Reference \\
\hline\noalign{\smallskip}
0.327  & 2520 &  100 & WENSS, this work \\ 
1.43   & 1094 &   13 & NVSS, \citet{braun07} \\ 
2.695  &  720 &   70 & \citet{haynes75} \\
5.000  &  472 &   25 & \citet{debruyn77} \\ 
6.630  &  360 &   30 & \citet{mccutcheon73} \\ 
10.625 &  239 &   10 & \citet{niklas95}, \citet{israel83} \\ 
24.500 &  167 &   10 & this work \\ 
\hline\noalign{\smallskip}
\end{tabular}
\end{flushleft}
\end{table*}

\begin{table*}[t]
\caption{NGC\,3034} 
\label{tab:ngc3034}
\begin{flushleft}
\begin{tabular}{rrrl}
\hline\noalign{\smallskip}
Frequency & Flux density & Error & Reference \\
\hline\noalign{\smallskip}
0.327  & 13830 &  690 & \citet{adebahr13} \\
0.750  & 10700 &  500 & \citet{kellerman69} \\ 
1.388  &  7805 &  385 & \citet{hummel80}, \citet{adebahr13} \\ 
2.695  &  5700 &  300 & \citet{kellerman69} \\
5.000  &  3900 &  200 & \citet{kellerman69} \\ 
10.700 &  2250 &   60 & \citet{klein88} \\ 
14.700 &  1790 &   40 & \citet{klein88} \\ 
24.500 &  1190 &   30 & \citet{klein88} \\ 
32.000 &  1020 &   60 & \citet{klein88} \\ 
\hline\noalign{\smallskip}
\end{tabular}
\end{flushleft}
\end{table*}

\begin{table*}[t]
\caption{NGC\,3079} 
\label{tab:ngc3079}
\begin{flushleft}
\begin{tabular}{rrrl}
\hline\noalign{\smallskip}

Frequency & Flux density & Error & Reference \\
\hline\noalign{\smallskip}
0.327  & 2440 &  100 & WENSS, this work \\ 
0.615  & 1740 &   90 & \citet{irwin03} \\ 
1.365  &  800 &   10 & \citet{braun07} \\ 
2.640  &  526 &   30 & this work \\
4.750  &  377 &   20 & \citet{gioia82} \\ 
10.650 &  205 &   11 & \citet{gioia82} \\ 
24.500 &  123 &    8 & this work \\ 
\hline\noalign{\smallskip}
\end{tabular}
\end{flushleft}
\end{table*}

\begin{table*}[t]
\caption{NGC\,3310} 
\label{tab:ngc3310}
\begin{flushleft}
\begin{tabular}{rrrl}
\hline\noalign{\smallskip}
Frequency & Flux density & Error & Reference \\
\hline\noalign{\smallskip}
0.327  &  904 &  30 & WENSS, this work \\ 
0.610  &  650 &  50 & \citet{vdkruit76} \\ 
1.433  &  357 &  10 & \citet{condon87}, \citet{hummel85} \\ 
2.695  &  240 &  30 & \citet{vdkruit76} \\
4.750  &  146 &   8 & \citet{gioia82} \\ 
10.650 &  100 &   6 & \citet{gioia82}, \citet{israel88}, \citet{niklas95} \\ 
24.500 &   80 &   5 & this work \\ 
\hline\noalign{\smallskip}
\end{tabular}
\end{flushleft}
\end{table*}

\begin{table*}[t]
\caption{NGC\,4038/39} 
\label{tab:ngc4038}
\begin{flushleft}
\begin{tabular}{rrrl}
\hline\noalign{\smallskip}
Frequency & Flux density & Error & Reference \\
\hline\noalign{\smallskip}
0.408  & 1250 &  60 & \citet{slee95} \\ 
1.410  &  544 &  30 & \citet{vdhulst79}, \citet{hummel80}, this work \\ 
2.700  &  353 &  30 & \citet{dejong67}, \citet{tovmassian68}, \citet{kazes70} \\
4.850  &  246 &  17 & \citet{griffith94} \\ 
10.550 &  146 &   7 & \citet{niklas95}, \citet{chyzy04} re-computed \\ 
24.500 &   92 &   9 & this work \\ 
\hline\noalign{\smallskip}
\end{tabular}
\end{flushleft}
\end{table*}

\begin{table*}[t]
\caption{NGC\,4449} 
\label{tab:ngc4449}
\begin{flushleft}
\begin{tabular}{rrrl}
\hline\noalign{\smallskip}
Frequency & Flux density & Error & Reference \\
\hline\noalign{\smallskip}
0.376  & 514 & 30 & \citet{purkayastha14} \\ 
0.609  & 450 & 40 & \citet{klein96} \\ 
1.427  & 278 & 15 & \citet{condon87}, NVSS, this work \citet{klein81} \\ 
2.695  & 204 & 12 & this work \\
4.875  & 137 &  9 & \citet{klein86}, \citet{sramek75} \\ 
10.650 &  87 &  5 & \citet{israel83}, \citet{klein86}, \citet{klein96} \\ 
24.500 &  67 & 10 & \citet{klein86} \\ 
\hline\noalign{\smallskip}
\end{tabular}
\end{flushleft}
\end{table*}

\begin{table*}[t]
\caption{NGC\,4490/85} 
\label{tab:ngc4490}
\begin{flushleft}
\begin{tabular}{rrrl}
\hline\noalign{\smallskip}
Frequency & Flux density & Error & Reference \\
\hline\noalign{\smallskip}
0.327  & 2040 &  50 & WENSS, this work \\ 
0.408  & 2099 &  50 & \citet{gioia80} \\ 
1.430  &  835 &  16 & \citet{lequeux71}, \citet{viallefond80}, \citet{condon87} \\ 
2.695  &  520 &  50 & \citet{kazes70} \\
4.81   &  331 &  13 & \citet{gioia82}, \citet{nikiel16} \\ 
6.630  &  262 &  40 & \citet{mccutcheon73} \\ 
10.700 &  185 &  27 & \citet{klein81} \\ 
24.500 &  106 &  10 & \citet{klein83b} \\ 
\hline\noalign{\smallskip}
\end{tabular}
\end{flushleft}
\end{table*}

\begin{table*}[t]
\caption{NGC\,4631} 
\label{tab:ngc4631}
\begin{flushleft}
\begin{tabular}{rrrl}
\hline\noalign{\smallskip}
Frequency & Flux density & Error & Reference \\
\hline\noalign{\smallskip}
0.327  & 2993 & 110 & \citet{hummel90}, WENSS, this work \\ 
0.419  & 2900 &  71 & \citet{gioia80}, \citet{israel83} \\ 
0.610  & 2200 & 100 & \citet{ekers77}, \citet{werner88} \\ 
0.835  & 2000 & 100 & \citet{israel88} \\ 
1.423  & 1282 &  20 & \citet{ekers77}, \citet{hummel90}, \citet{braun07} \\ 
2.688  &  835 &  32 & \citet{wielebinski77}, \citet{werner88} \\
4.750  &  544 &  20 & \citet{wielebinski77}, \citet{israel83}, \citet{werner88} \\ 
8.35   &  310 &  16 & \citet{mora13} \\
10.625 &  260 &  10 & \citet{israel83}, \citet{dumke95} \\ 
24.500 &  172 &  15 & this work \\ 
\hline\noalign{\smallskip}
\end{tabular}
\end{flushleft}
\end{table*}

\begin{table*}[t]
\caption{NGC\,5194/5} 
\label{tab:ngc5194}
\begin{flushleft}
\begin{tabular}{rrrl}
\hline\noalign{\smallskip}
Frequency & Flux density & Error & Reference \\
\hline\noalign{\smallskip}
0.327  &  3812 &  150 & WENSS, this work \\ 
0.408  &  3640 &  700 & \citet{gioia80} \\ 
0.610  &  2790 &  200 & \citet{segalovitz77} \\ 
1.365  &  1420 &   10 & \citet{braun07} \\ 
2.695  &   780 &   50 & \citet{klein84b} \\
4.750  &   488 &   20 & \citet{gioia82}, \citet{israel83}, \citet{klein87} \\ 
8.35   &   306 &   26 & \citet{klein81} \\
10.700 &   239 &   11 & \citet{israel83}, \citet{klein84b} \\ 
14.700 &   190 &   20 & \citet{klein84b} \\ 
22.800 &   142 &   15 & \citet{klein84b} \\ 
\hline\noalign{\smallskip}
\end{tabular}
\end{flushleft}
\end{table*}

\begin{table*}[t]
\caption{NGC\,6052} 
\label{tab:ngc6052}
\begin{flushleft}
\begin{tabular}{rrrl}
\hline\noalign{\smallskip}
Frequency & Flux density & Error & Reference \\
\hline\noalign{\smallskip}
0.325  & 244   & 20  & \citet{deeg93} \\
1.445  & 105   & 5   & \citet{deeg93}; NVSS, this work \\
4.75   &  41.7 & 0.8 & \citet{klein84a}, \citet{klein91} \\
10.7   &  22.8 & 2.2 & \citet{klein91}, \citet{heidmann82}, \citet{maehara85}  \\
23.7   & 12.5  & 2.1 & \citet{heidmann82}, \citet{klein91} \\
32.0   & $<$12 &     & \citet{klein91} \\
\hline\noalign{\smallskip}
\end{tabular}
\end{flushleft}
\end{table*}

\clearpage
\Online
\section{Fit plots} 
\label{app:fits}

In this appendix, we  show the best fits of all four models (constant, curved, 
break, and cutoff) for an easier comparison, along with the parameter spaces and 
we give the best-fit values for the free parameters in Tab.~\ref{tab:all_fit_parameters}.
The fitting method has been explained in Sect.~\ref{sect:fitmethod}. In short, we have 
tested the four different models presented in Sect.~\ref{subsect:radio_spec} for each 
galaxy (which are represented by the solid red line): (i) simple power-law (i.e. with 
constant log-log slope), (ii) slightly curved law, (iii) power-law with a break, and 
(iv) power-law with an exponential decline. The only fixed parameter was the slope of 
the (optically thin) thermal emission, i.e.  $\alpha_{th} = 0.1$. The optimal values 
for the parameters are listed in the lower left part of the figures, along with the 
reduced chi-square, $\chi_\nu^2$.

Then, for a given model, we generate 1000 spectra similar to the observed one. At each 
frequency, the generated value is allowed to vary within the error of the observational 
data. This variation follows a Gaussian distribution 
around the observed data, and 
 the observational error represents $3\sigma$ of the Gaussian 
distribution. The 1000 spectra generated randomly this way are represented by faint grey 
lines in the main plots of the App.~\ref{app:fits}. Each generated spectrum is then fitted 
in the same way as the observational data (see Sect.~\ref{sect:fitmethod}) and yields a 
value for each free parameter. For each model, the distribution of the possible 1000 values 
taken by the parameters are shown in two parameter space plots under the main plot, allowing 
an assessment of the robustness of the fits.

In case of IC\,10 (IIZw 40) , the least-squares minimization fails for the break model (cutoff model) so that this 
model cannot be shown.

\begin{landscape}

\begin{table}
\begin{center}
\begin{tabular}{| l | rccc| rcccc | rcccc | rcccc |}
\hline
& \multicolumn{4}{|c|}{constant} & \multicolumn{5}{|c|}{curved} & \multicolumn{5}{|c|}{break} & \multicolumn{5}{|c|}{cutoff} \\
Gal & $S_{tot}$ & $f_{th}$ & $\alpha_{nth}$ & $\chi^2_\nu$ & $S_{tot}$ & $f_{th}$ & $\alpha_{nth}$ & $\nu_b$ & $\chi^2_\nu$ & $S_{tot}$ & $f_{th}$ & $\alpha_{nth}$ & $\nu_b$ & $\chi^2_\nu$ & $S_{tot}$ & $f_{th}$ & $\alpha_{nth}$ & $\nu_b$ & $\chi^2_\nu$ \\
      & [mJy] & & & & [mJy] & & & [GHz] & & [mJy] & & & [GHz] & & [mJy] & & & [GHz] & \\
\hline
II\,Zw\,40 & 32 & 0.0 & 0.2 & 0.837 & 60 & 0.24 & -0.04 & 0.42 & 1.266 & 34 & 0.68 & 0.4 & 1.46 & 0.361 & -- & -- & -- & -- & -- \\
II\,Zw\,70 & 5 & 0.62 & 1.15 & 0.332 & 5 & 0.63 & 1.07 & 14.69 & 0.659 & 5 & 0.66 & 1.00 & 3.18 & 0.592 & 5 & 0.65 & 1.01 & 4.26 & 0.589 \\
IC\,10 & 446 & 0.21 & 0.58 & 0.180 & 446 & 0.21 & 0.58 & 230274.07 & 0.270 & -- & -- & -- & -- & -- & 447 & 0.27 & 0.62 & 86.58 & 0.255 \\
NGC\,1569 & 473 & 0.00 & 0.47 & 1.218 & 497 & 0.00 & 0.19 & 1.30 & 0.112 & 481 & 0.16 & 0.50 & 5.48 & 0.513 & 494 & 0.22 & 0.42 & 12.40 & 0.118 \\
NGC\,4449 & 325 & 0.00 & 0.53 & 1.385 & 333 & 0.04 & 0.26 & 0.86 & 1.060 & 333 & 0.20 & 0.52 & 2.29 & 0.306 & 342 & 0.27 & 0.40 & 4.74 & 0.171 \\
\hdashline
NGC\,4490 & 995 & 0.00 & 0.69 & 4.307 & 1034 & 0.00 & 0.42 & 1.03 & 1.628 & 1039 & 0.08 & 0.61 & 2.92 & 0.264 & 1051 & 0.13 & 0.57 & 6.75 & 0.201 \\
NGC\,4631 & 1570 & 0.00 & 0.71 & 7.867 & 1614 & 0.00 & 0.44 & 1.31 & 4.246 & 1617 & 0.07 & 0.63 & 3.59 & 2.597 & 1637 & 0.14 & 0.57 & 6.59 & 2.082 \\
NGC\,5194 & 1807 & 0.00 & 0.81 & 4.301 & 1798 & 0.00 & 0.52 & 1.08 & 1.320 & 1780 & 0.07 & 0.67 & 2.58 & 0.698 & 1788 & 0.11 & 0.67 & 7.16 & 0.624 \\
\hdashline
NGC\,4038 & 683 & 0.04 & 0.70 & 0.077 & 688 & 0.09 & 0.62 & 16.81 & 0.073 & 683 & 0.13 & 0.71 & 8.16 & 0.006 & 686 & 0.15 & 0.71 & 17.99 & 0.036 \\
NGC\,6052 & 124 & 0.00 & 0.70 & 2.846 & 127 & 0.00 & 0.40 & 0.80 & 2.145 & 129 & 0.08 & 0.56 & 2.45 & 0.320 & 130 & 0.14 & 0.55 & 6.49 & 1.425 \\
\hdashline
NGC\,2146 & 1356 & 0.00 & 0.67 & 8.278 & 1361 & 0.01 & 0.39 & 0.98 & 5.388 & 1336 & 0.10 & 0.58 & 2.97 & 1.406 & 1359 & 0.16 & 0.51 & 6.22 & 0.470 \\
NGC\,3034 & 8627 & 0.00 & 0.59 & 9.917 & 8952 & 0.00 & 0.34 & 3.15 & 3.261 & 8880 & 0.06 & 0.46 & 3.87 & 1.399 & 9043 & 0.14 & 0.40 & 11.23 & 0.450 \\
NGC\,3079 & 1115 & 0.02 & 0.74 & 1.223 & 1119 & 0.07 & 0.63 & 5.41 & 1.508 & 1111 & 0.10 & 0.74 & 9.05 & 1.586 & 1114 & 0.12 & 0.75 & 20.36 & 1.607 \\
NGC\,3310 & 432 & 0.12 & 0.74 & 3.087 & 438 & 0.16 & 0.52 & 0.89 & 2.893 & 470 & 0.19 & 0.60 & 1.19 & 0.399 & 456 & 0.25 & 0.58 & 3.88 & 0.414 \\
\hline

\end{tabular}

\caption{Optimal parameters obtained by the best fit to each model (constant, curved, break,
        and cutoff).}
\label{tab:all_fit_parameters}
\end{center}
\end{table}

\end{landscape}

Apart from the reduced chi-squared, $\chi^2_\nu$, an important test whether the fits are 
physically meaningful is to look at the resulting thermal flux densities obtained at each 
frequency after subtracting the computed nonthermal flux density from the total (observed). 
A fit makes only sense if the resulting thermal flux densities so obtained obey to the 

expected optically thin spectrum with slope $-0.10$. This is shown in each diagram by the 
cyan data points, with the errors taken from those of the measured total flux densities. 
The resulting slopes are listed in Table~\ref{tab:slopes}.

\begin{table}[!h]
\begin{center}
\begin{tabular}{crrrr}
\hline
Galaxy & constant & curved & break & cutoff \\
\hline
II\,Zw\,40 & $ -0.87 $ & $ -0.09 $ & $ -0.09 $ & -- \\
II\,Zw\,70 & $ -0.10 $ & $ -0.10 $ & $ -0.10 $ & $ -0.10 $ \\
IC\,10 & $ -0.10 $ & $ -0.10 $ & -- & $ -0.10 $ \\
NGC\,1569 & $ -0.21 $ & $ -0.07 $ & $ -0.11 $ & $ -0.10 $ \\
NGC\,4449 & $ -0.29 $ & $ -0.10 $ & $ -0.11 $ & $ -0.11 $ \\
\hdashline
NGC\,4490 & $ -0.54 $ & $ -0.28 $ & $ -0.12 $ & $ -0.10 $ \\
NGC\,4631 & $ -0.95 $ & $ -0.35 $ & $ -0.19 $ & $ -0.15 $ \\
NGC\,5194 & $ -2.12 $ & $ -0.50 $ & $ -0.13 $ & $ -0.11 $ \\
\hdashline
NGC\,4038 & $ -0.09 $ & $ -0.10 $ & $ -0.10 $ & $ -0.10 $ \\
NGC\,6052 & $ 0.97 $ & $ -0.83 $ & $ -0.10 $ & $ -0.10 $ \\
\hdashline
NGC\,2146 & $ -0.38 $ & $ 0.07 $ & $ -0.09 $ & $ -0.10 $ \\
NGC\,3034 & $ -0.88 $ & $ -0.75 $ & $ -0.25 $ & $ -0.11 $ \\
NGC\,3079 & $ -0.39 $ & $ -0.15 $ & $ -0.13 $ & $ -0.12 $ \\
NGC\,3310 & $ -0.13 $ & $ -0.11 $ & $ -0.10 $ & $ -0.10 $ \\
\hline
\end{tabular}
\caption{Slopes of the thermal flux densities. These linear fits are depicted by solid magenta lines 
in the figures of the present appendix.}
\label{tab:slopes}
\end{center}
\end{table}

{\bf Caption for the plots of this appendix:}
In the main plots, the caption is as follows: the best fit of the radio continuum 
spectrum is represented by a solid red line. The best parameters are listed in the 
lower left part of the figure along with the reduced chi-square ($\chi^2_\nu$). The 
faint grey lines represent the randomly 1000 spectra generated in the fitting process.
The free-free (thermal) and synchrotron (nonthermal) components of the models are 
depicted by the dotted cyan line and dashed green line, respectively. The green 
crosses represent the observed nonthermal component (i.e., the modeled thermal 
component removed from the observational data). The cyan stars delineate the 
observed thermal component (i.e., the modeled nonthermal component subtracted 
from the observational data). The thermal component is drawn only if the fitted 
coefficient, $S_{th, \rm 1 GHz}$, is strictly positive. If present (curved, break, 
and cutoff models), the vertical dash-dotted yellow line depicts the break frequency, 
$\nu_b$. The magenta line delineates a linear fit to the thermal flux densities.


\begin{figure*}
\im{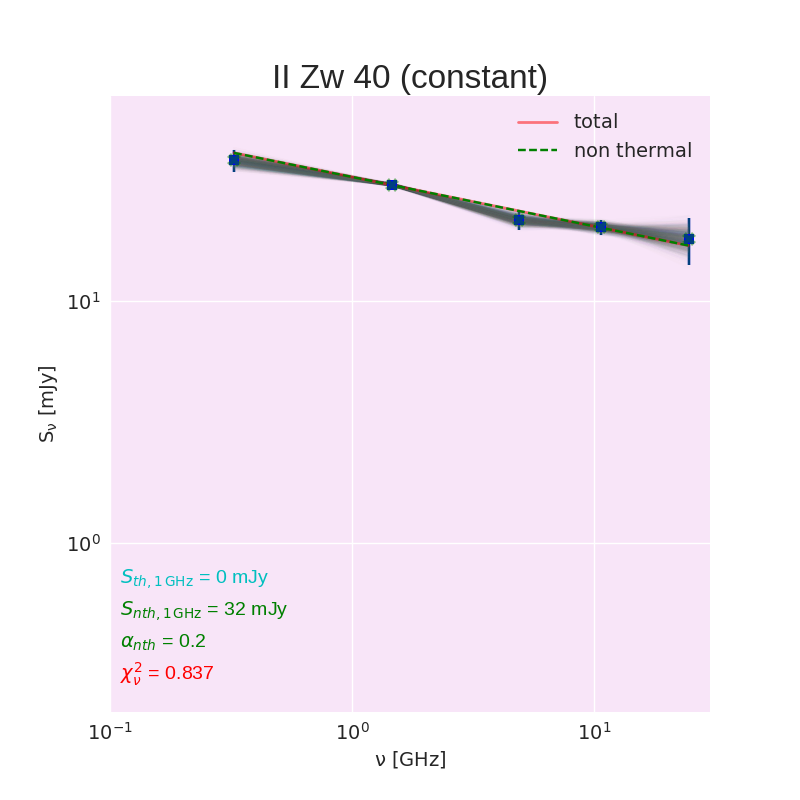} \hfill
\im{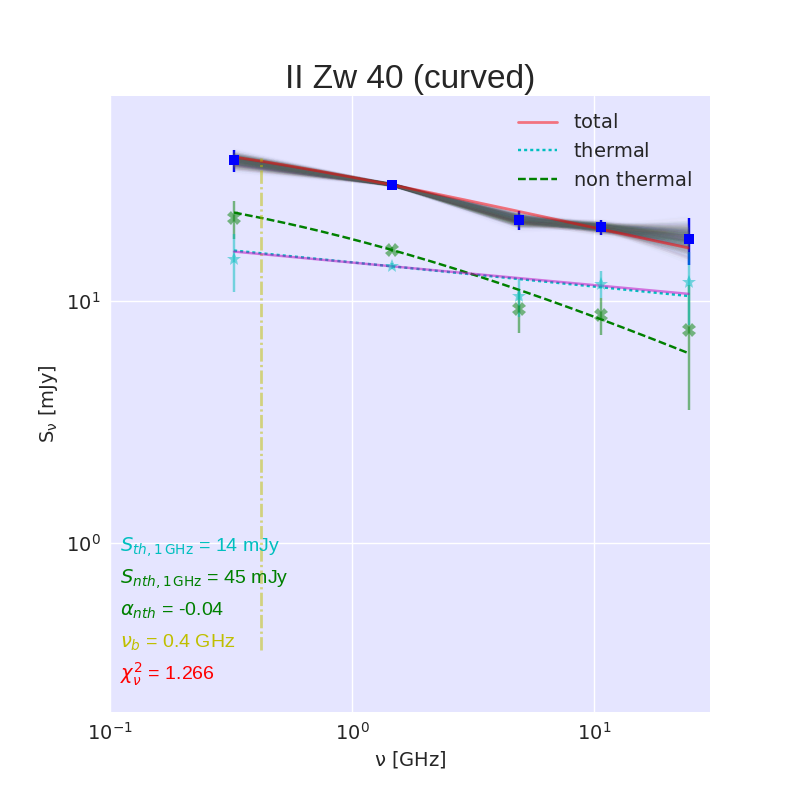}\\

\image{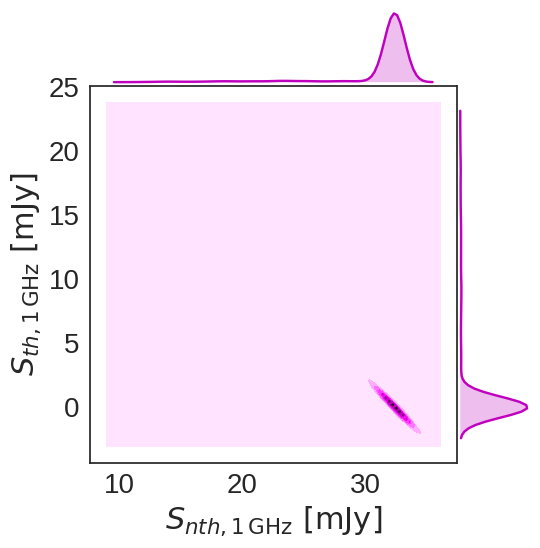}
\image{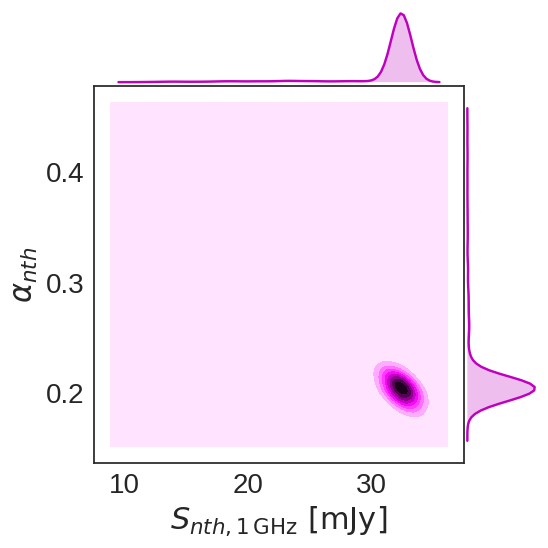} \hfill
\image{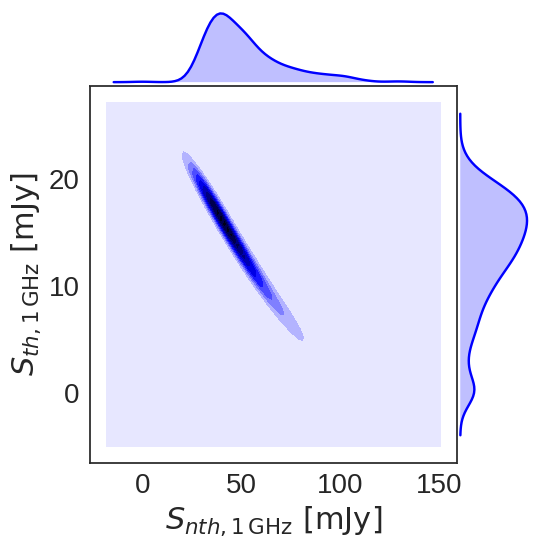}
\image{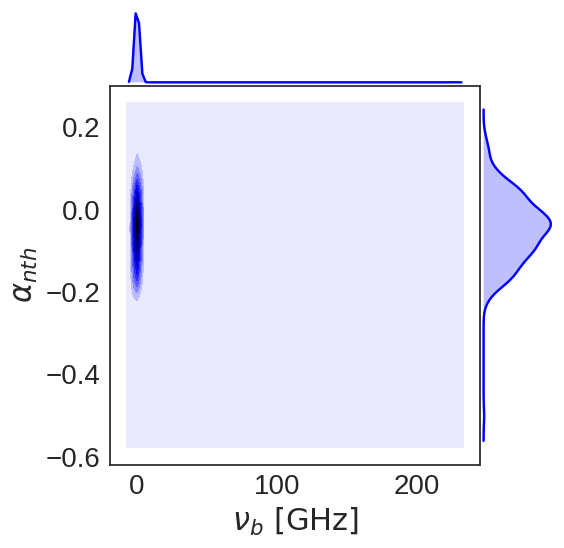}\\

\im{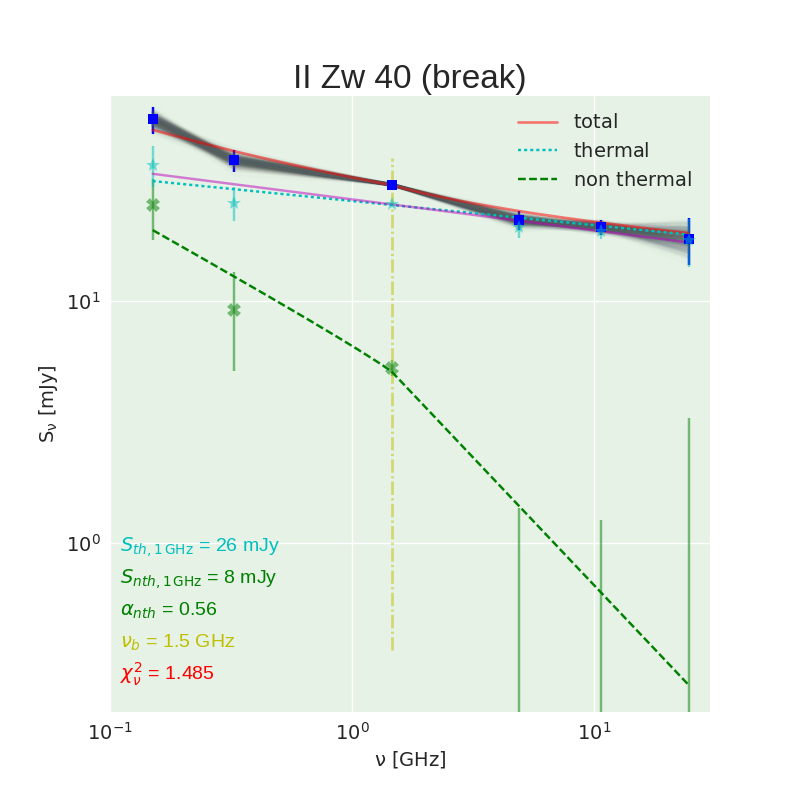} \hfill

\image{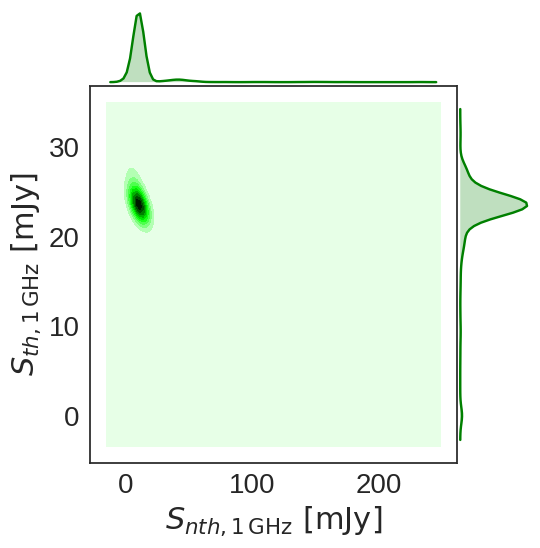}
\image{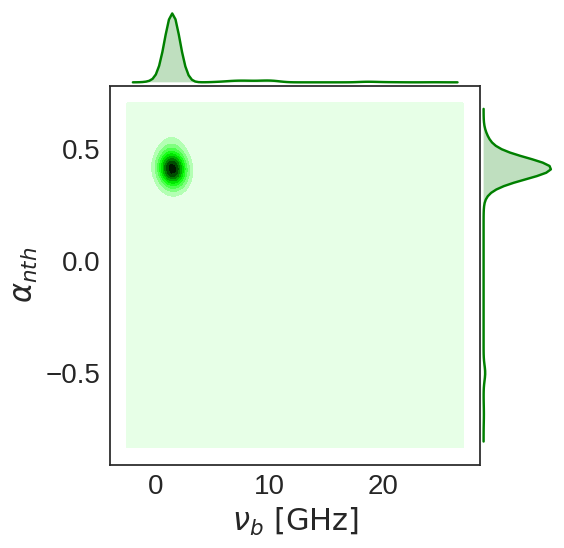} \hfill
\end{figure*}


\begin{figure*}
\im{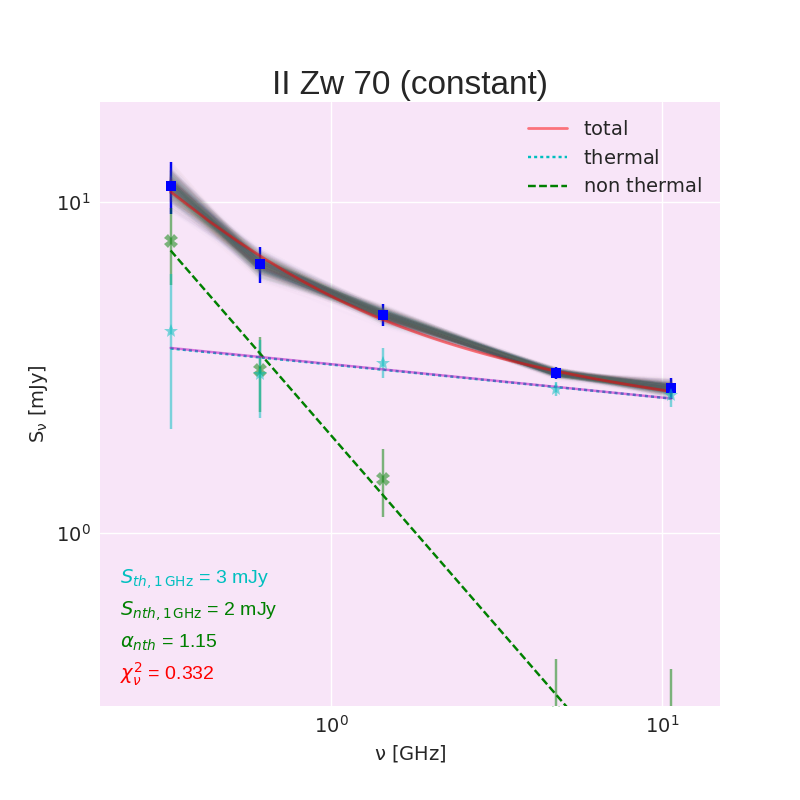} \hfill
\im{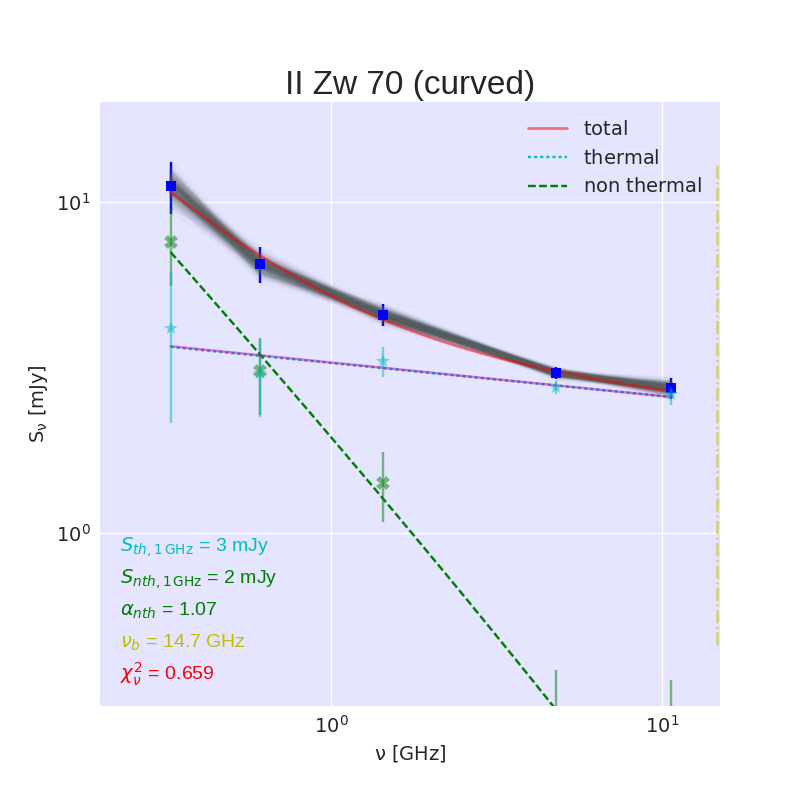}\\

\image{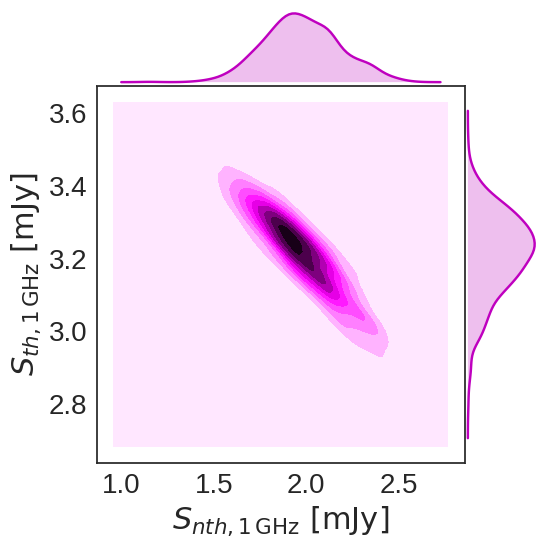}
\image{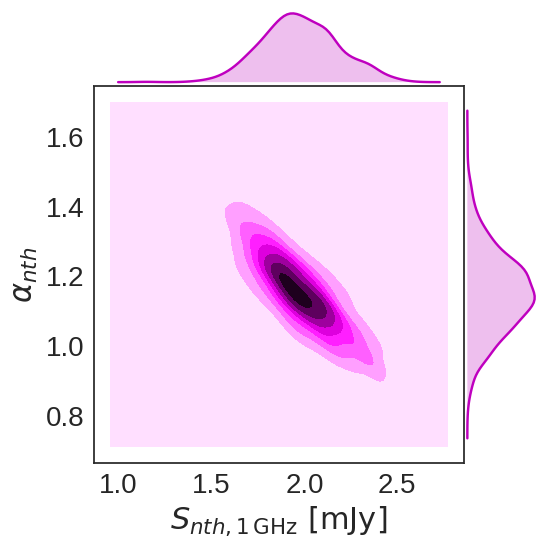} \hfill
\image{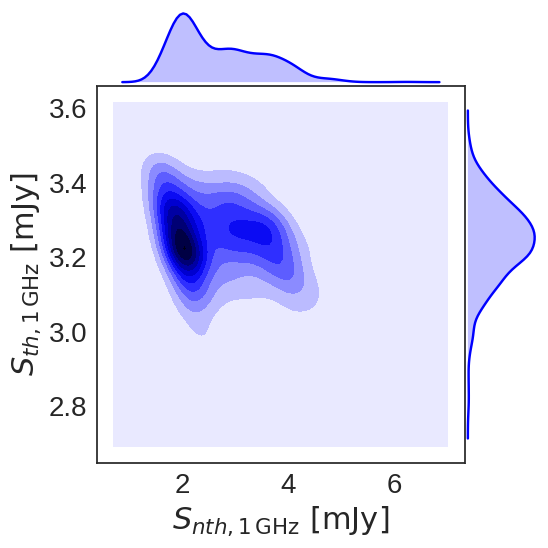}
\image{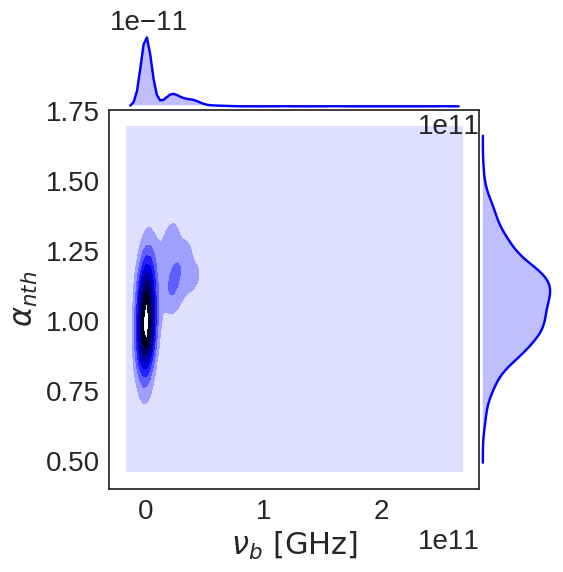}\\

\im{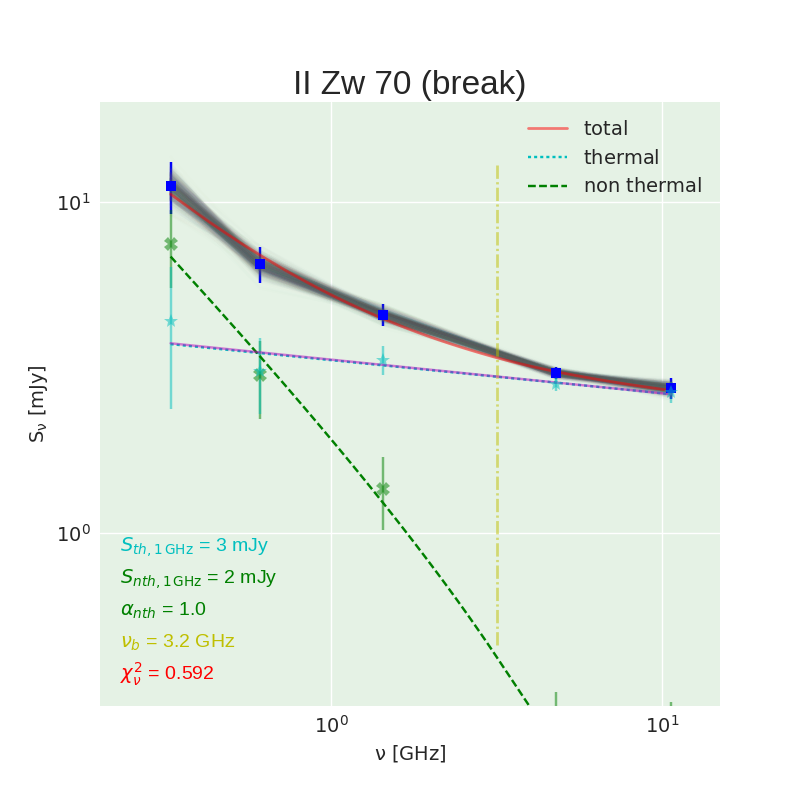} \hfill
\im{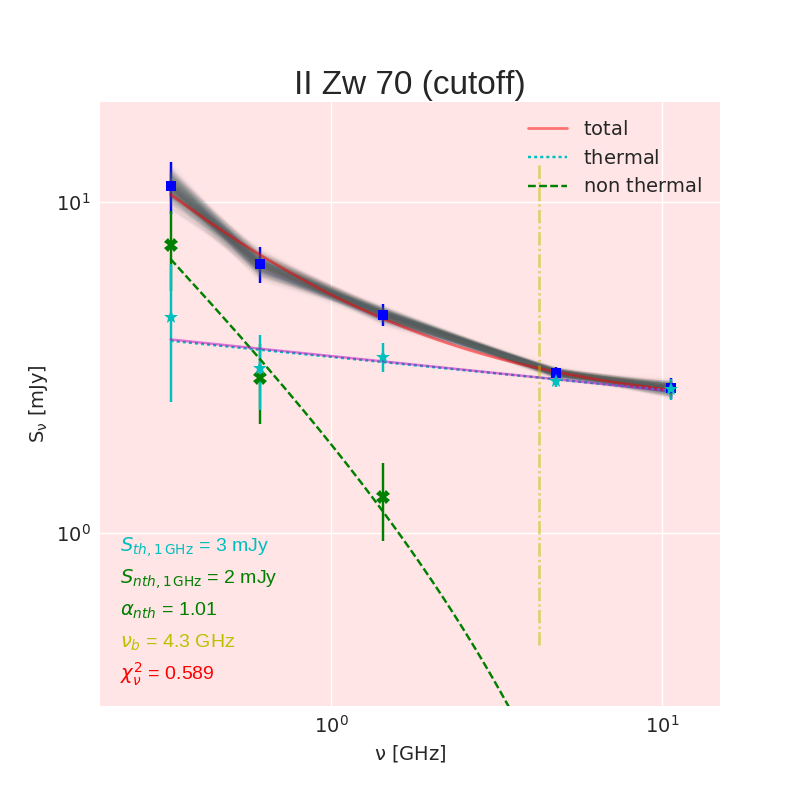}\\

\image{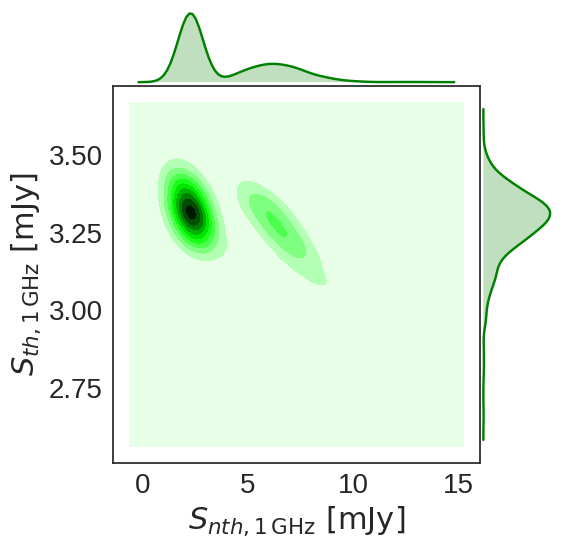}
\image{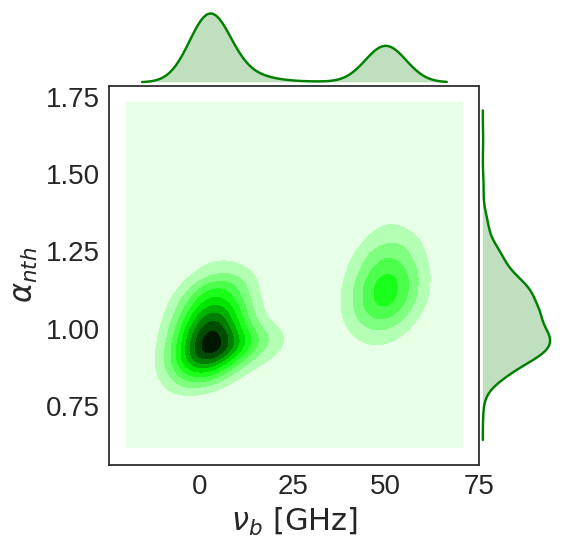} \hfill
\image{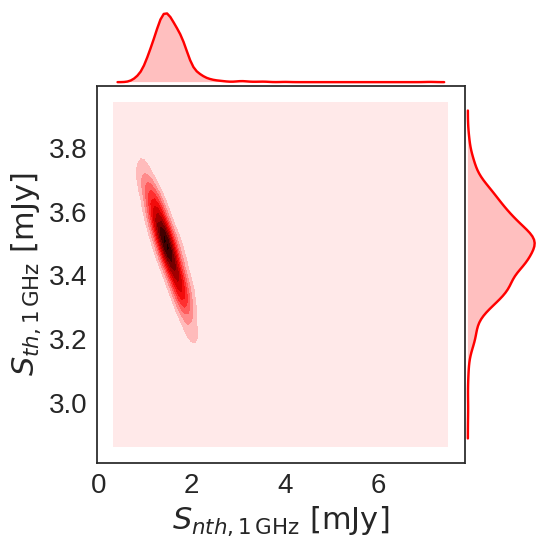}
\image{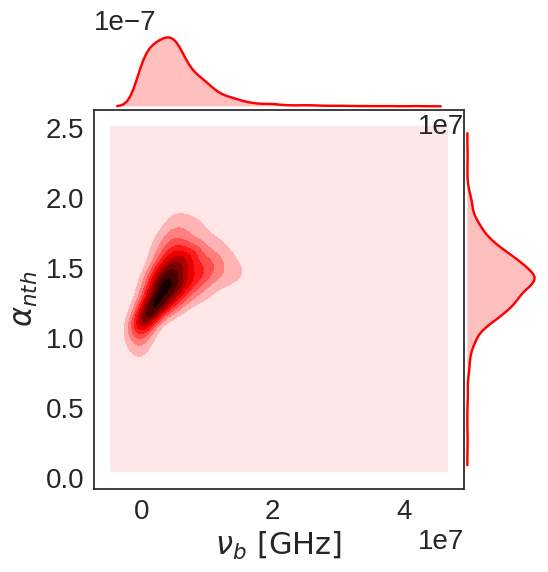}
\end{figure*}

\begin{center}
\begin{figure*}
\im{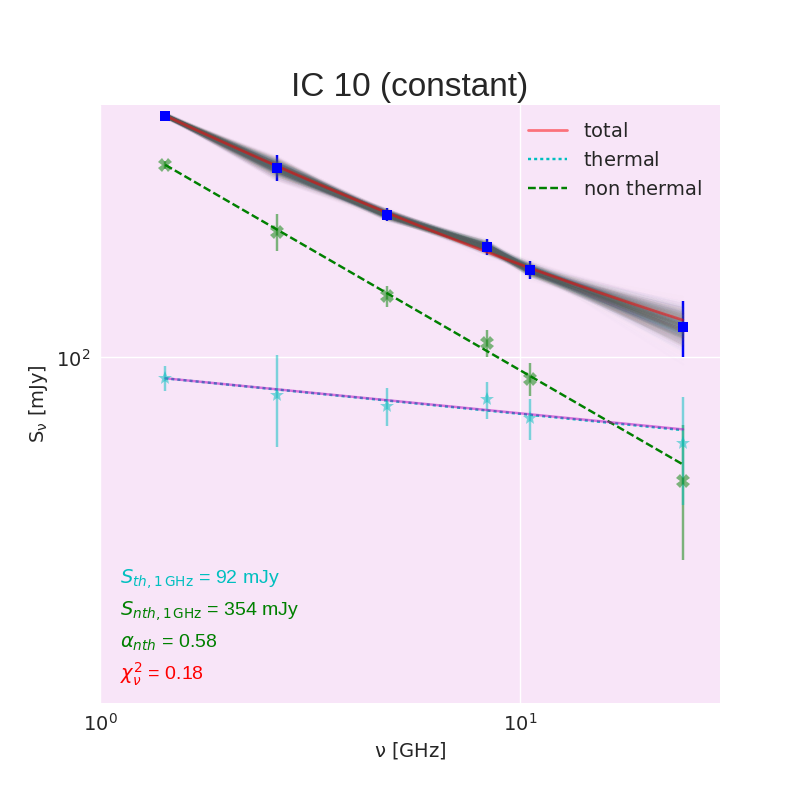} \hfill
\im{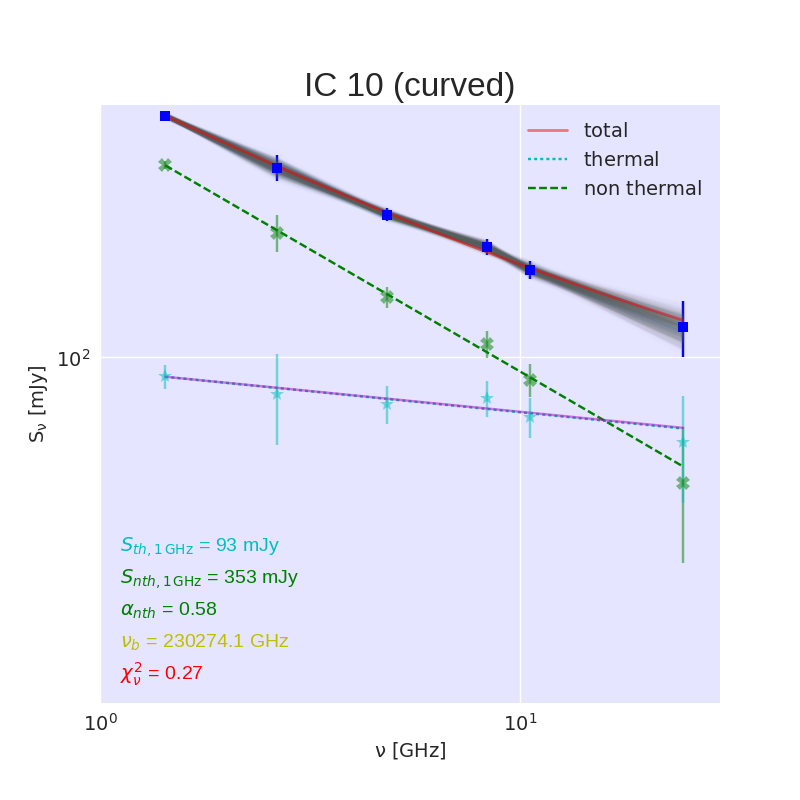}\\

\image{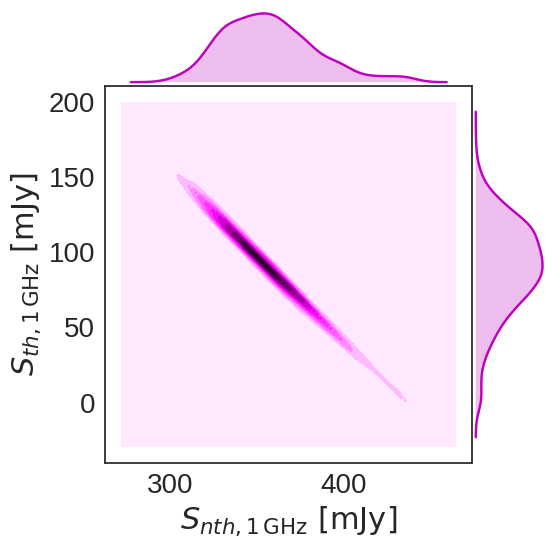}
\image{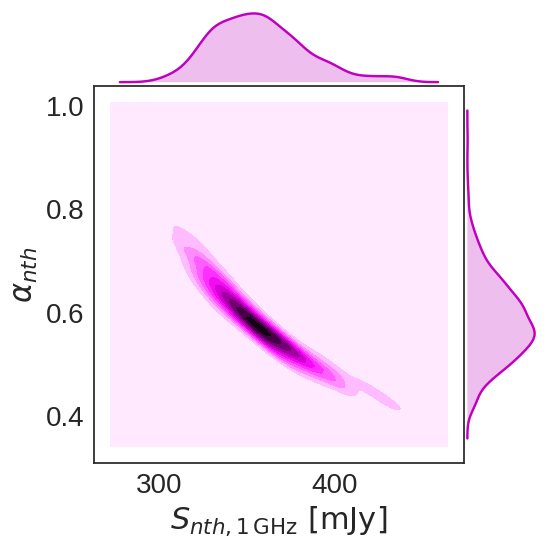} \hfill
\image{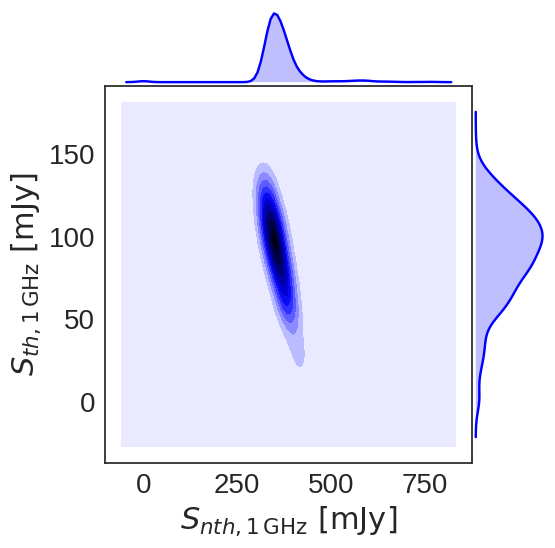}
\image{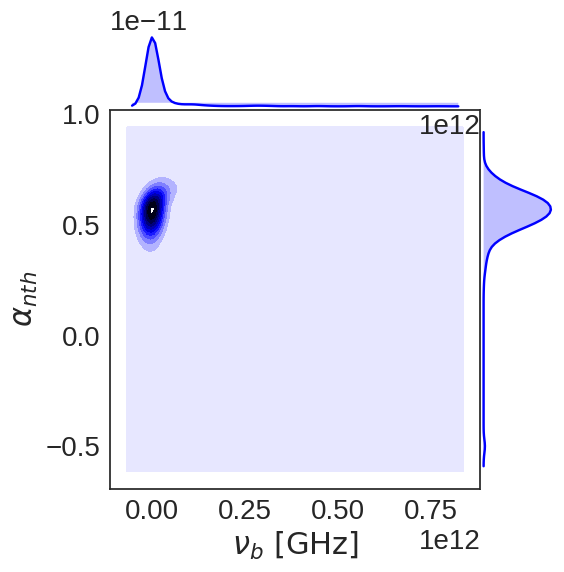}\\

\hfill \im{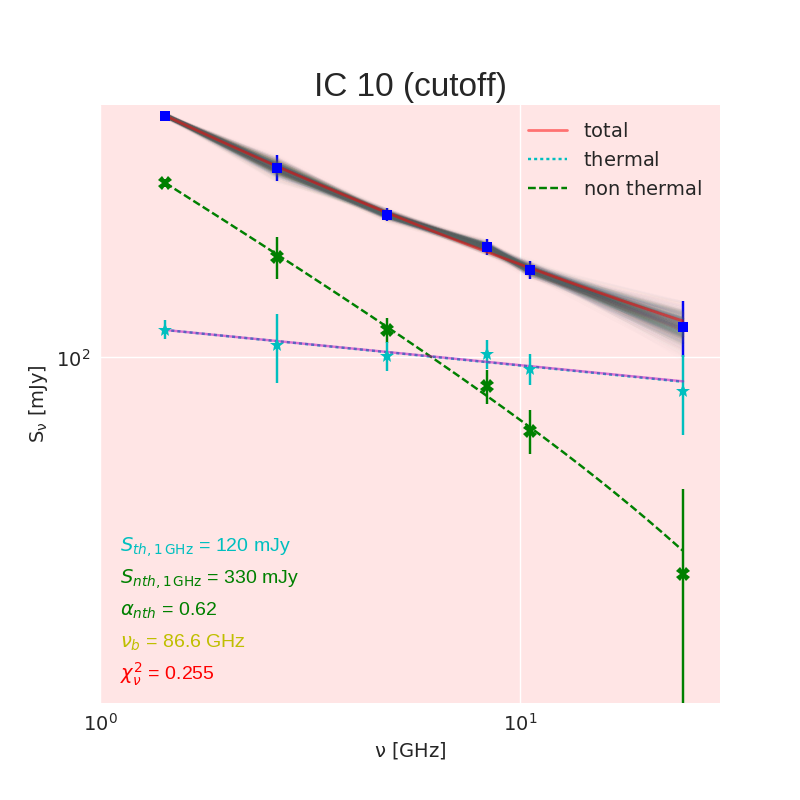}\\

\hfill \image{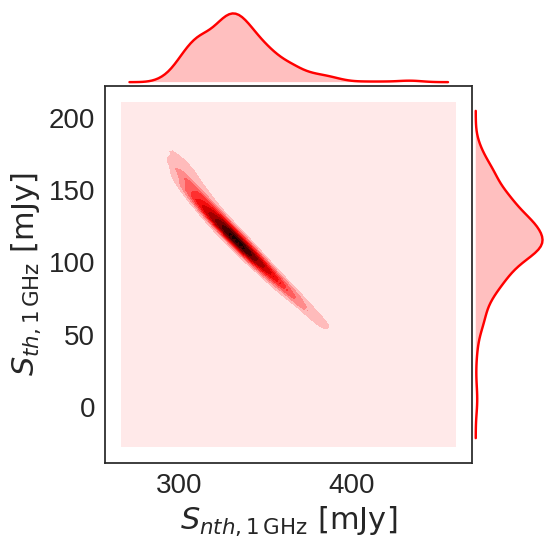}
\image{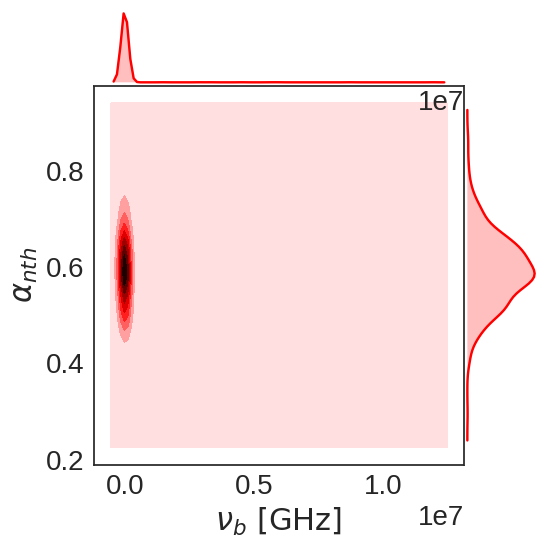}
\end{figure*}
\newpage
\clearpage

\begin{figure*}
\im{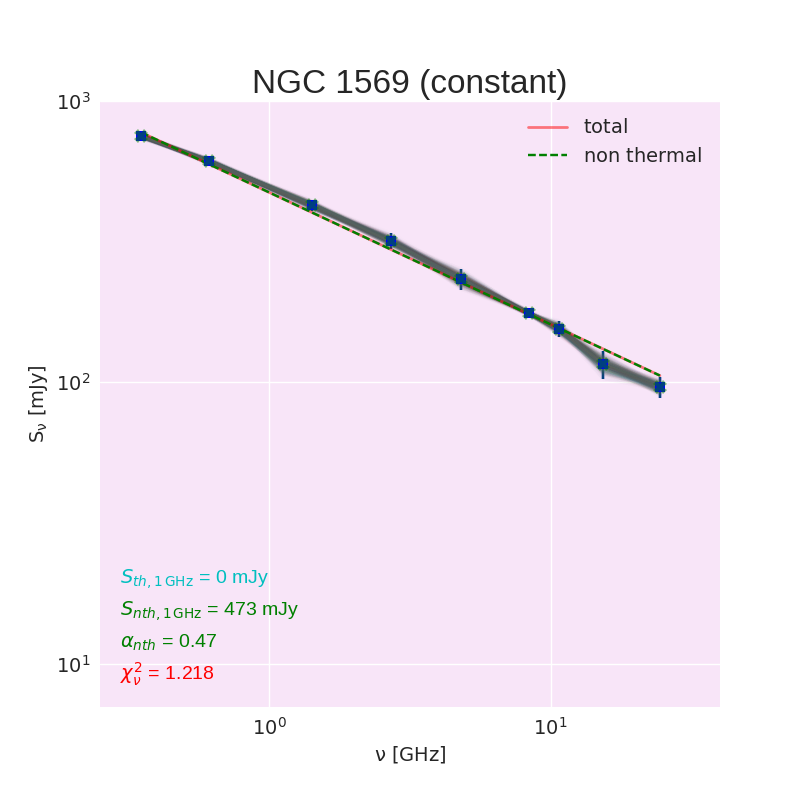} \hfill
\im{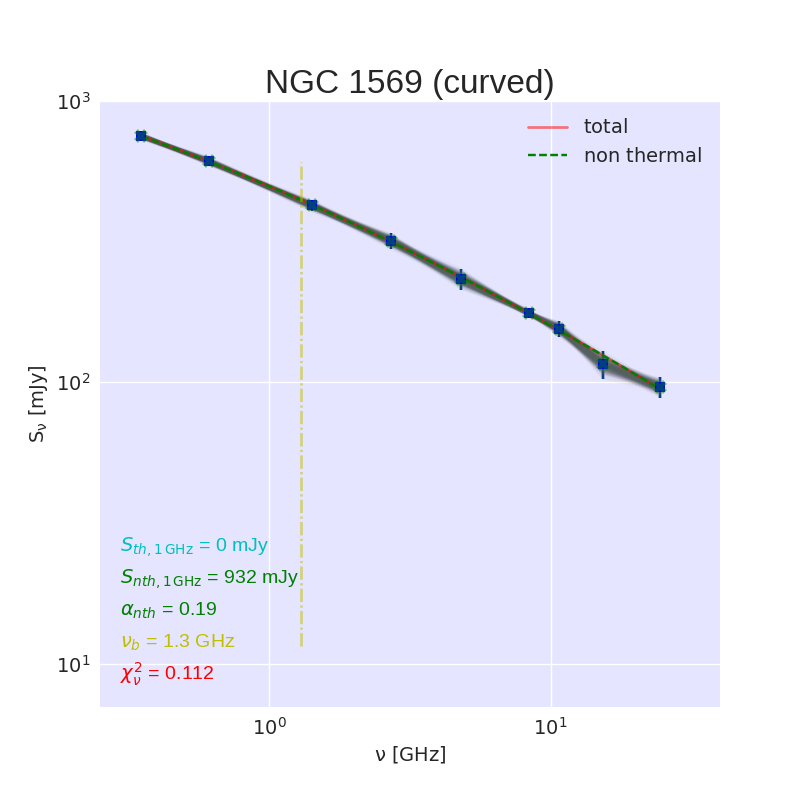}\\

\image{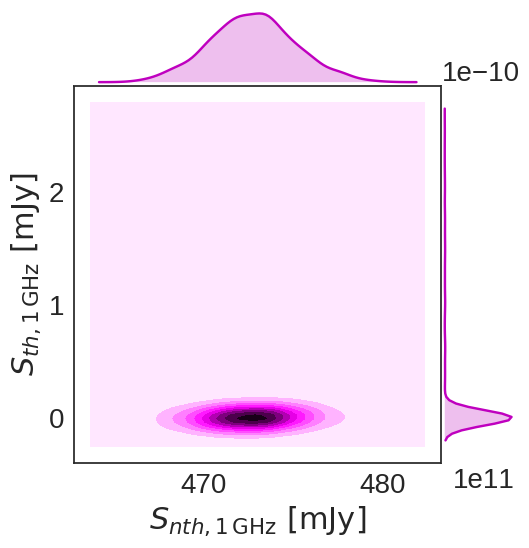}
\image{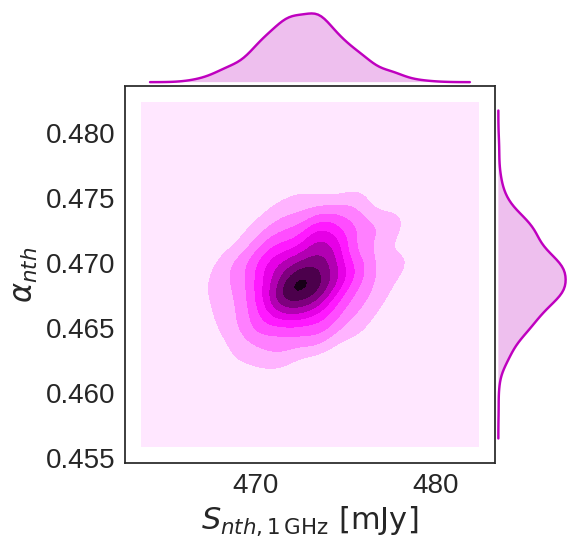} \hfill
\image{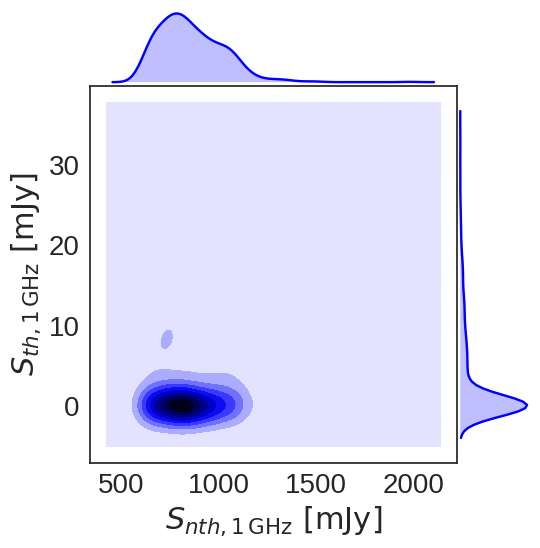}
\image{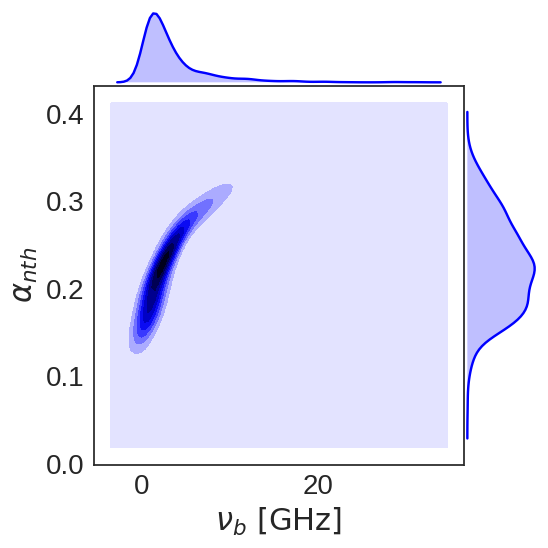}\\

\im{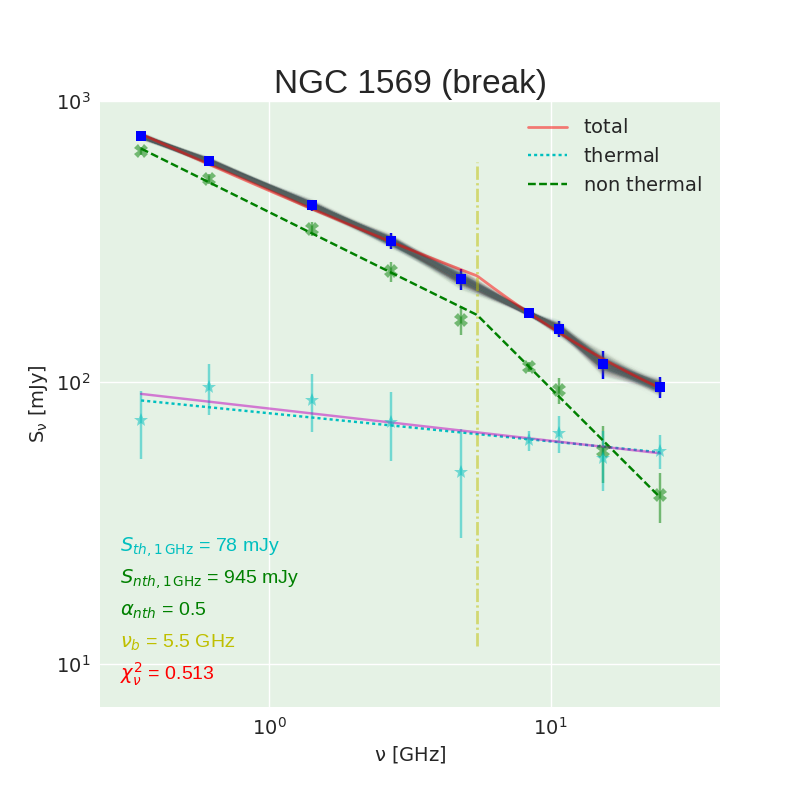}  \hfill
\im{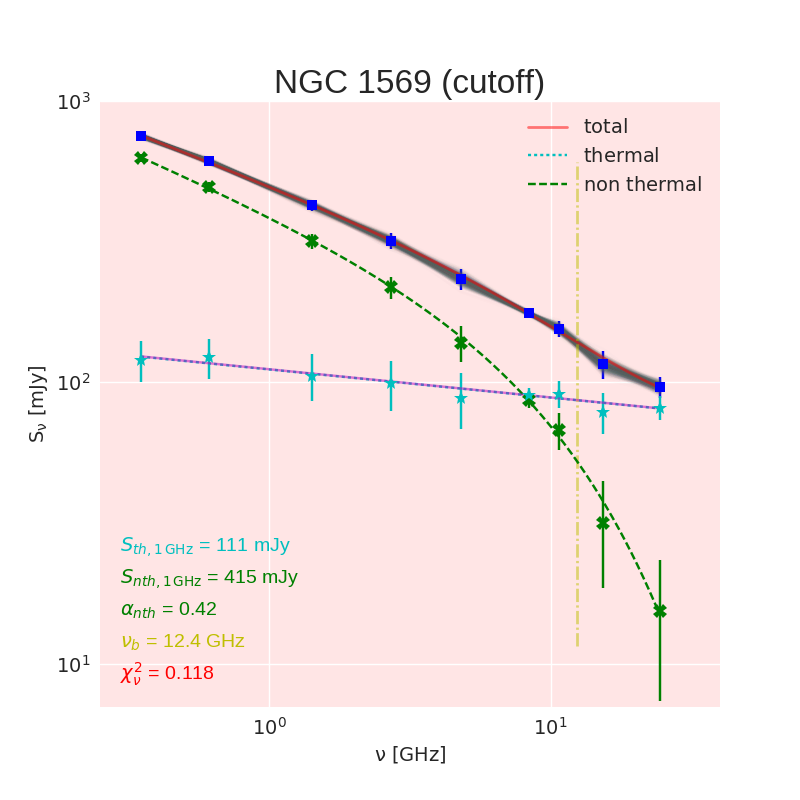}\\

\image{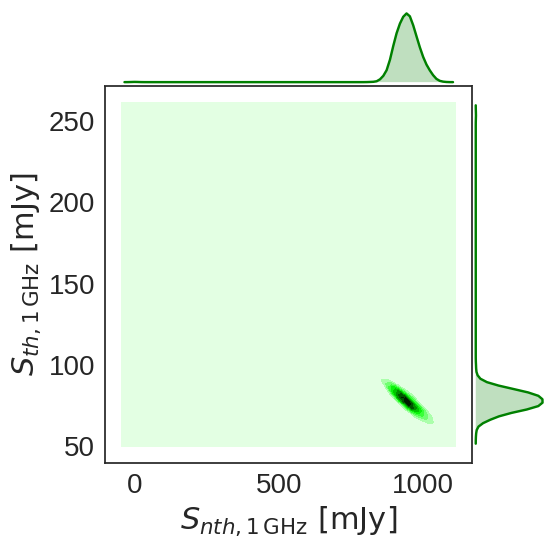} 
\image{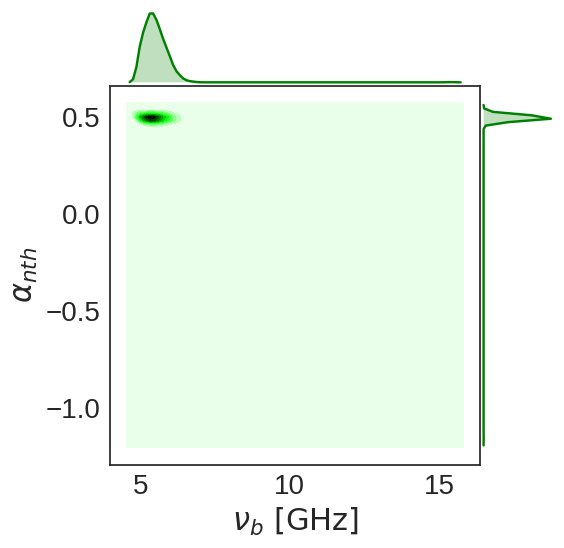}  \hfill
\image{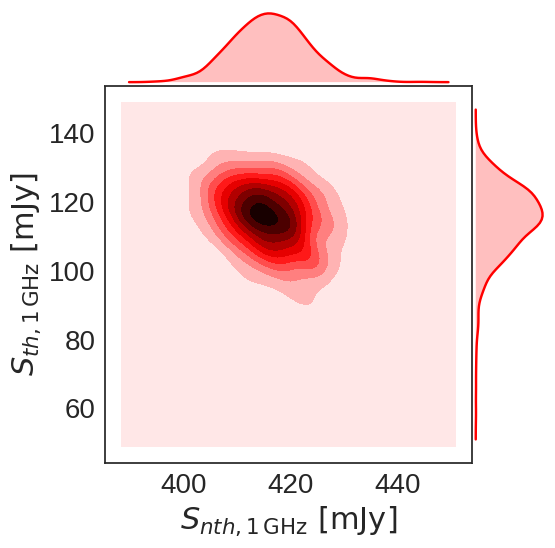}
\image{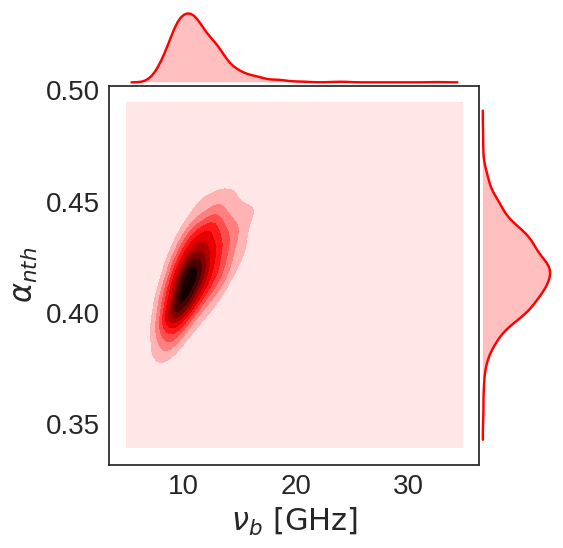}
\end{figure*}
\newpage
\clearpage

\begin{figure*}
\im{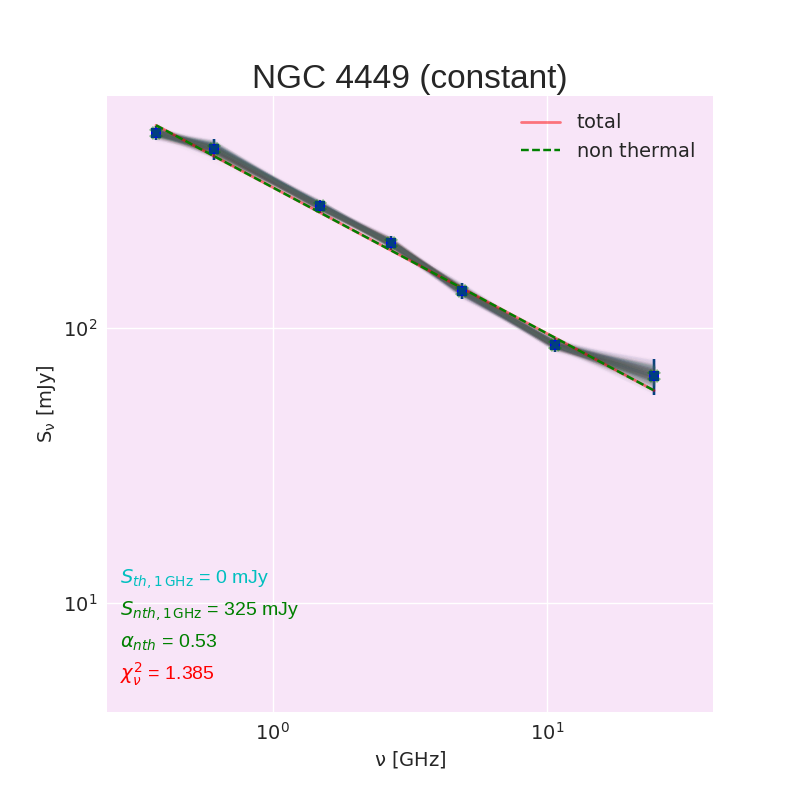} \hfill
\im{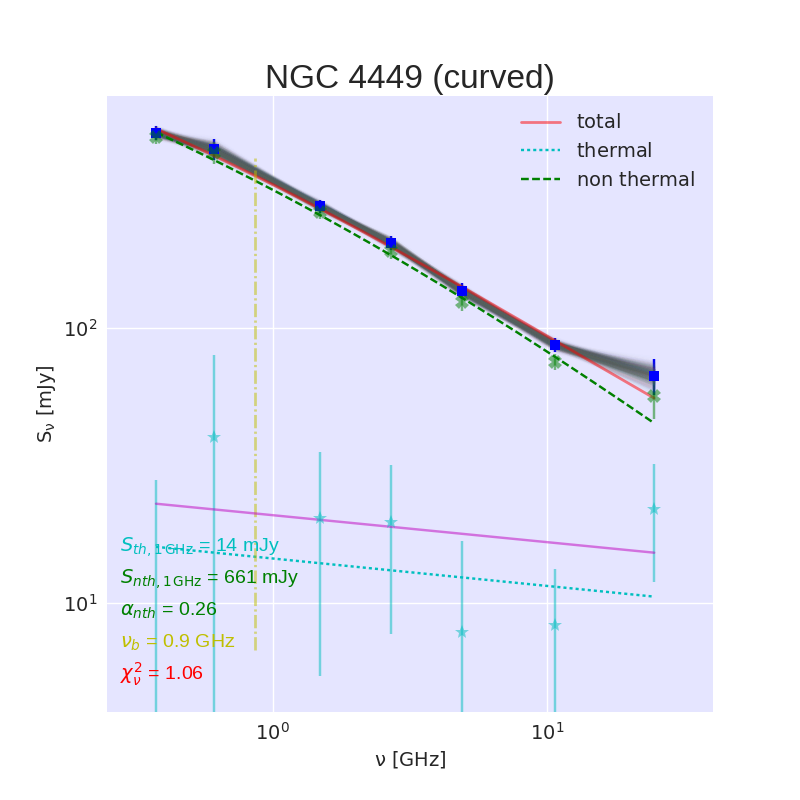}\\

\image{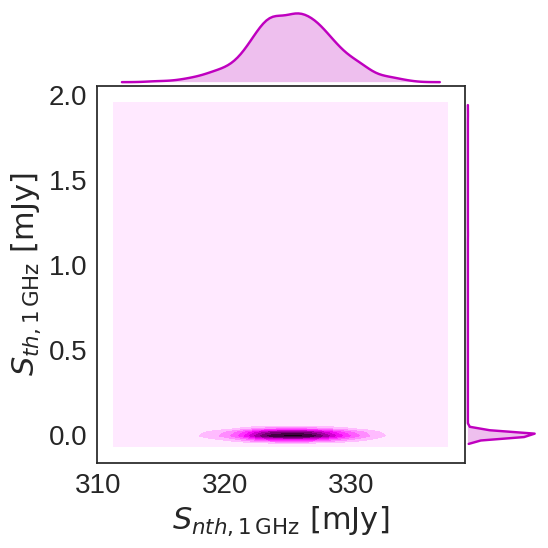}
\image{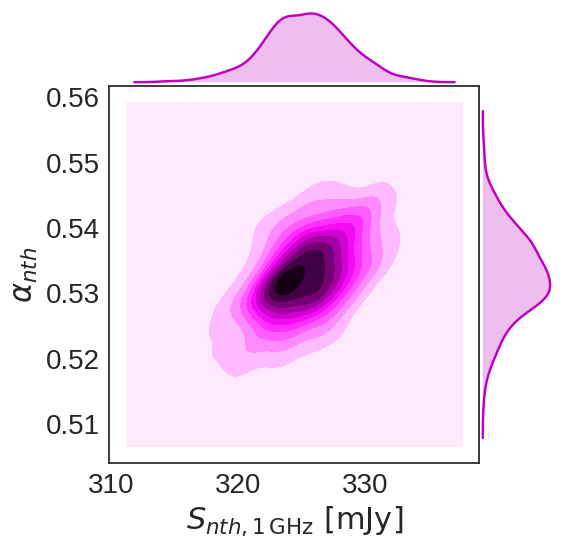} \hfill
\image{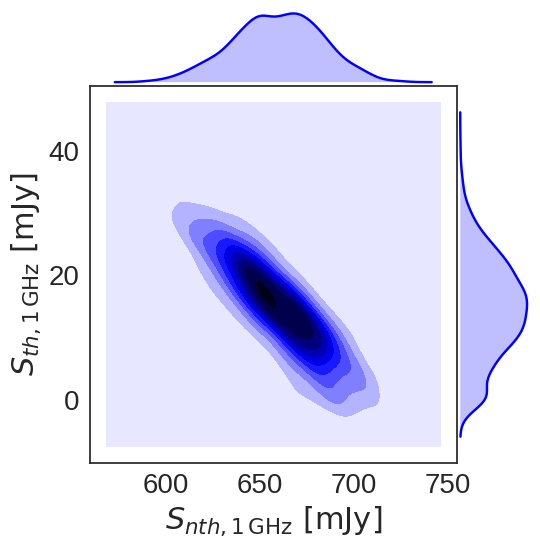}
\image{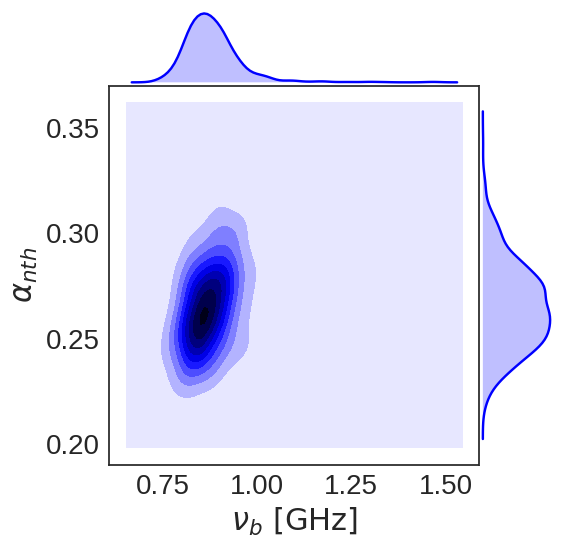}\\

\im{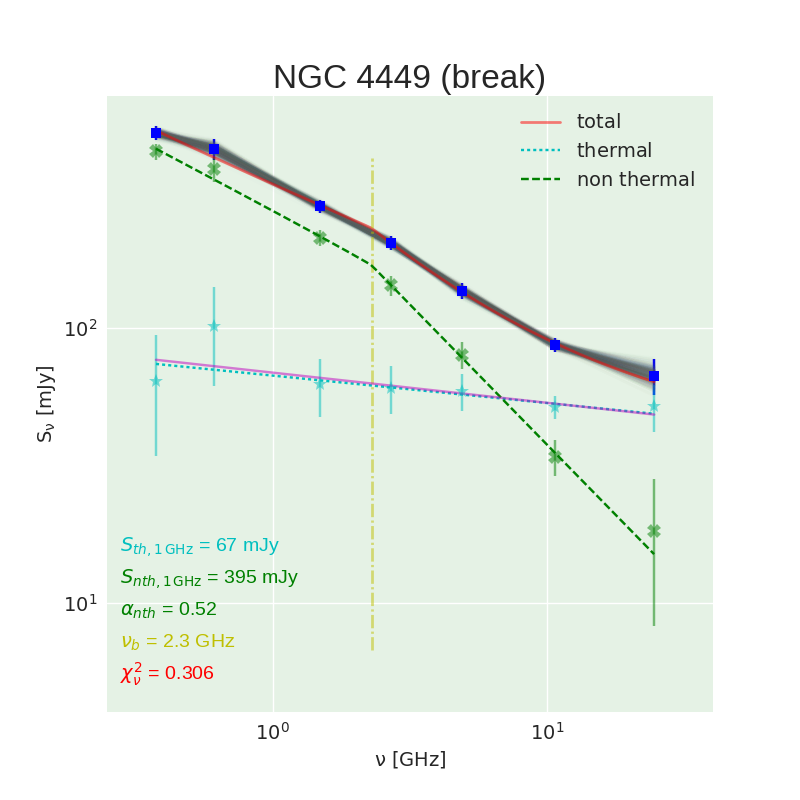} \hfill
\im{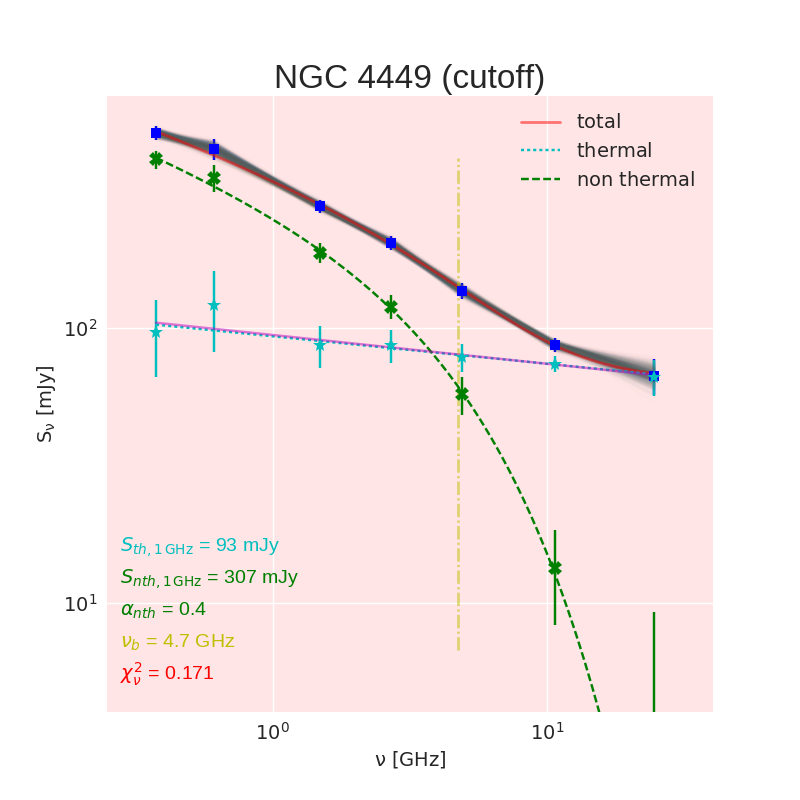}\\

\image{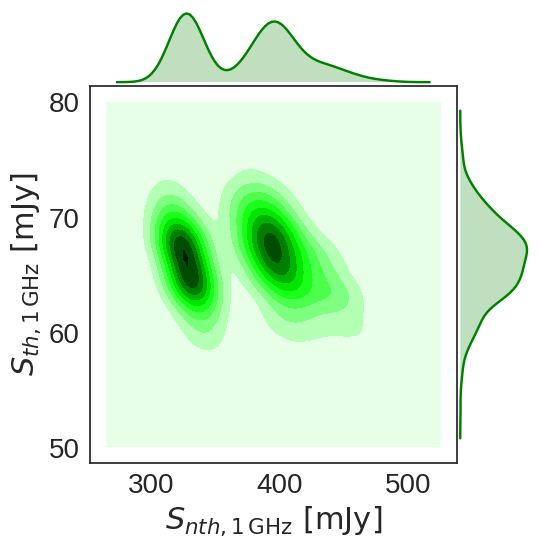}
\image{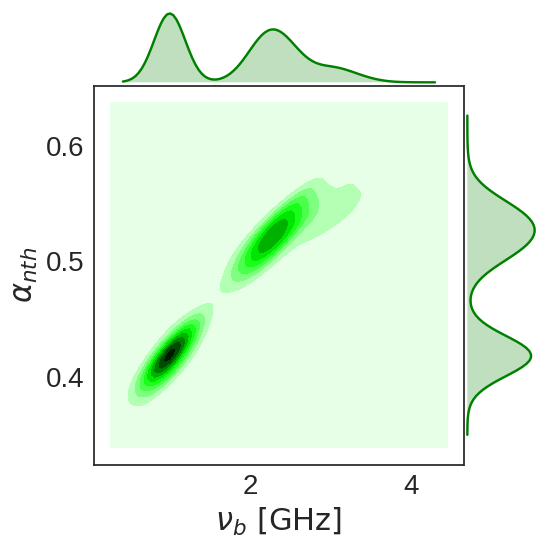} \hfill
\image{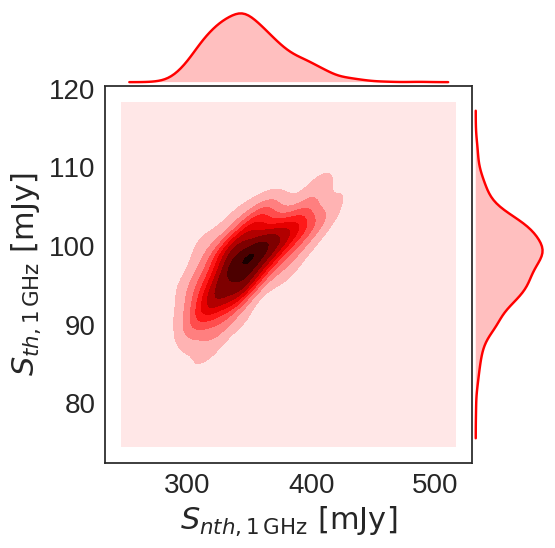}
\image{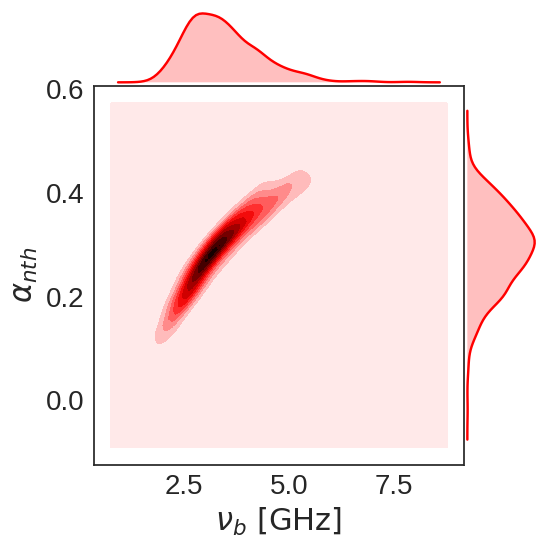}
\end{figure*}
\newpage
\clearpage

\begin{figure*}
\im{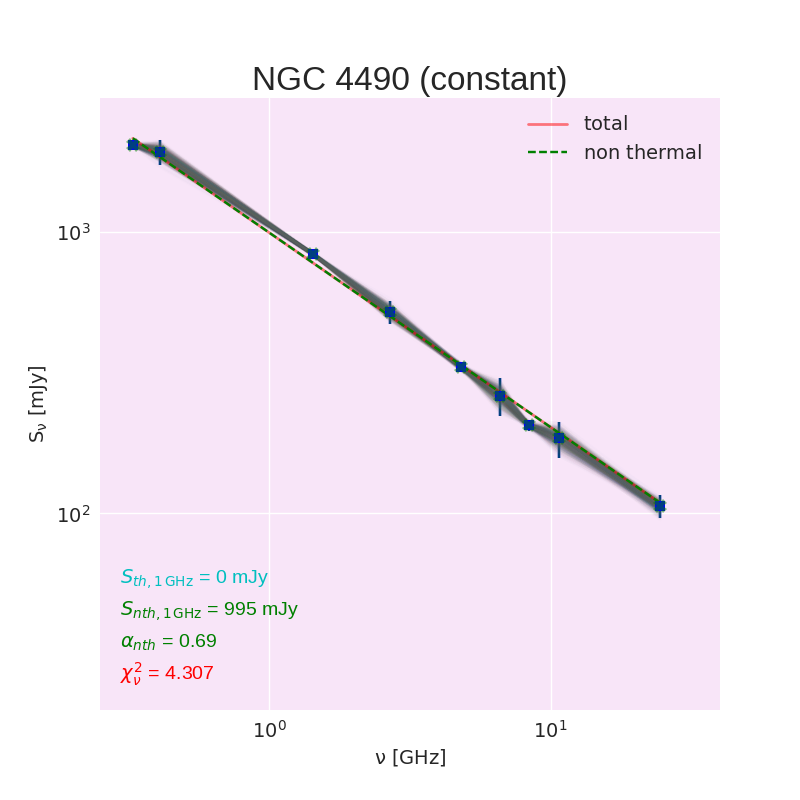} \hfill
\im{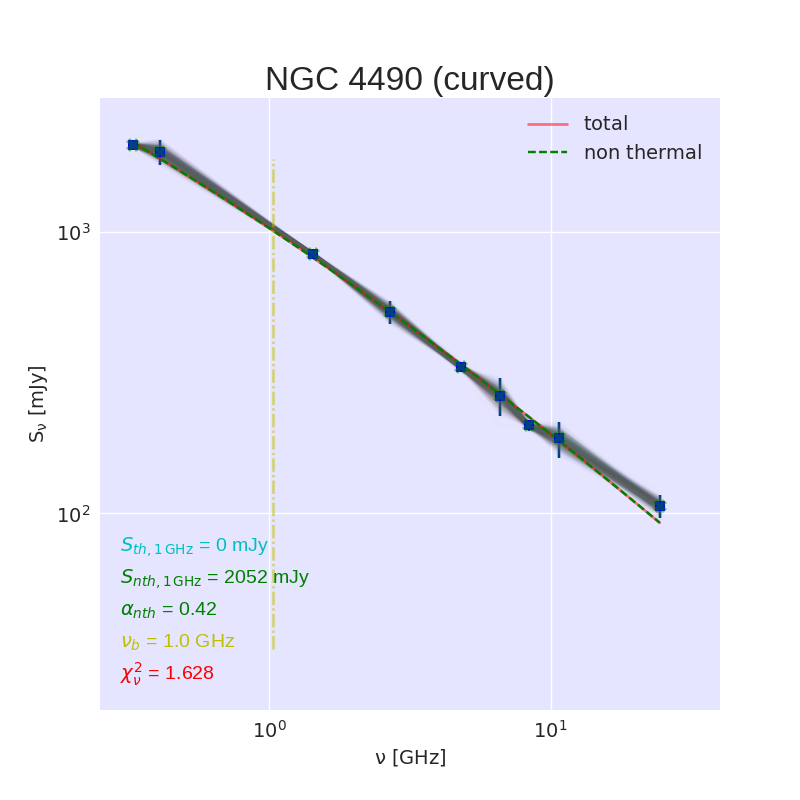}\\

\image{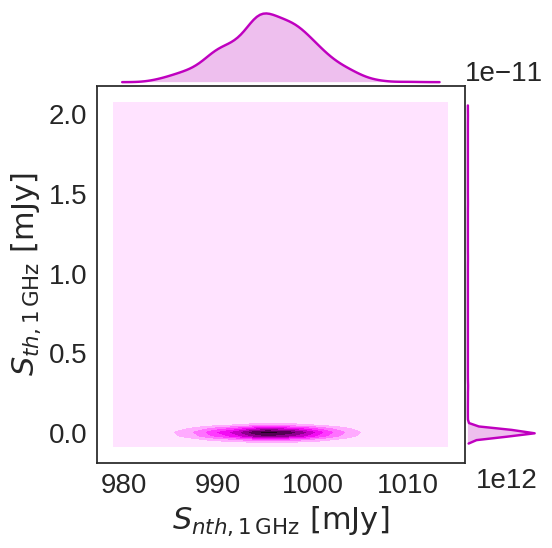}
\image{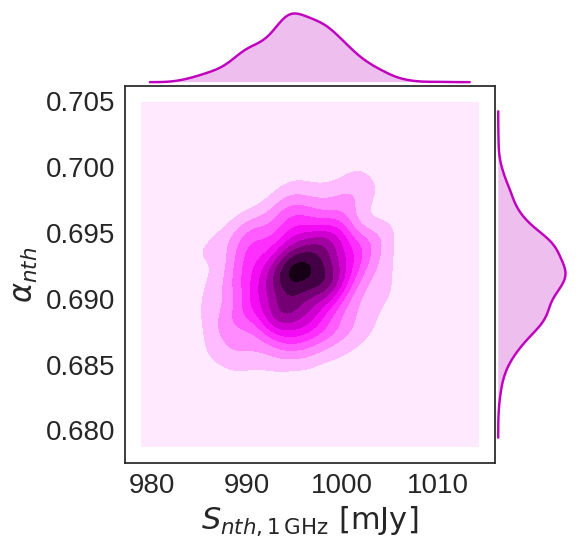} \hfill
\image{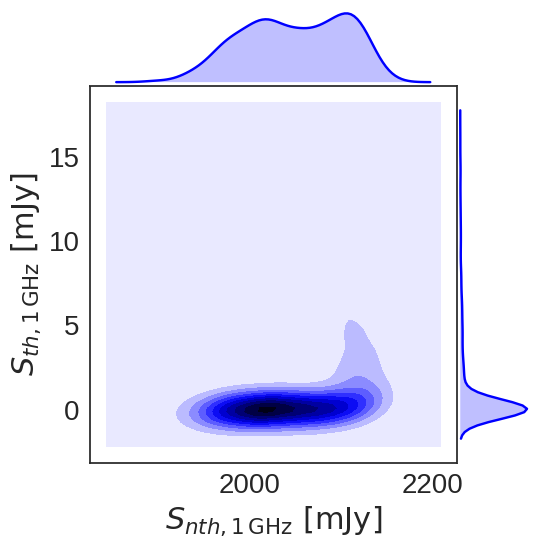}
\image{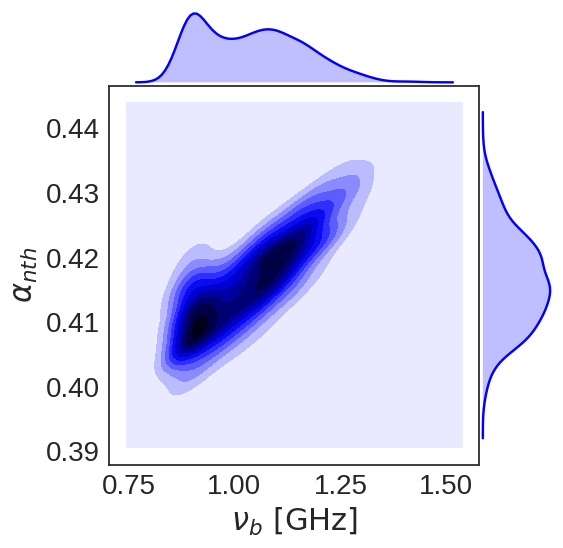}\\

\im{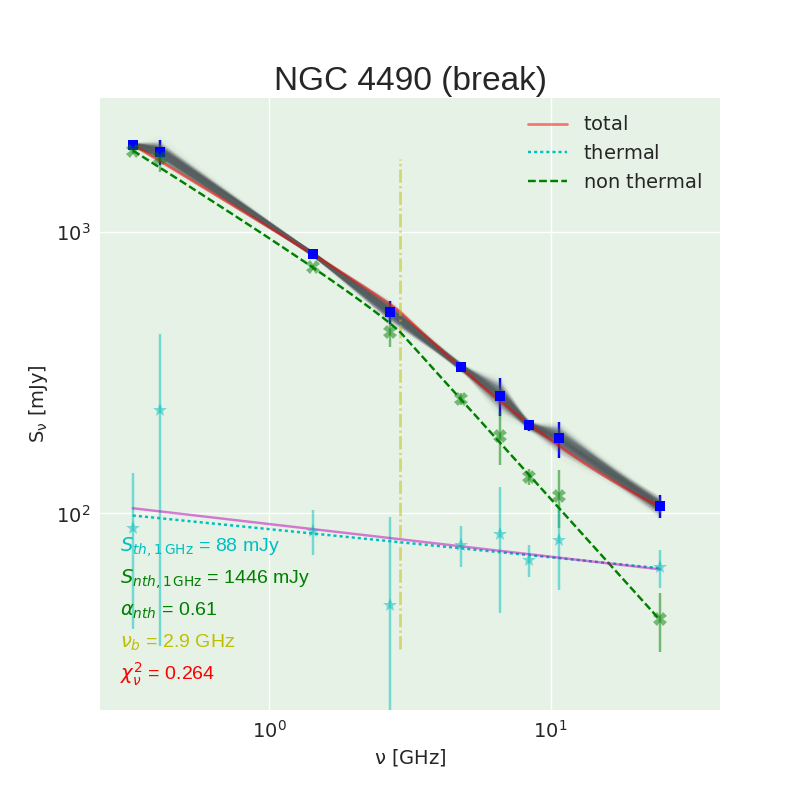} \hfill
\im{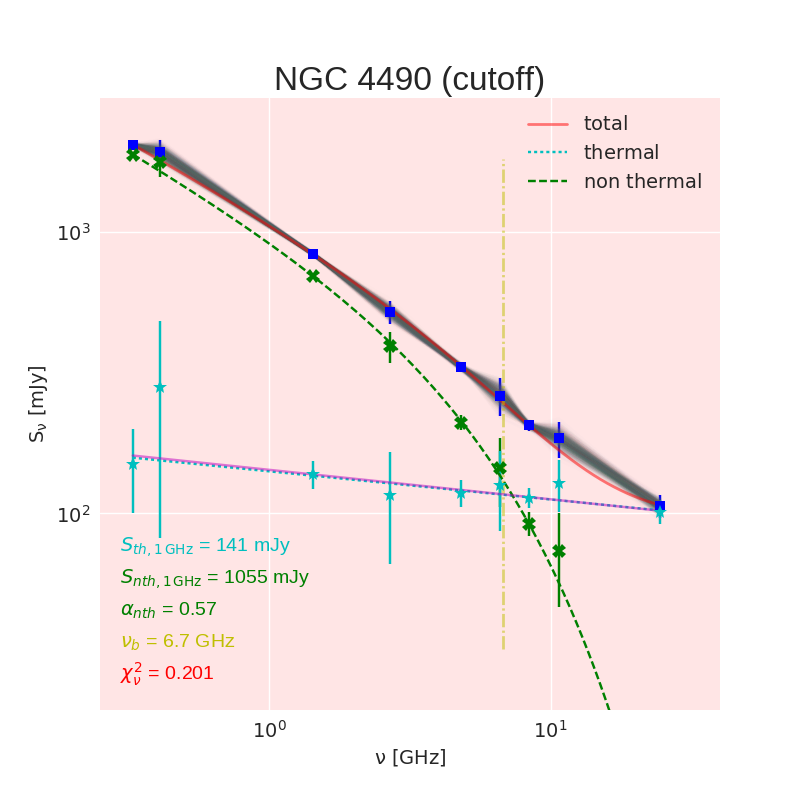}\\

\image{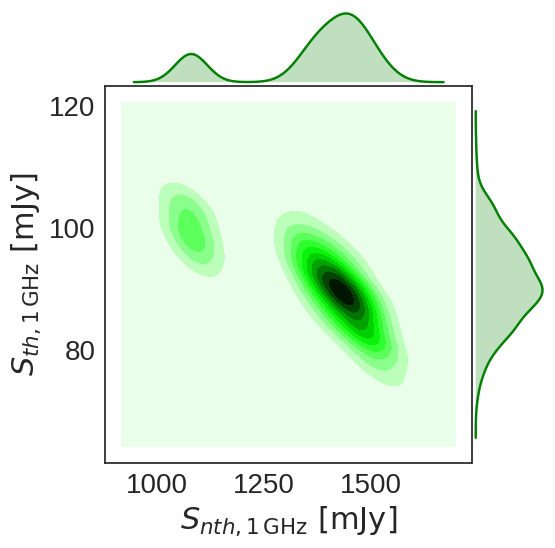}
\image{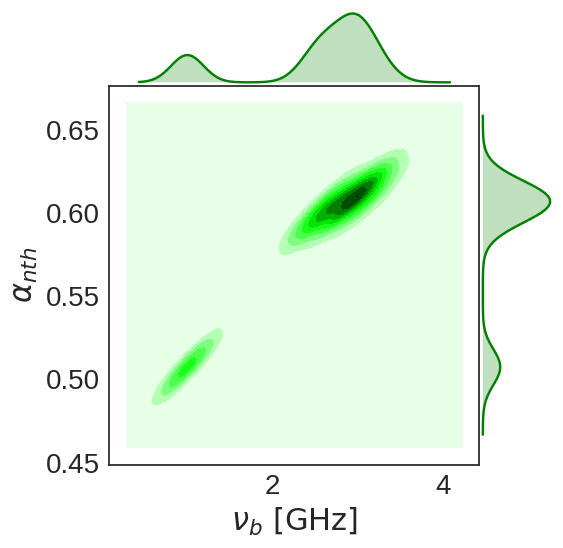} \hfill
\image{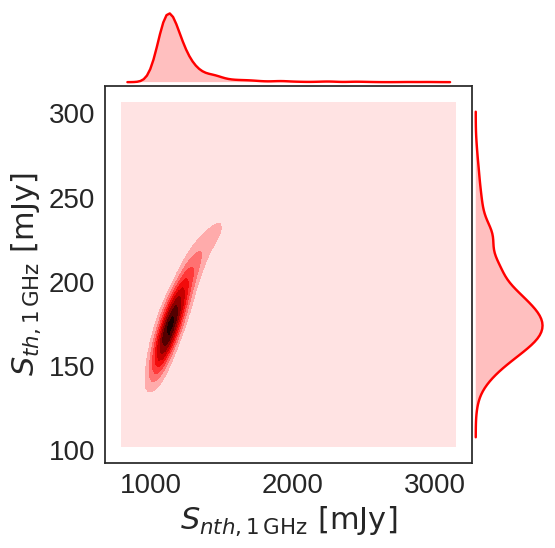}
\image{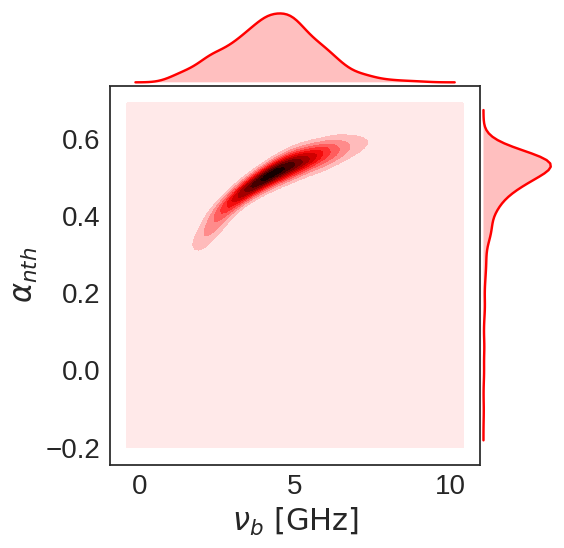}
\end{figure*}
\newpage
\clearpage

\begin{figure*}
\im{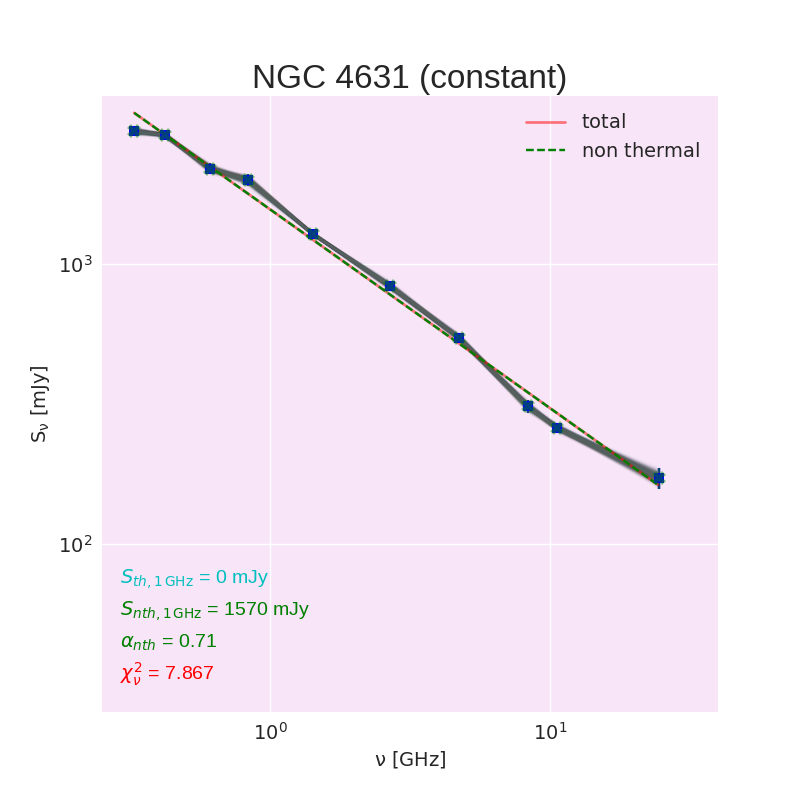} \hfill
\im{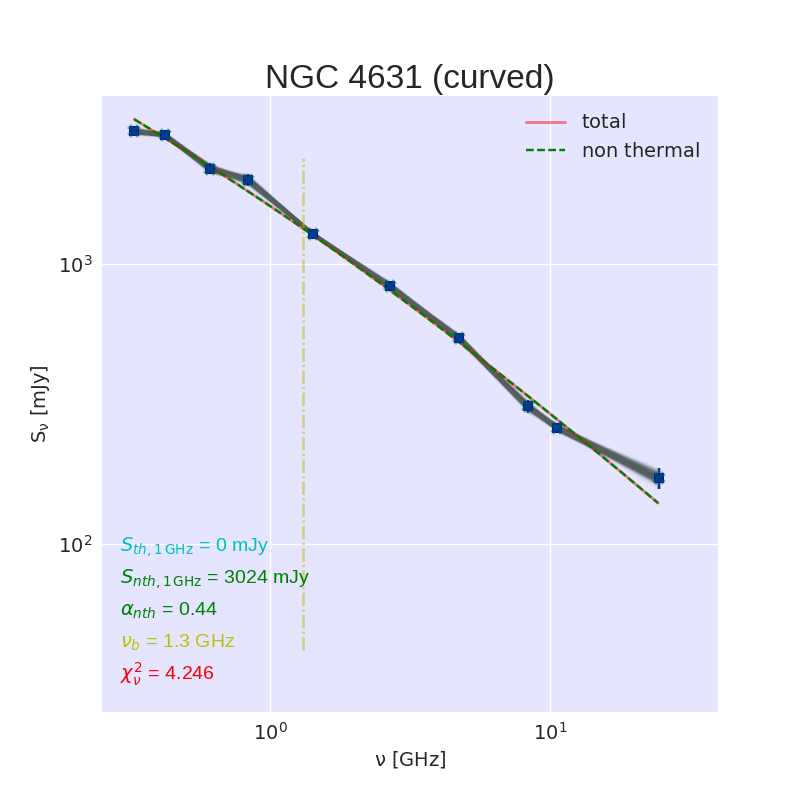}\\

\image{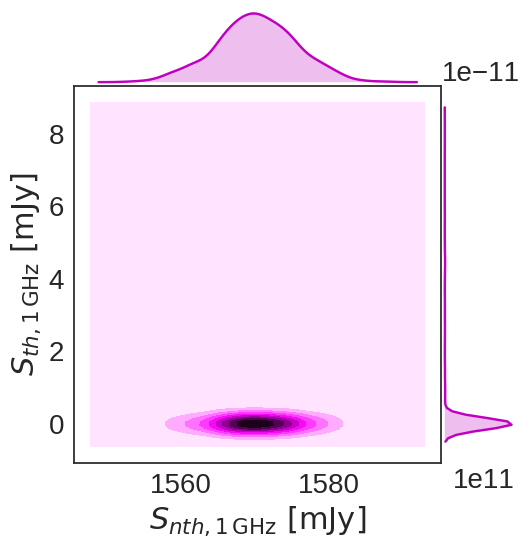}
\image{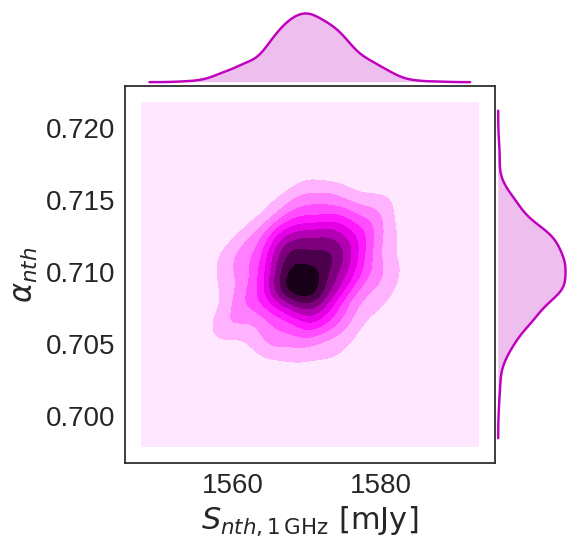} \hfill
\image{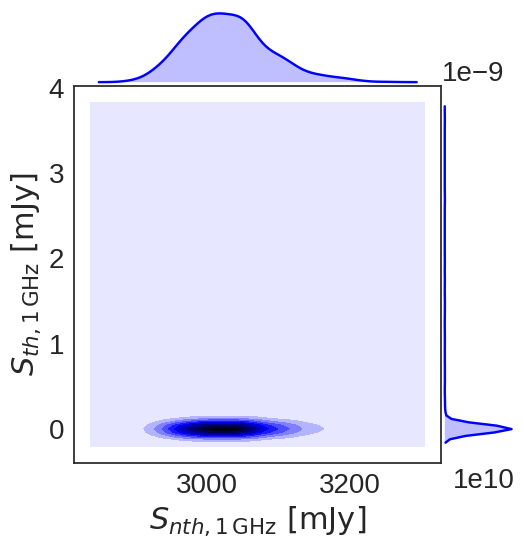}
\image{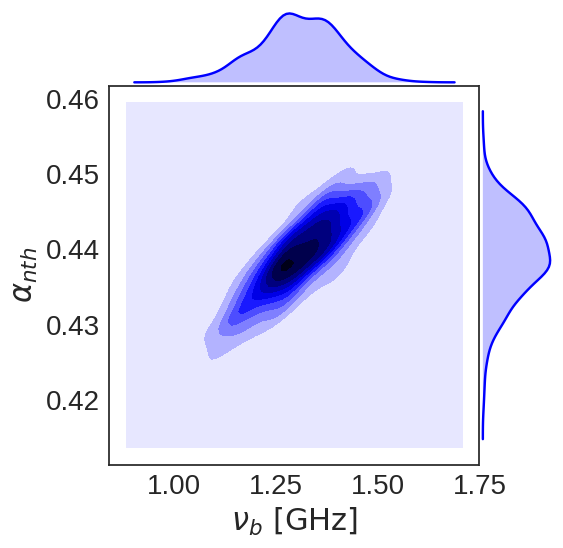}\\

\im{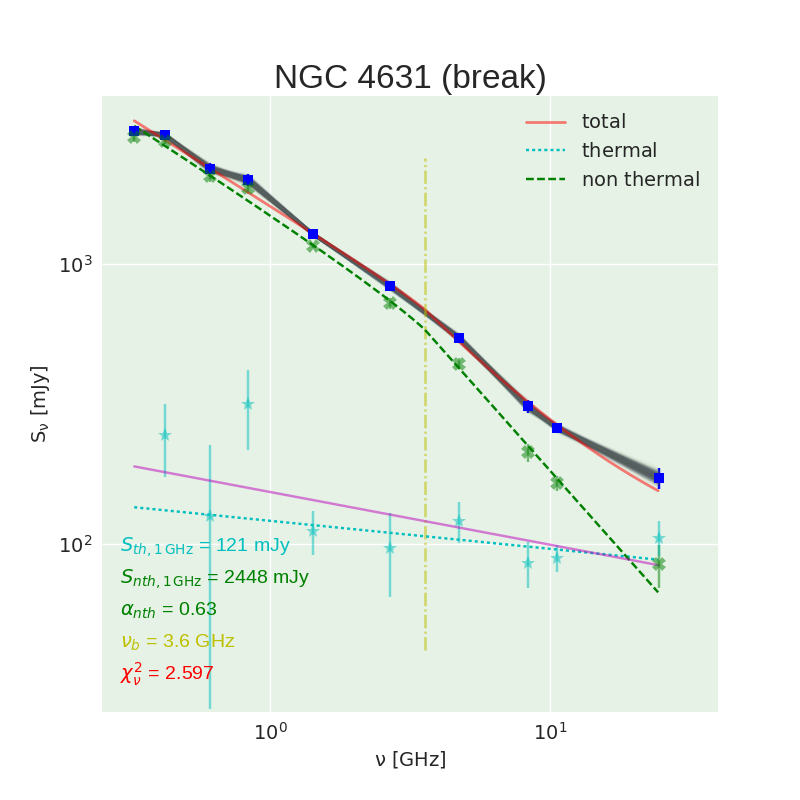} \hfill 
\im{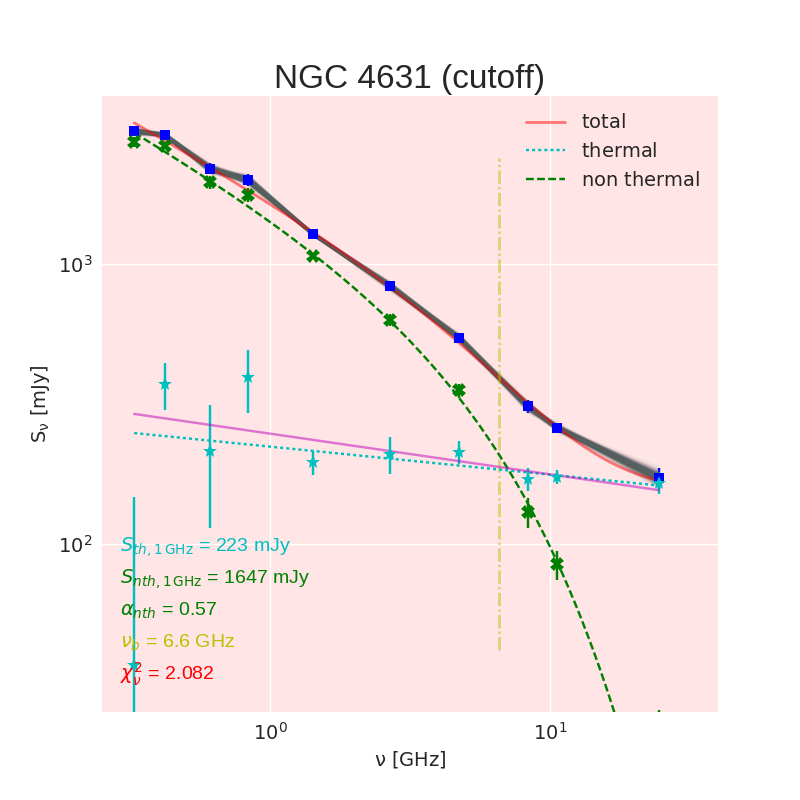}\\

\image{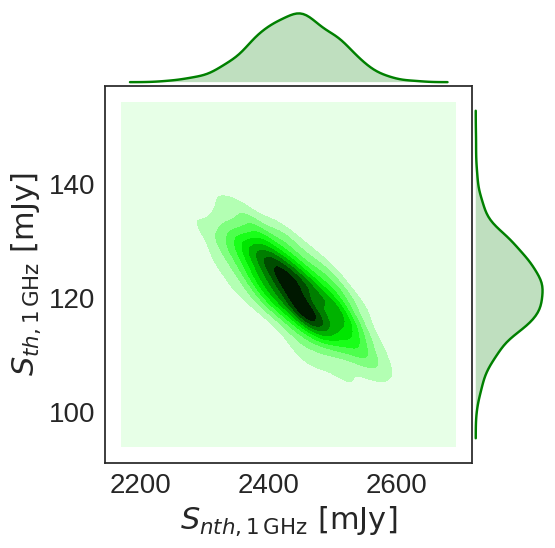} 
\image{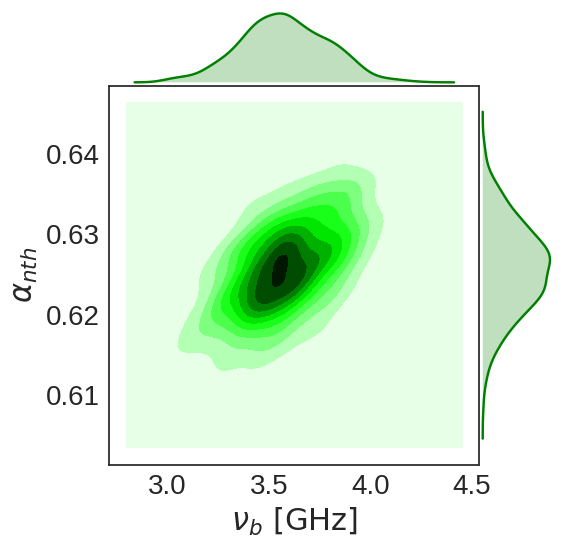}  \hfill
\image{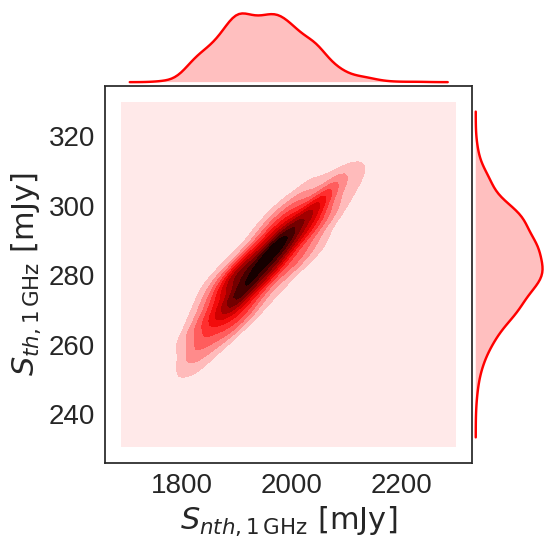}
\image{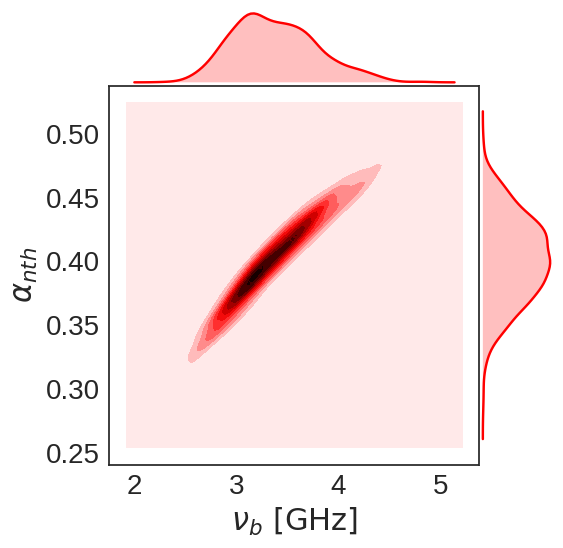}
\end{figure*}
\newpage
\clearpage

\begin{figure*}
\im{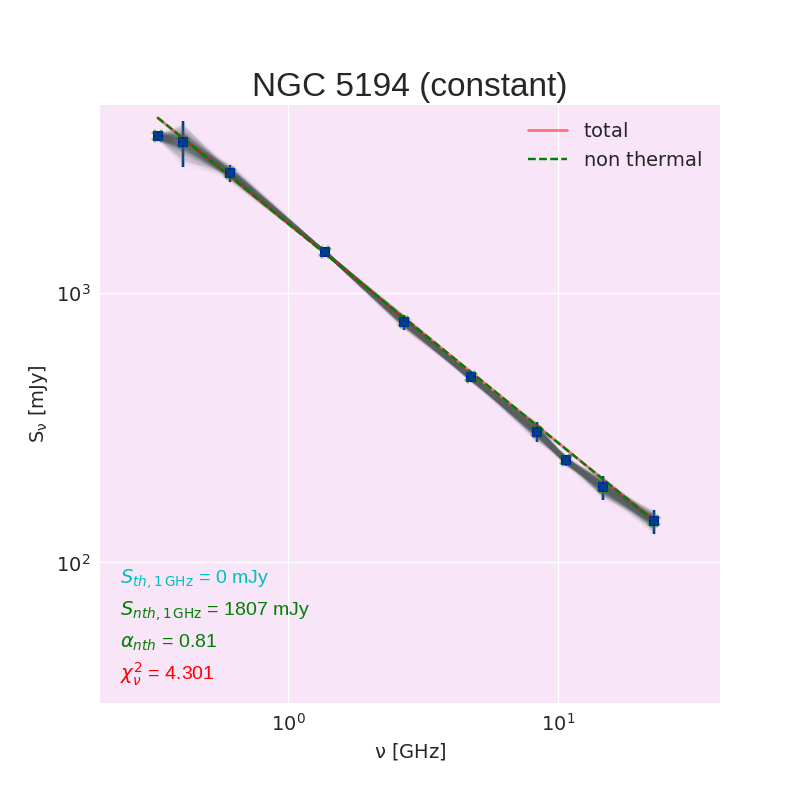} \hfill
\im{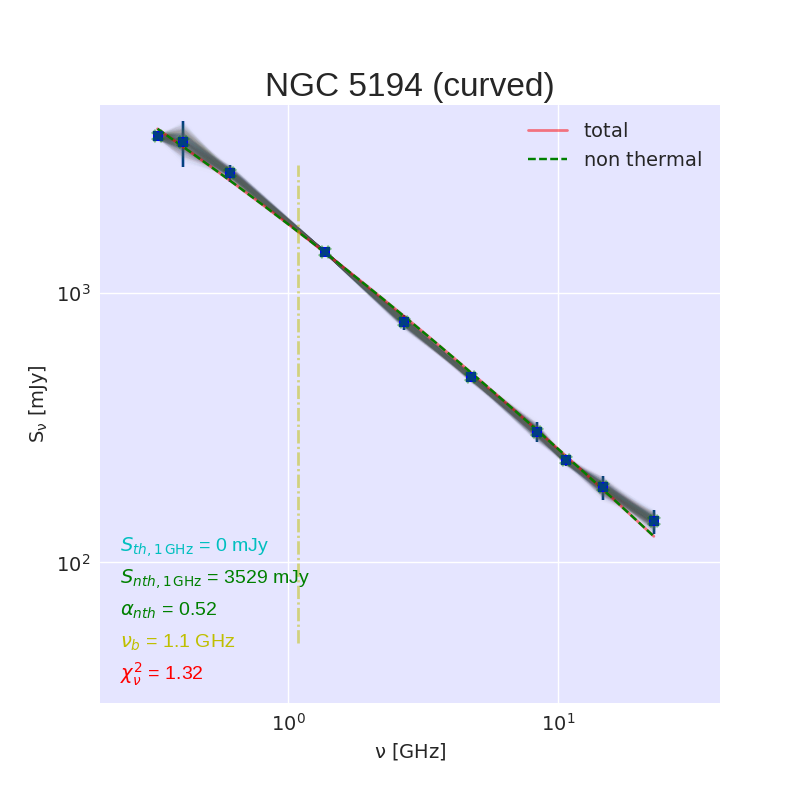}\\

\image{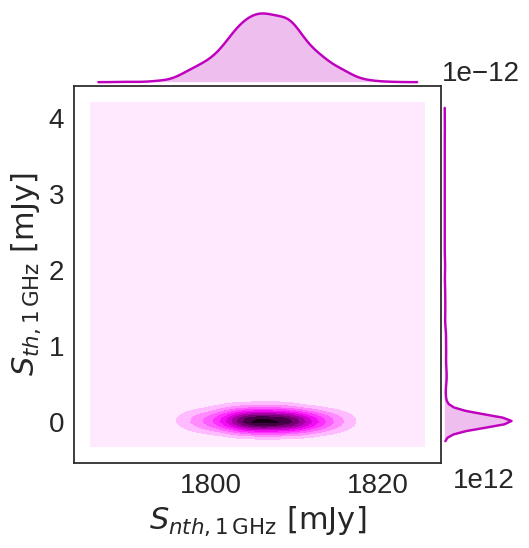}
\image{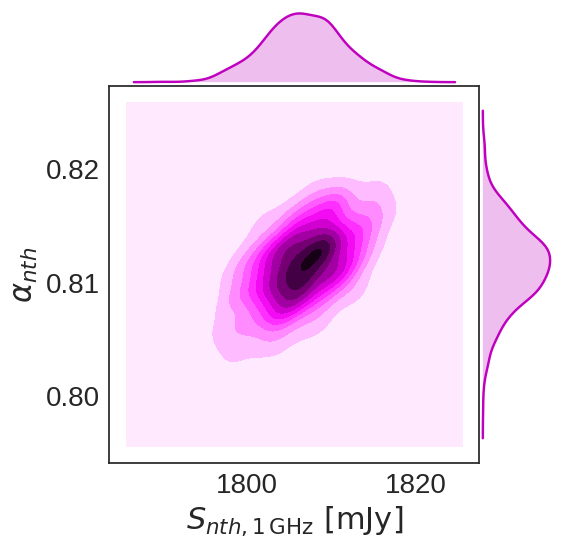} \hfill
\image{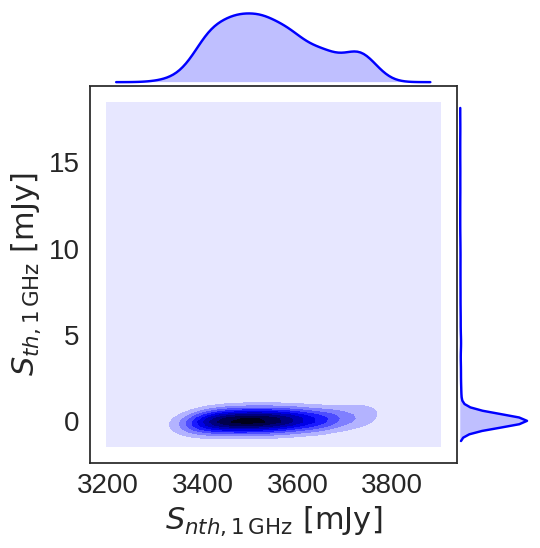}
\image{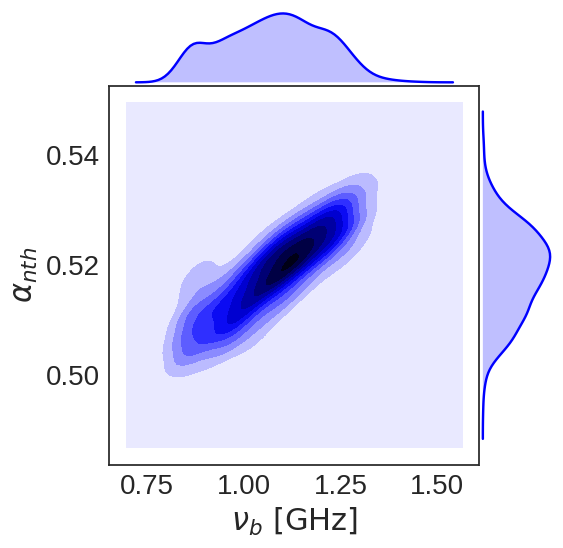}\\

\im{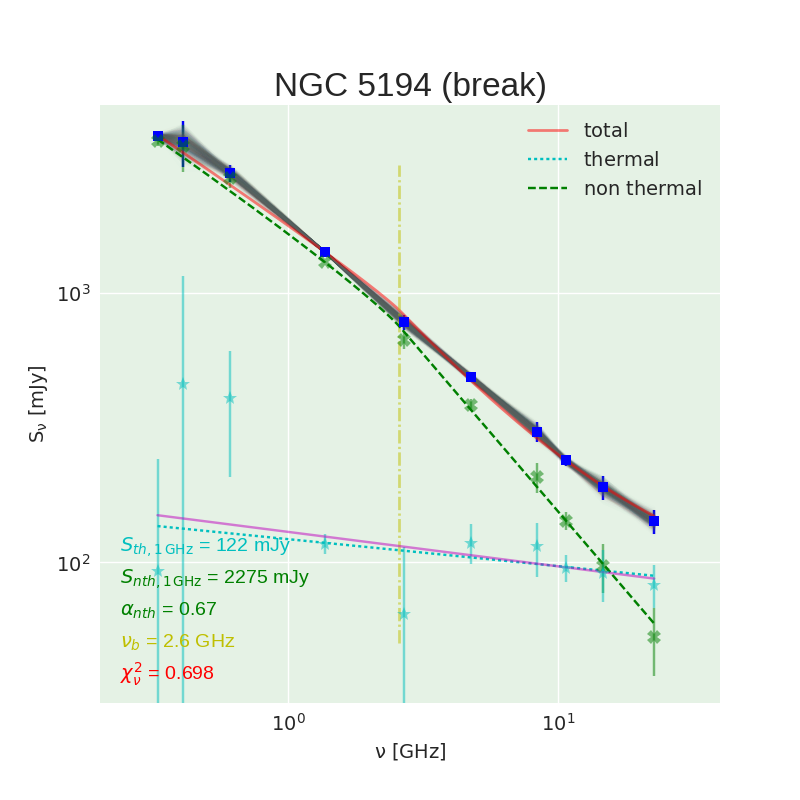}  \hfill
\im{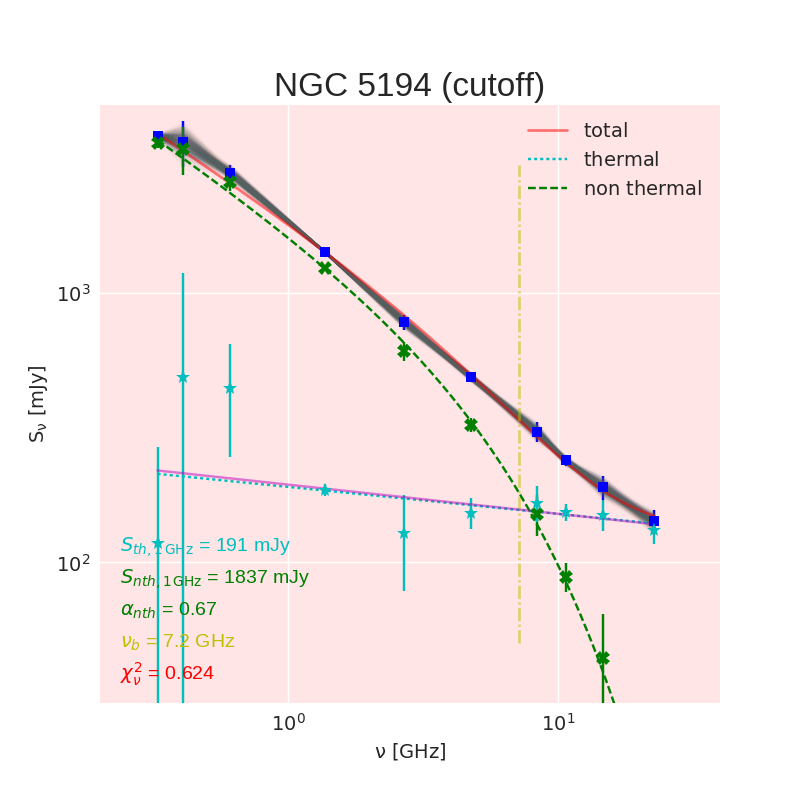}\\

\image{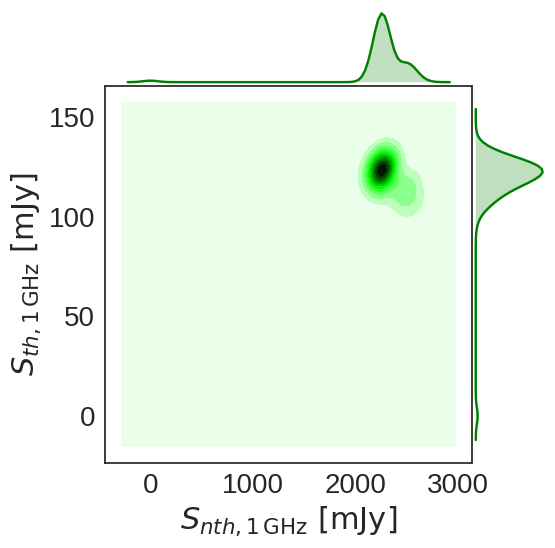} 
\image{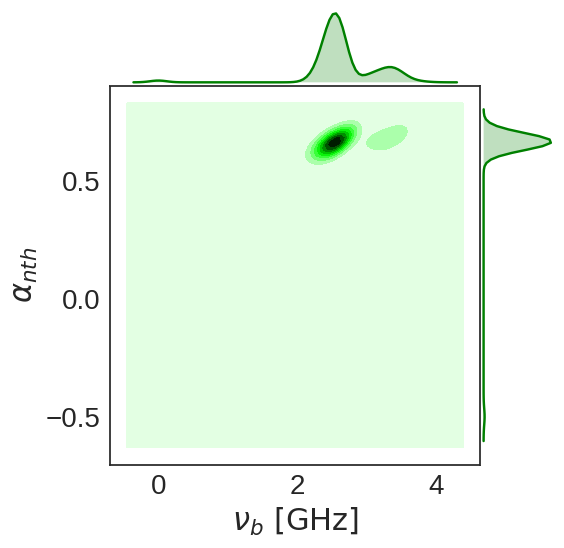}  \hfill
\image{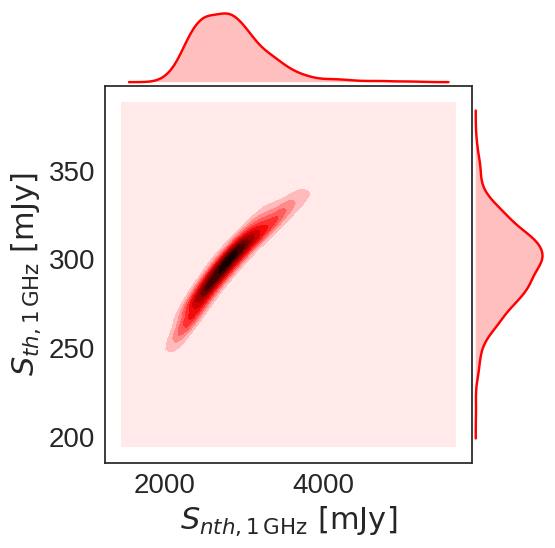}
\image{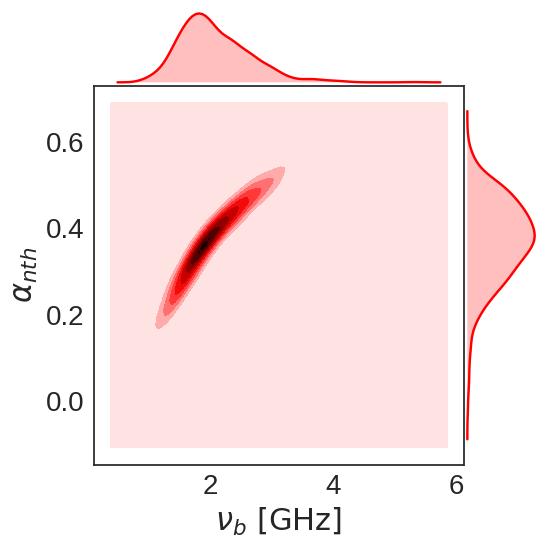}
\end{figure*}
\newpage

\begin{figure*}
\im{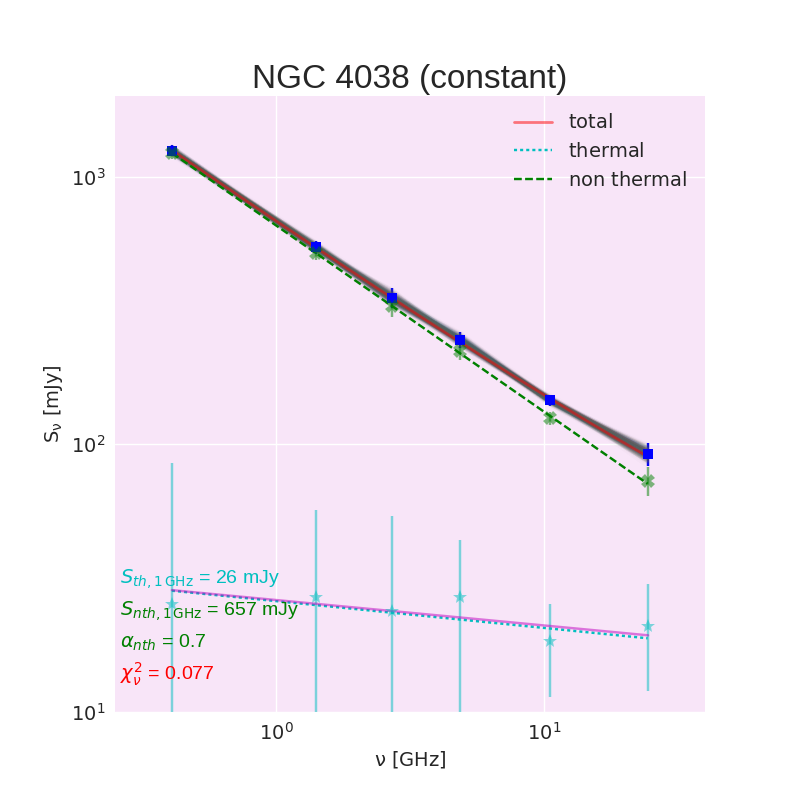} \hfill
\im{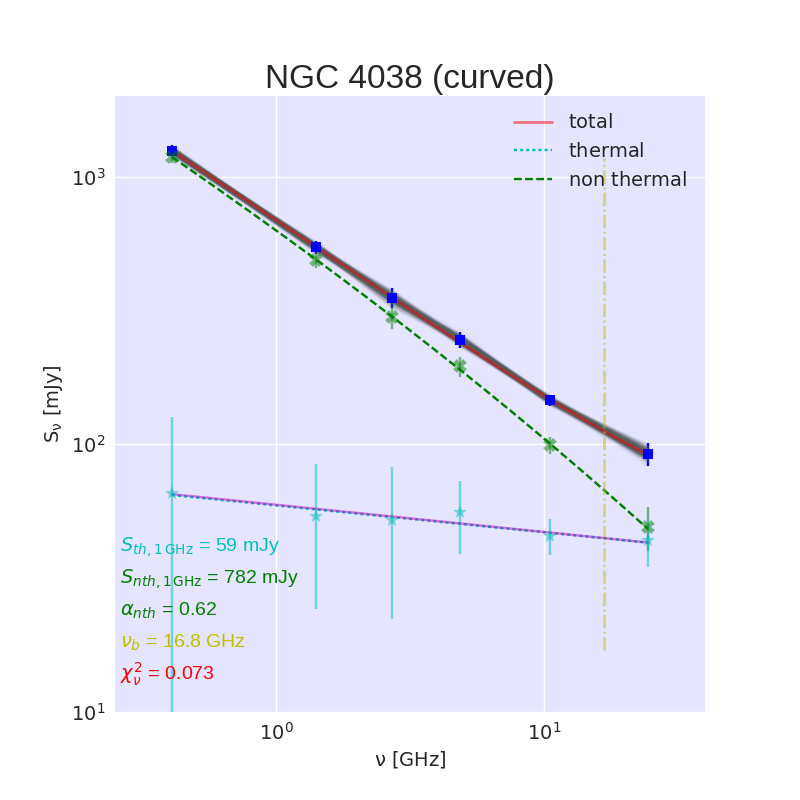}\\

\image{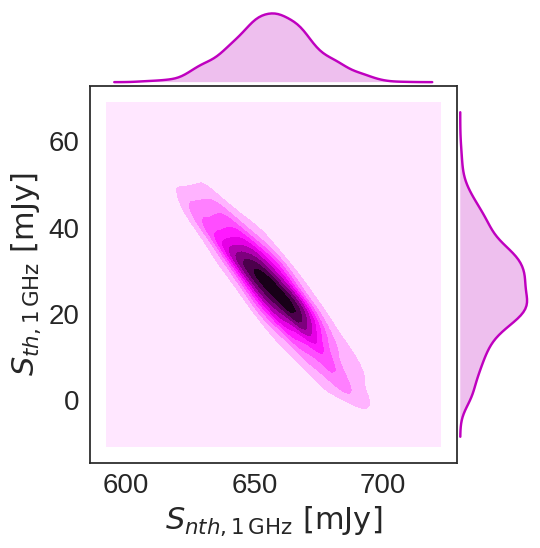}
\image{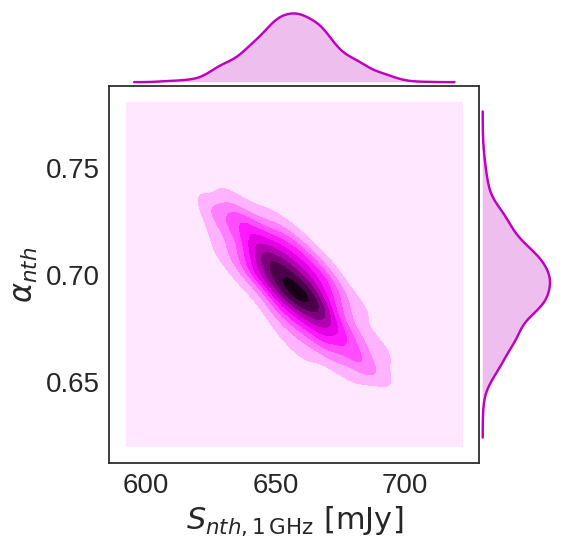} \hfill
\image{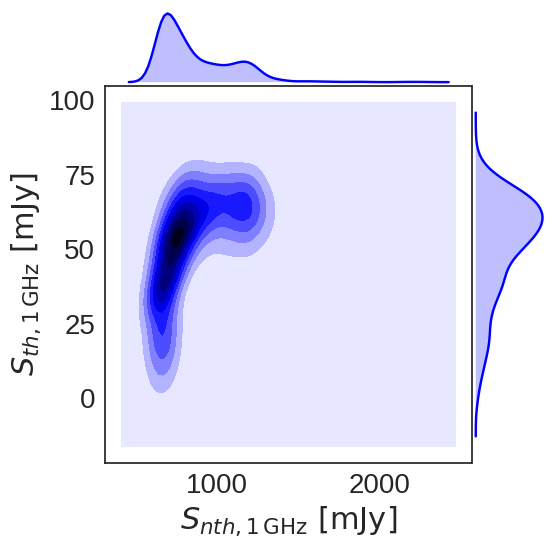}
\image{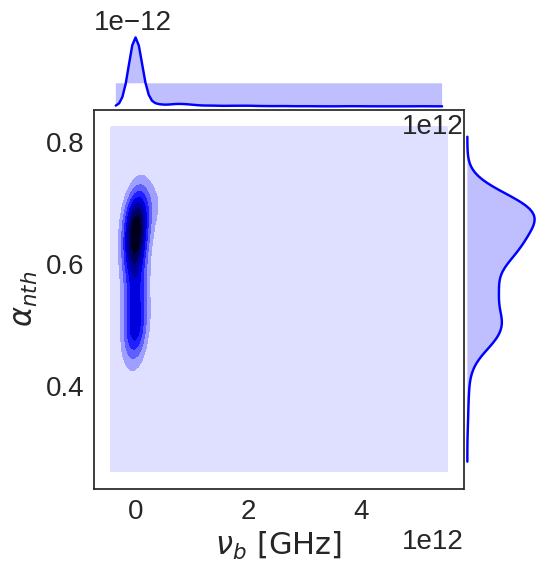}\\

\im{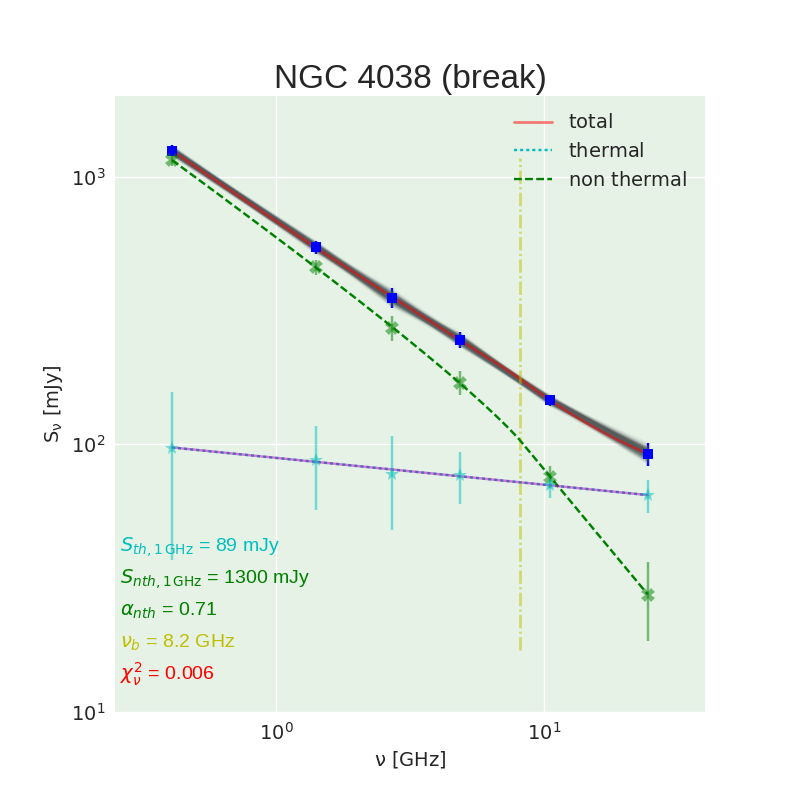} \hfill
\im{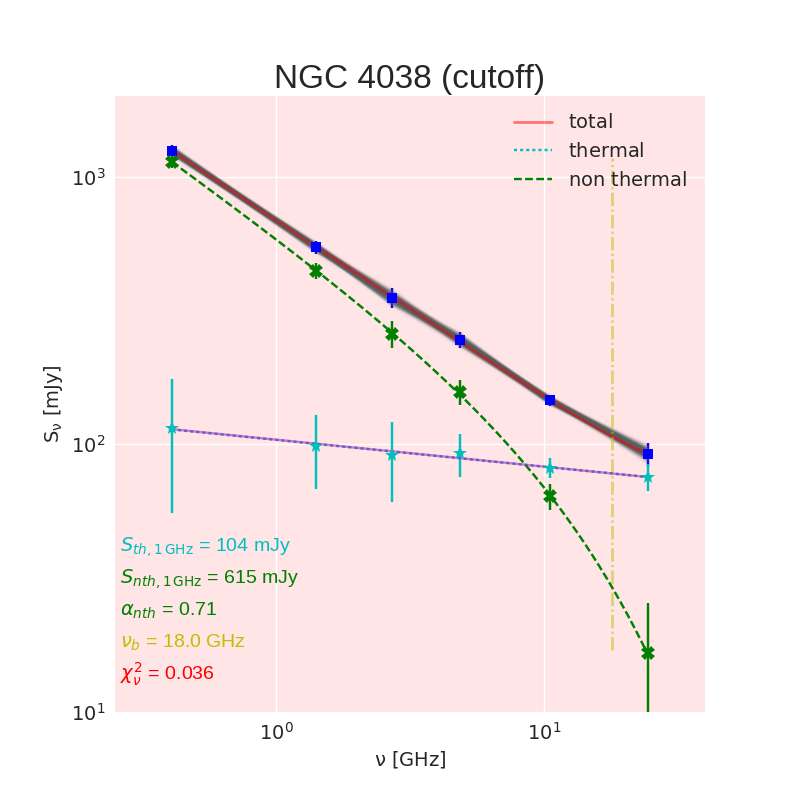}\\

\image{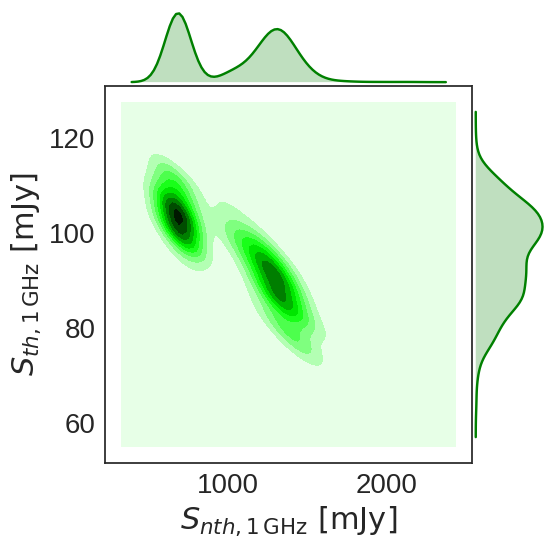}
\image{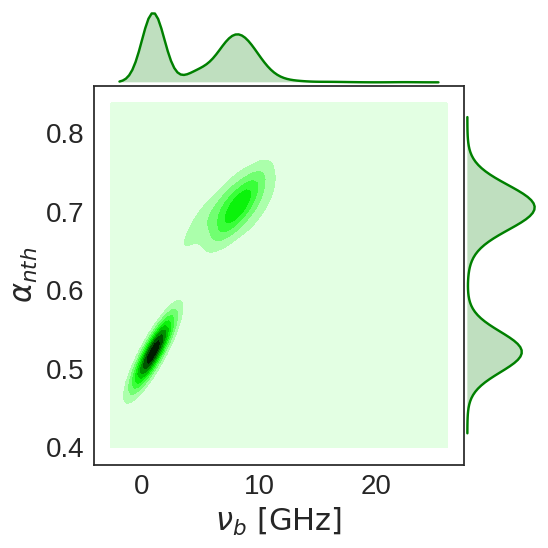} \hfill
\image{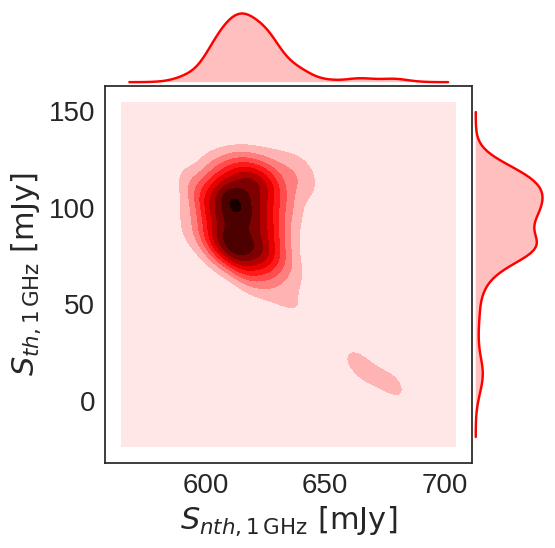}
\image{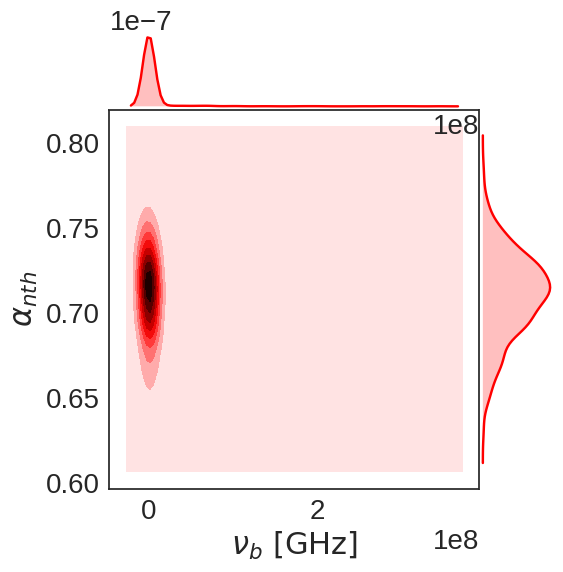}
\end{figure*}
\newpage
\clearpage

\begin{figure*}
\im{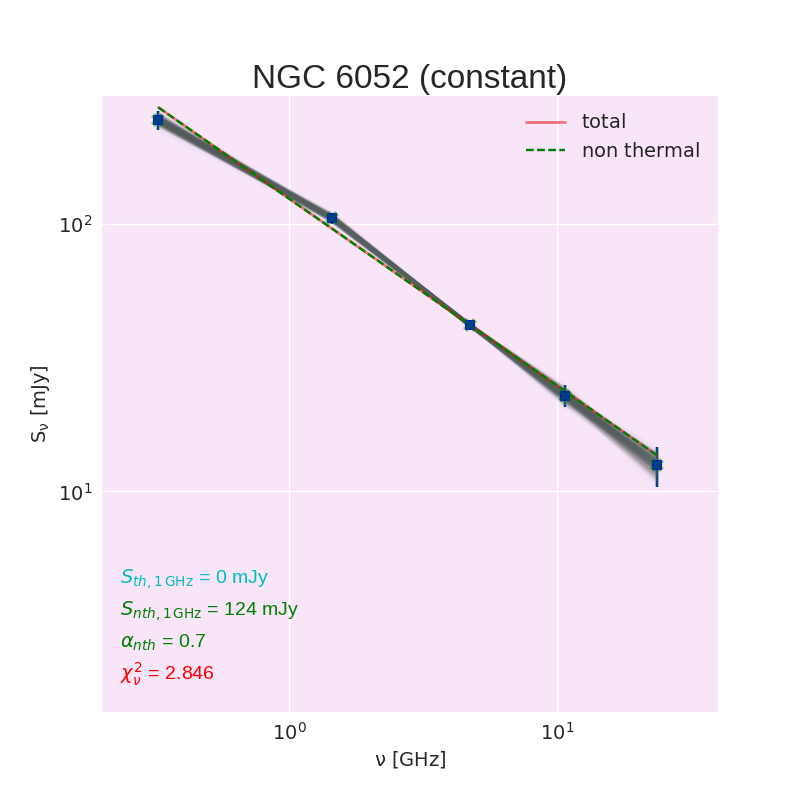} \hfill
\im{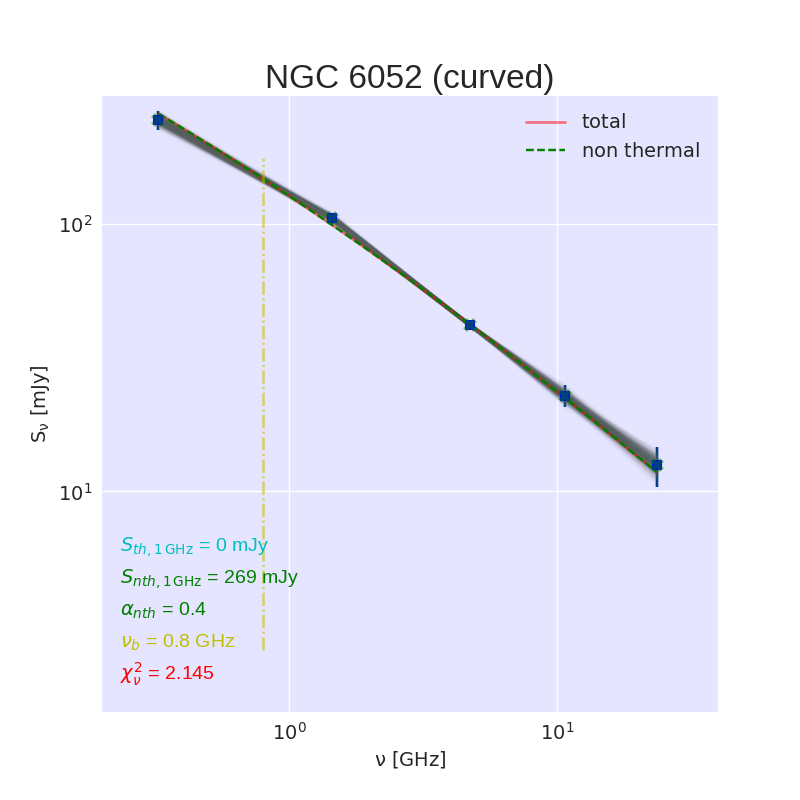}\\

\image{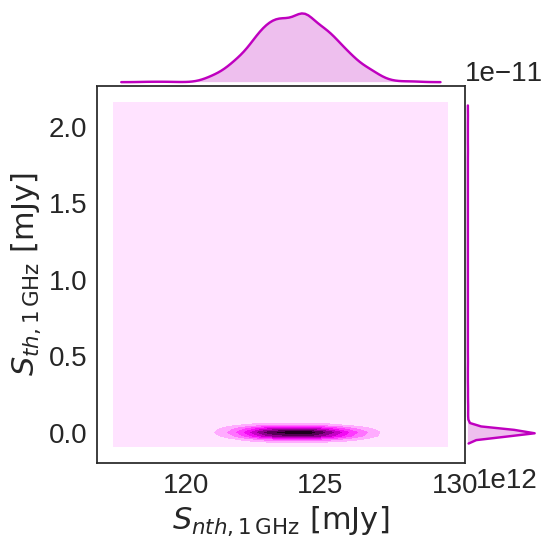}
\image{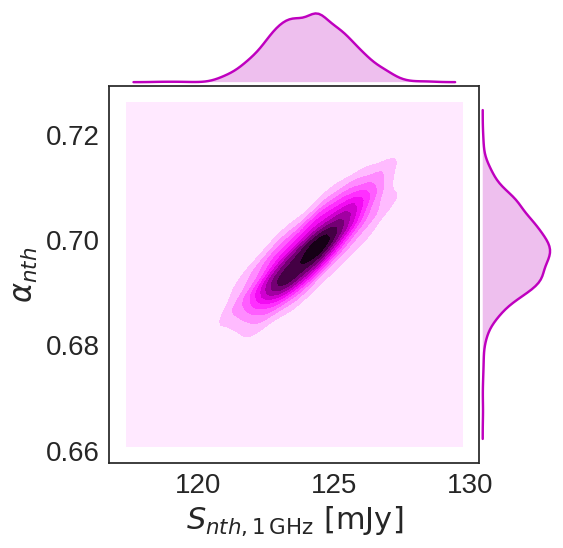} \hfill
\image{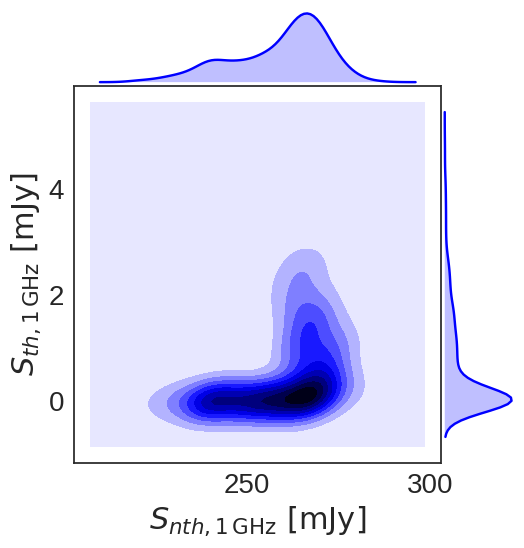}
\image{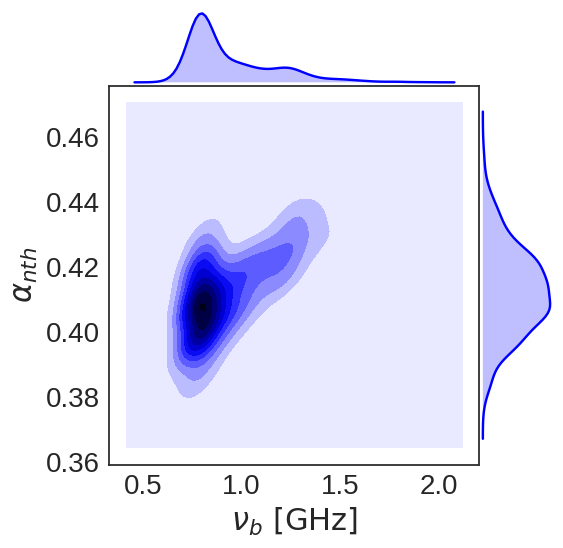}\\

\im{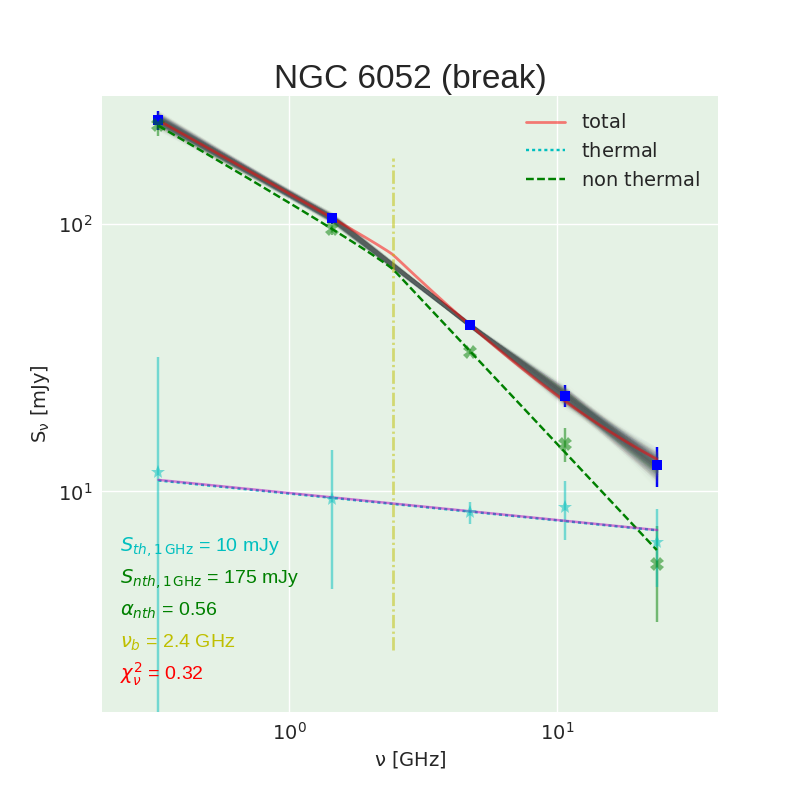} \hfill
\im{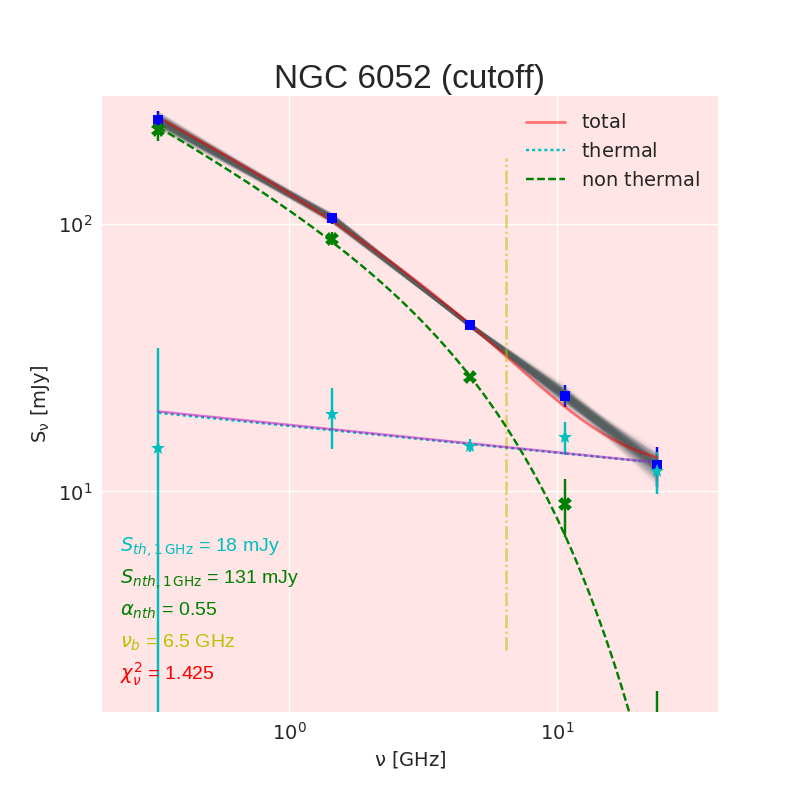}\\

\image{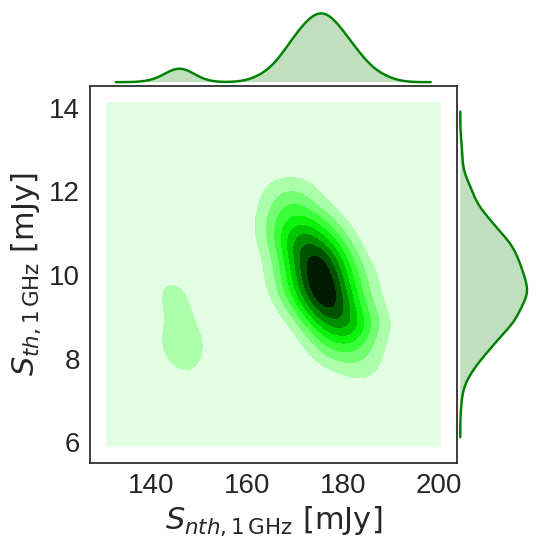}
\image{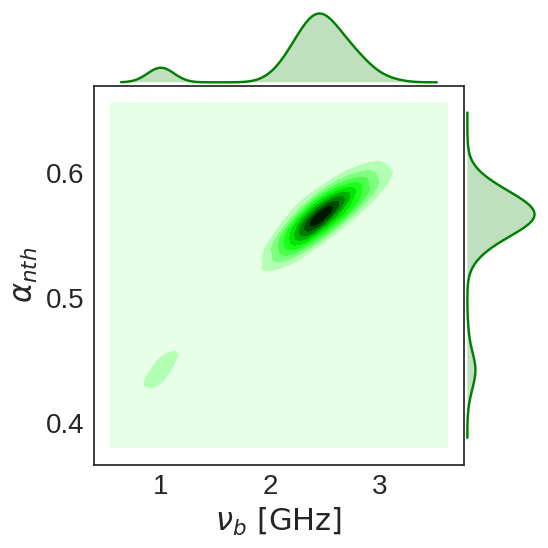} \hfill
\image{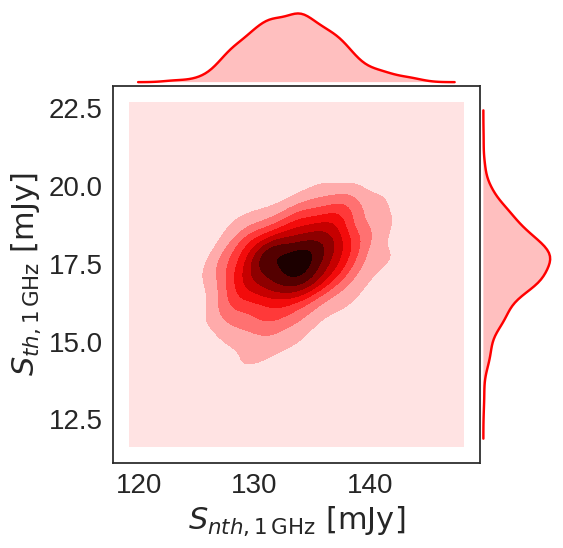}
\image{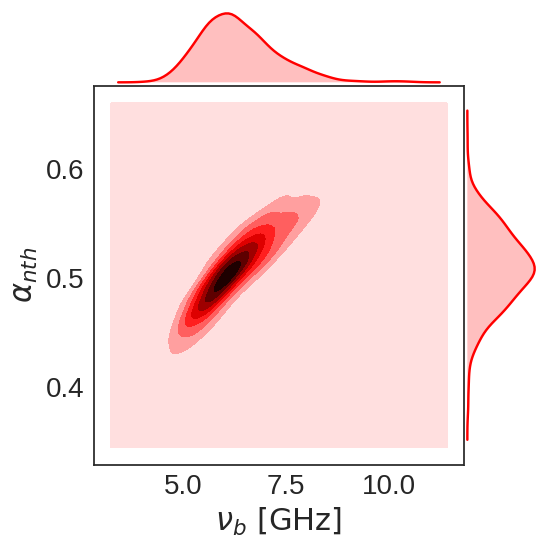}
\end{figure*}
\end{center}

\begin{figure*}
\im{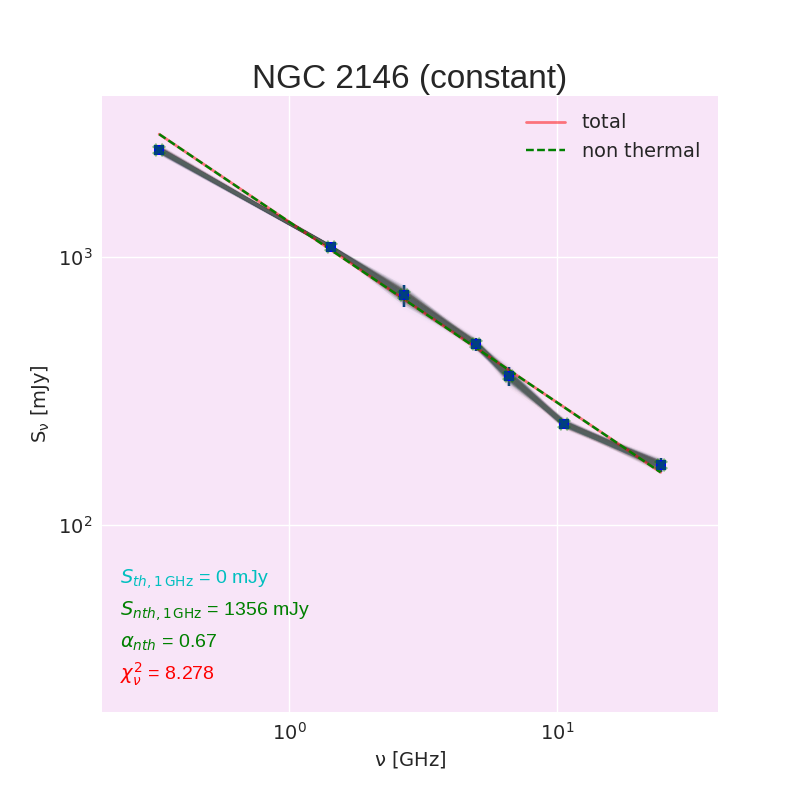} \hfill
\im{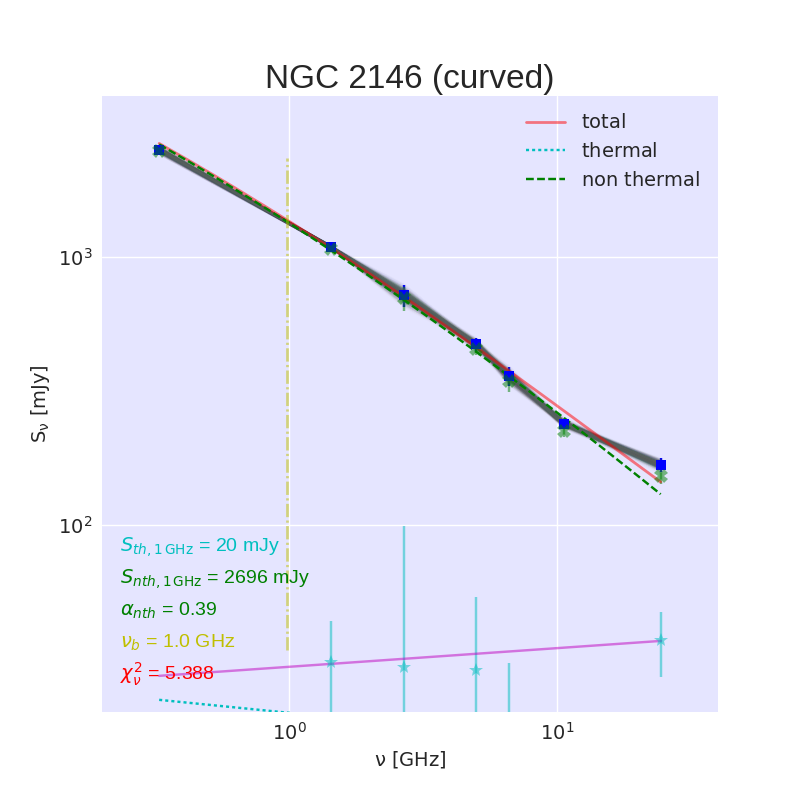}\\

\image{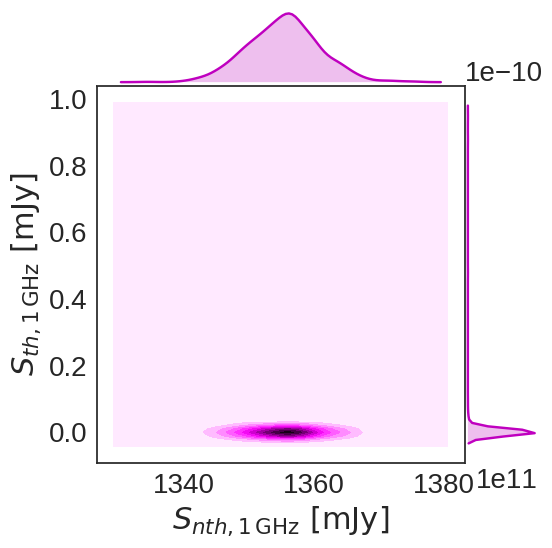}
\image{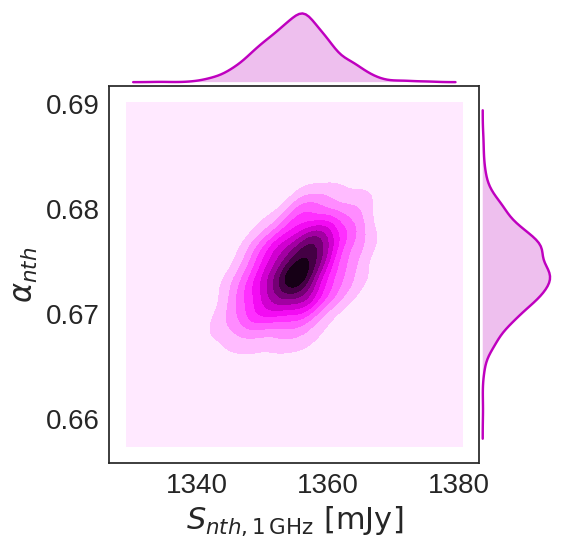} \hfill
\image{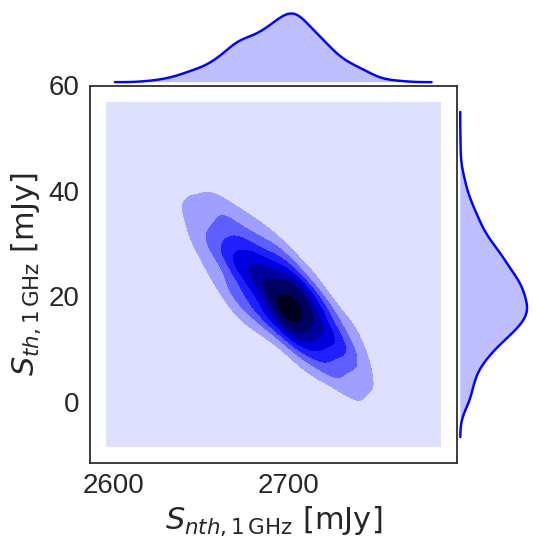}
\image{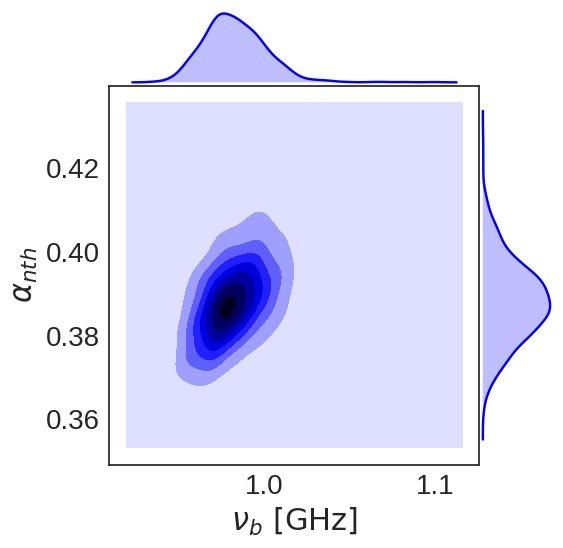}\\

\im{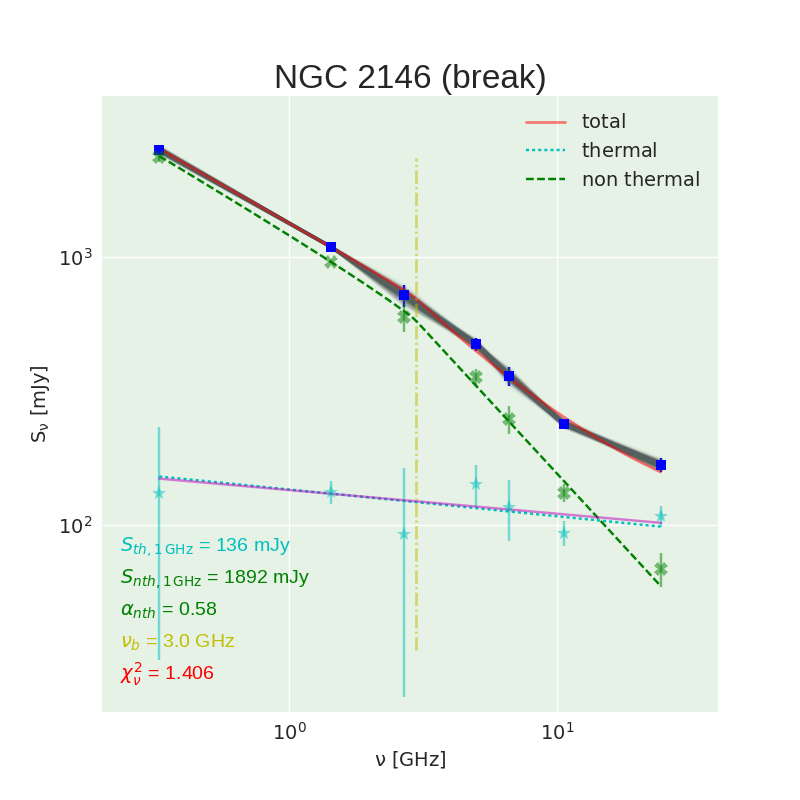}  \hfill
\im{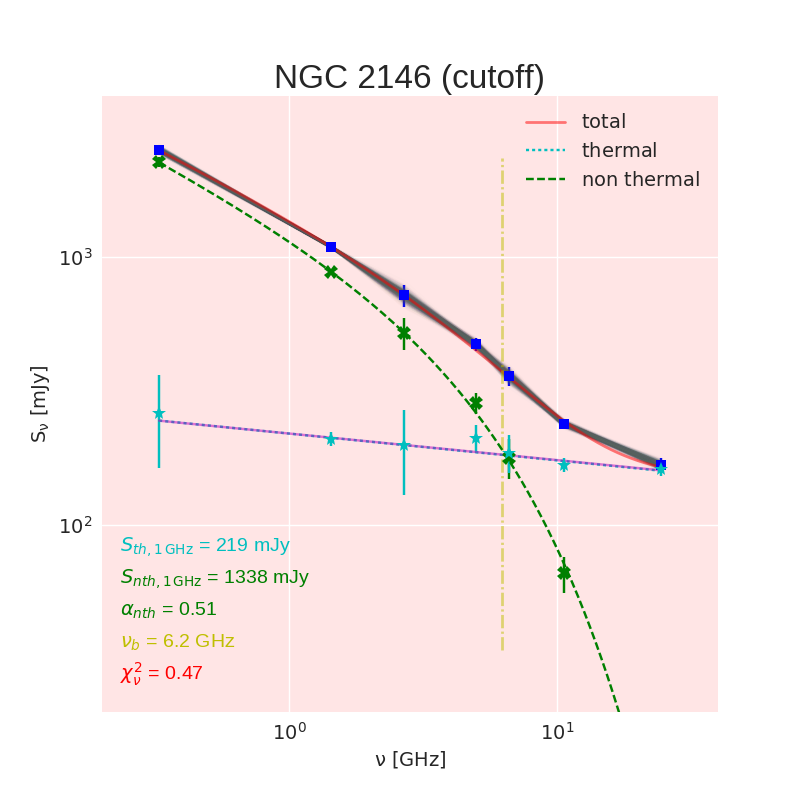}\\

\image{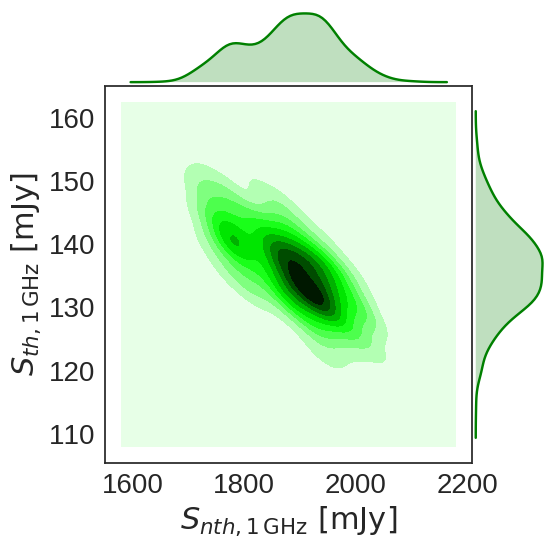} 
\image{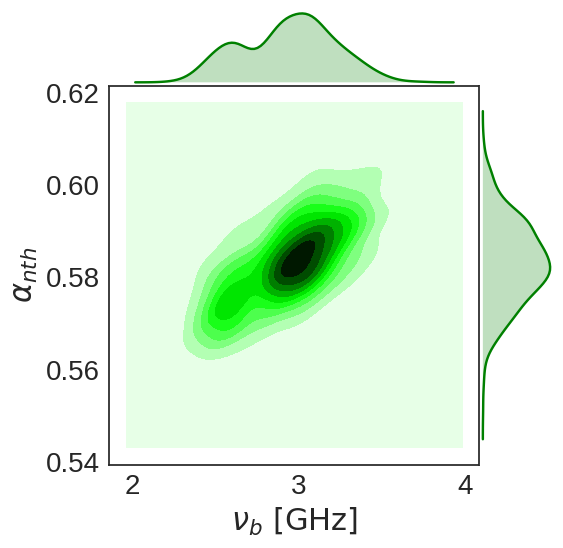}  \hfill
\image{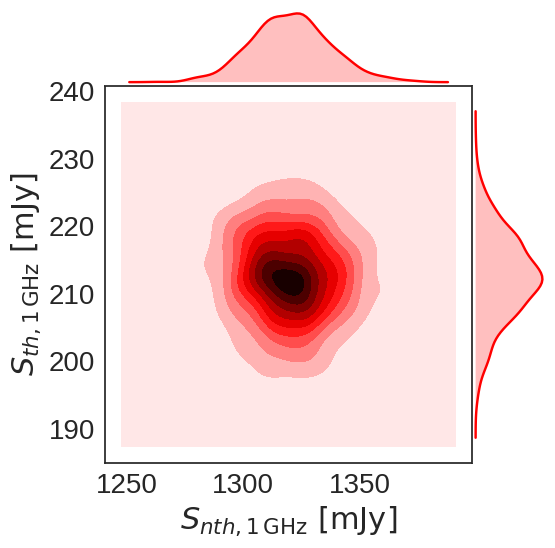}
\image{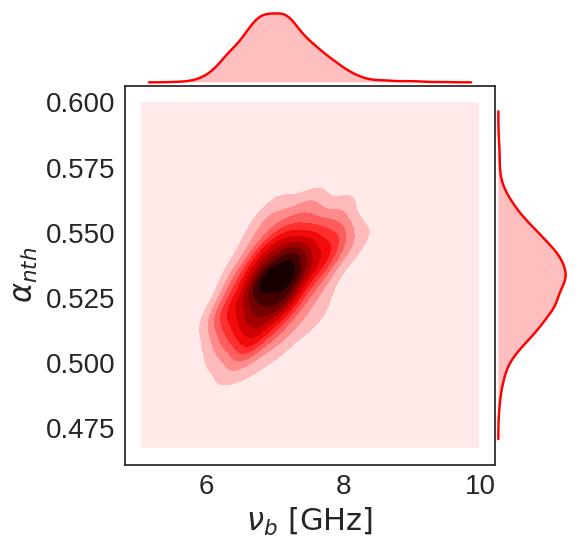}
\end{figure*}
\newpage
\clearpage

\begin{figure*}
\im{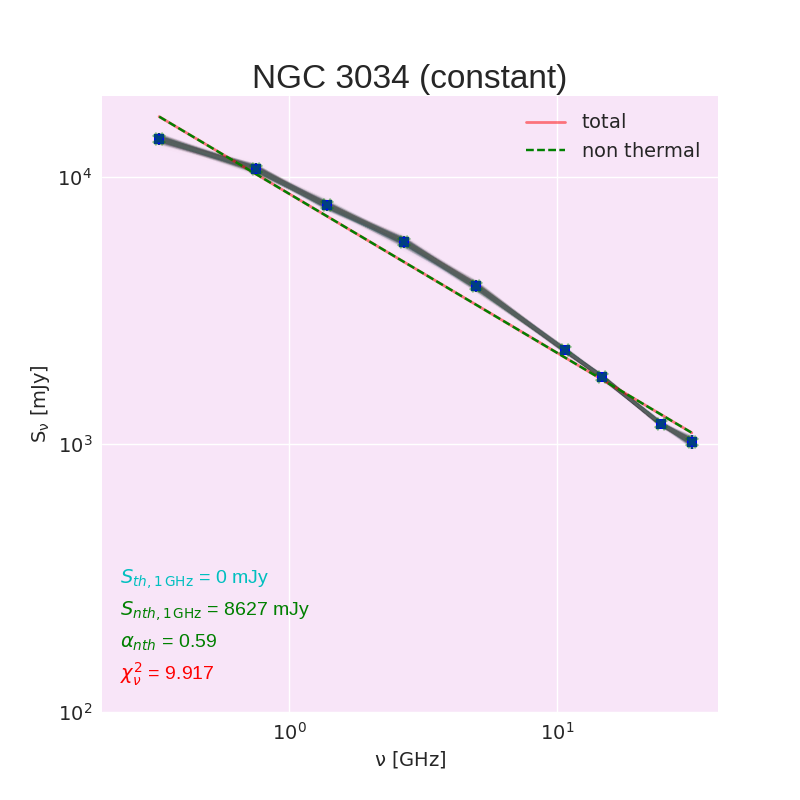} \hfill
\im{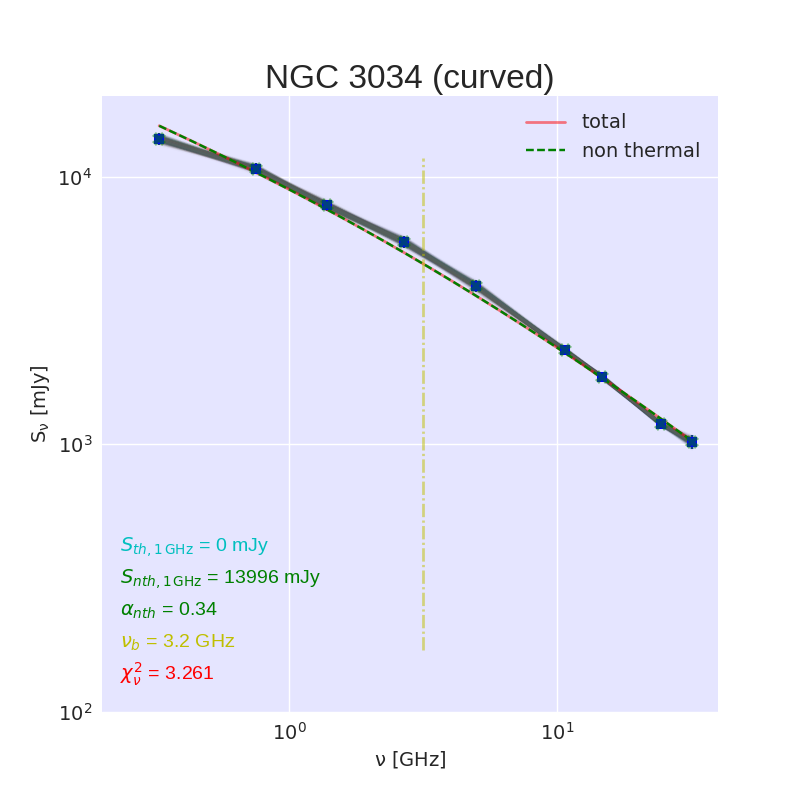}\\

\image{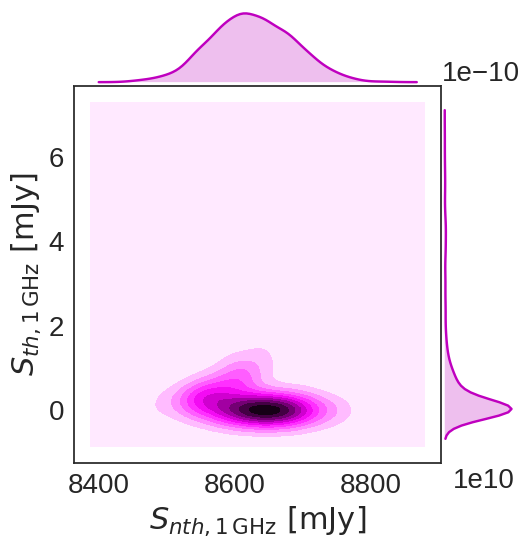}
\image{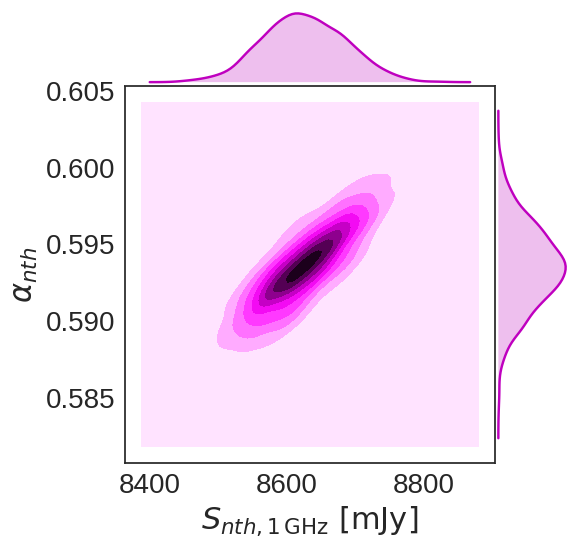} \hfill
\image{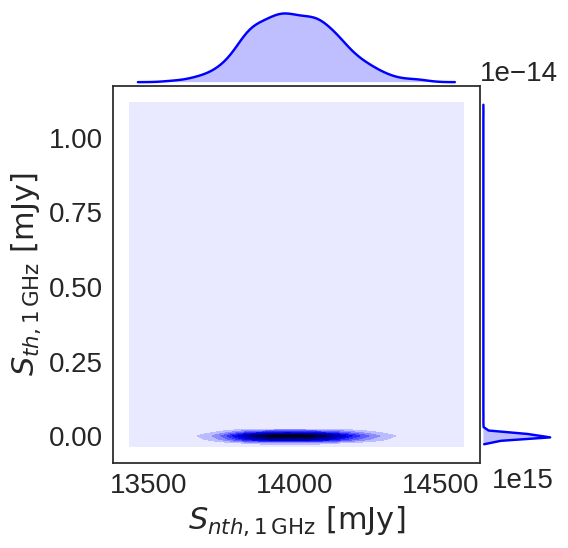}
\image{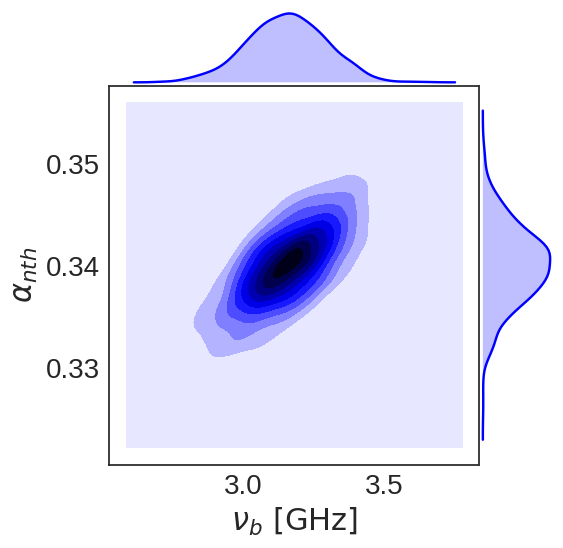}\\

\im{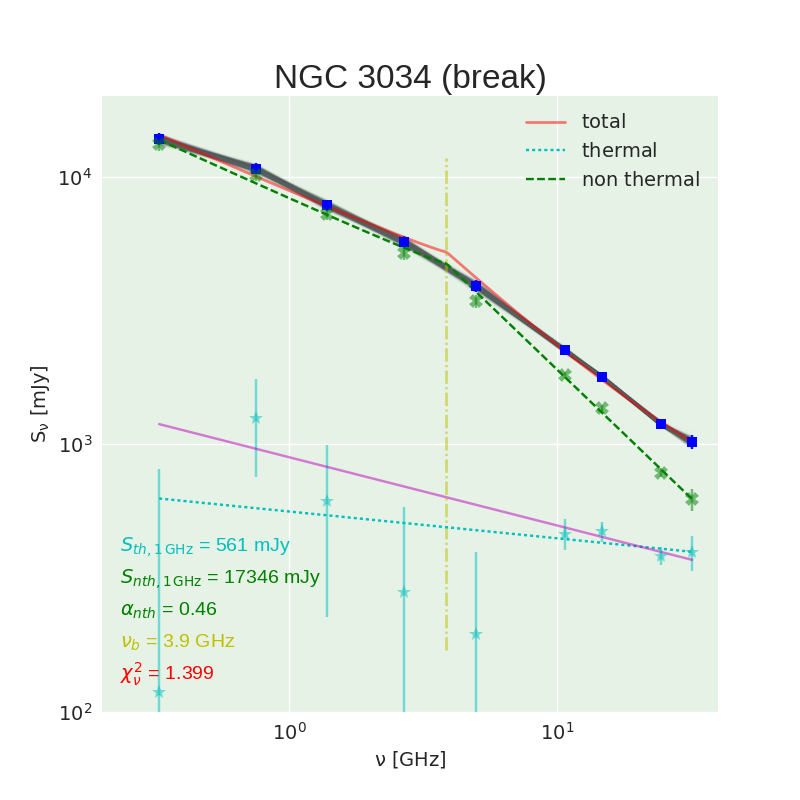}  \hfill
\im{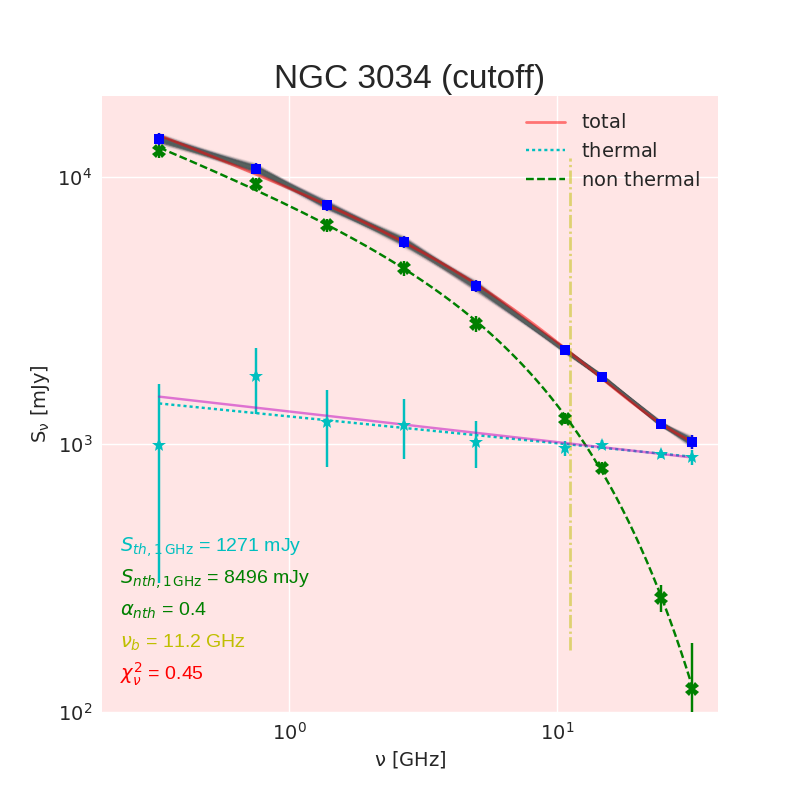}\\

\image{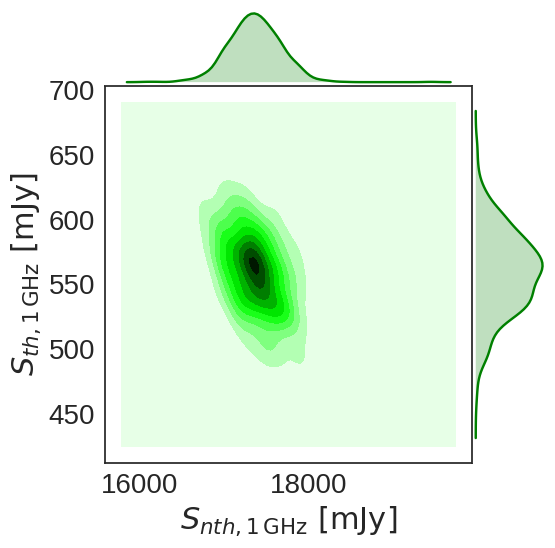} 
\image{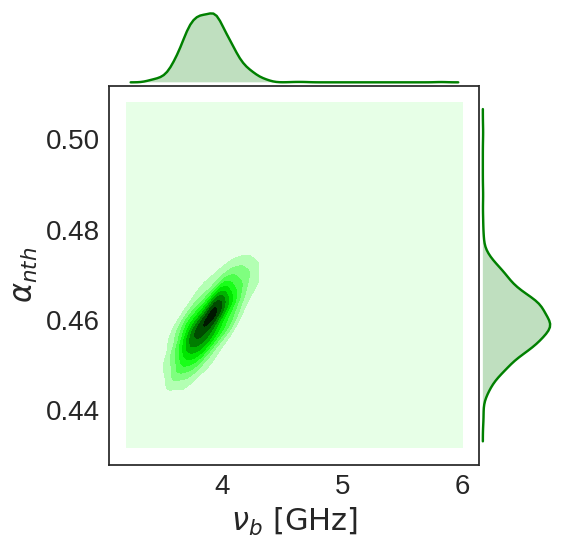}  \hfill
\image{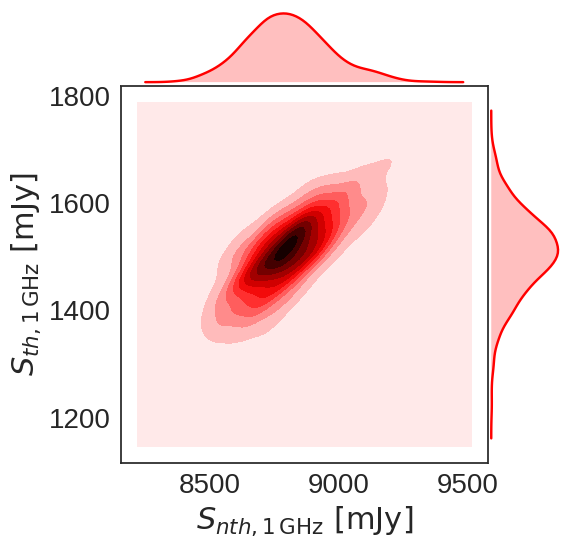}
\image{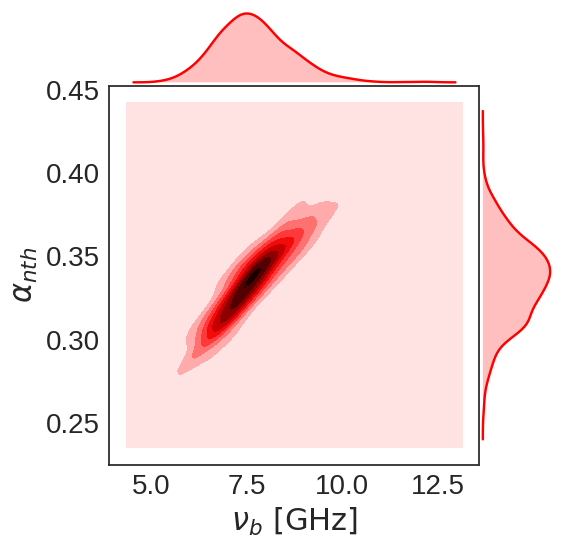}
\end{figure*}
\newpage
\clearpage

\begin{figure*}
\im{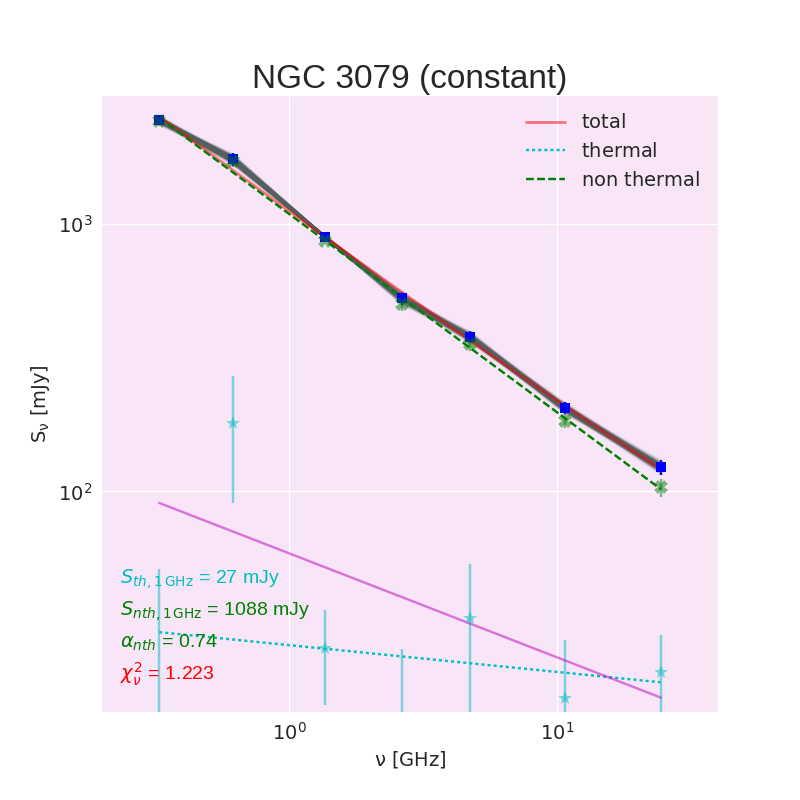} \hfill
\im{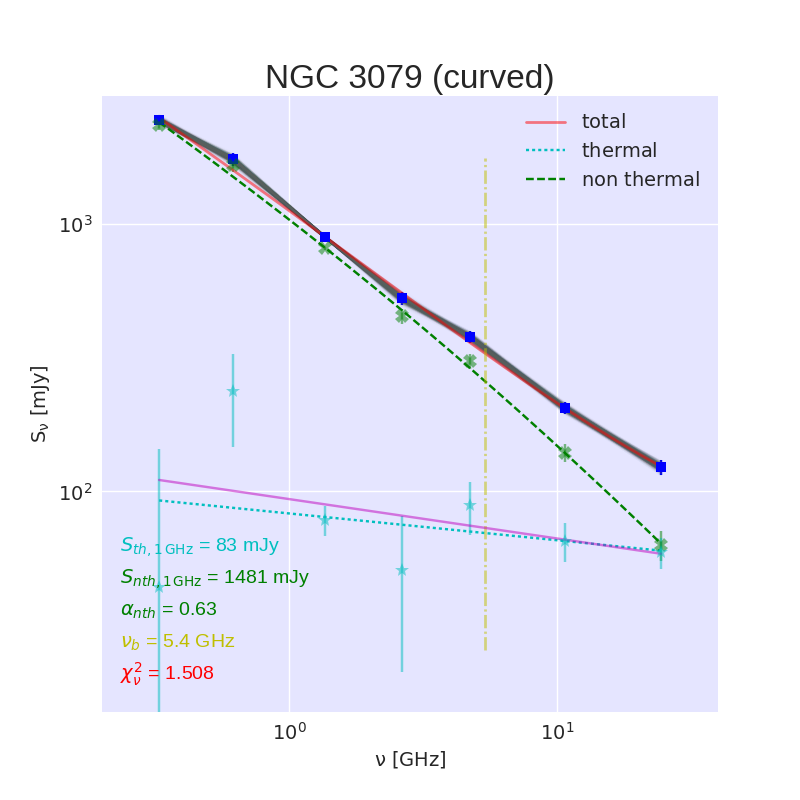}\\

\image{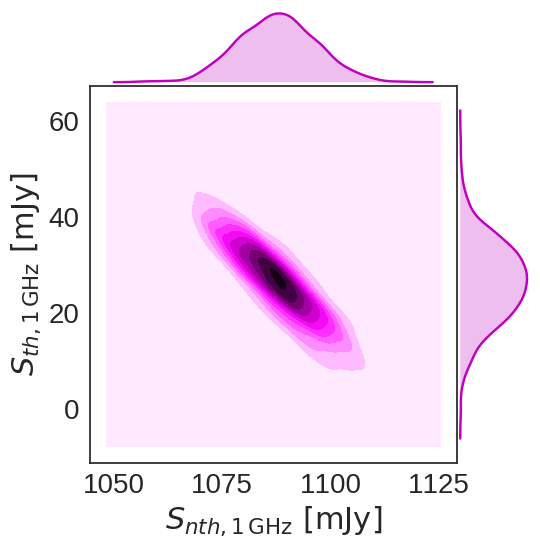}
\image{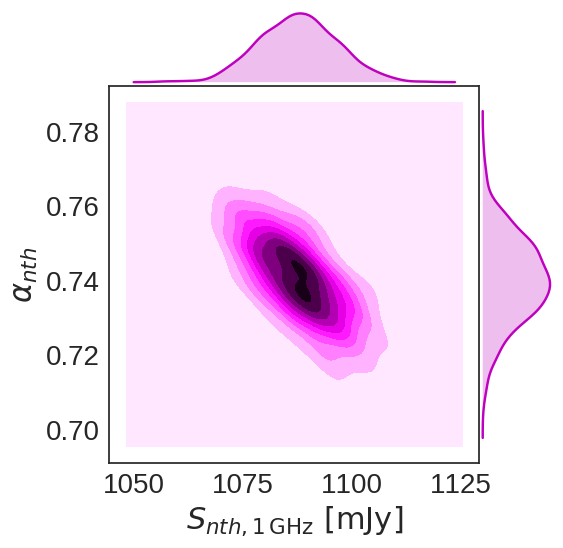} \hfill
\image{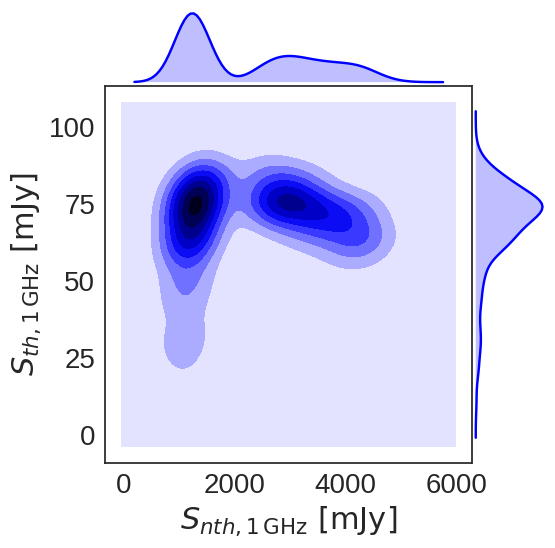}
\image{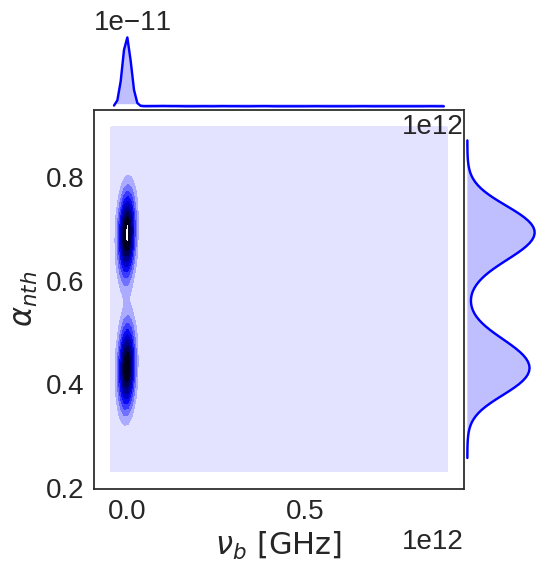}\\

\im{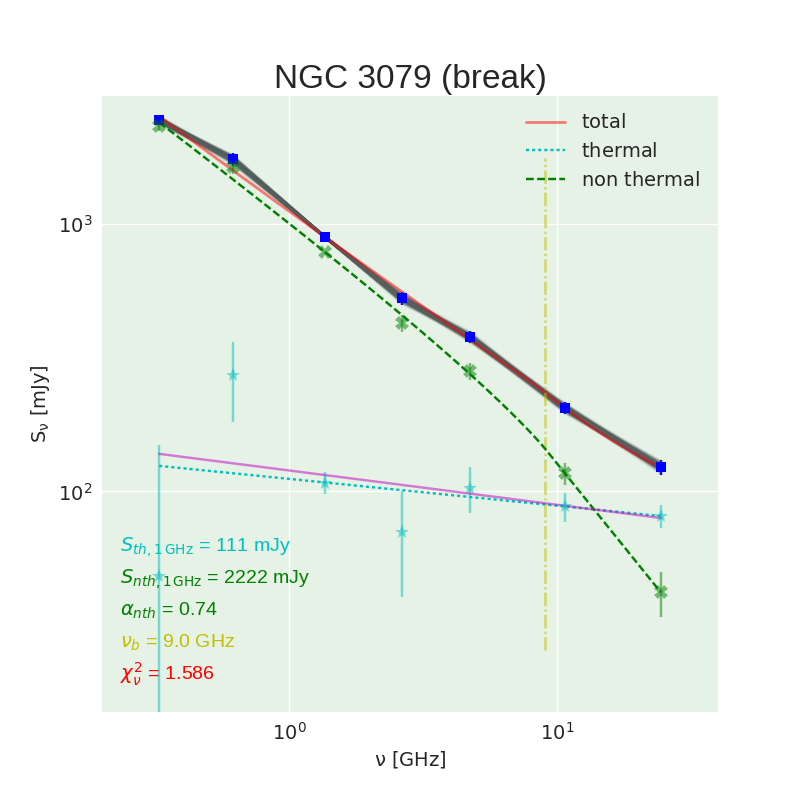} \hfill
\im{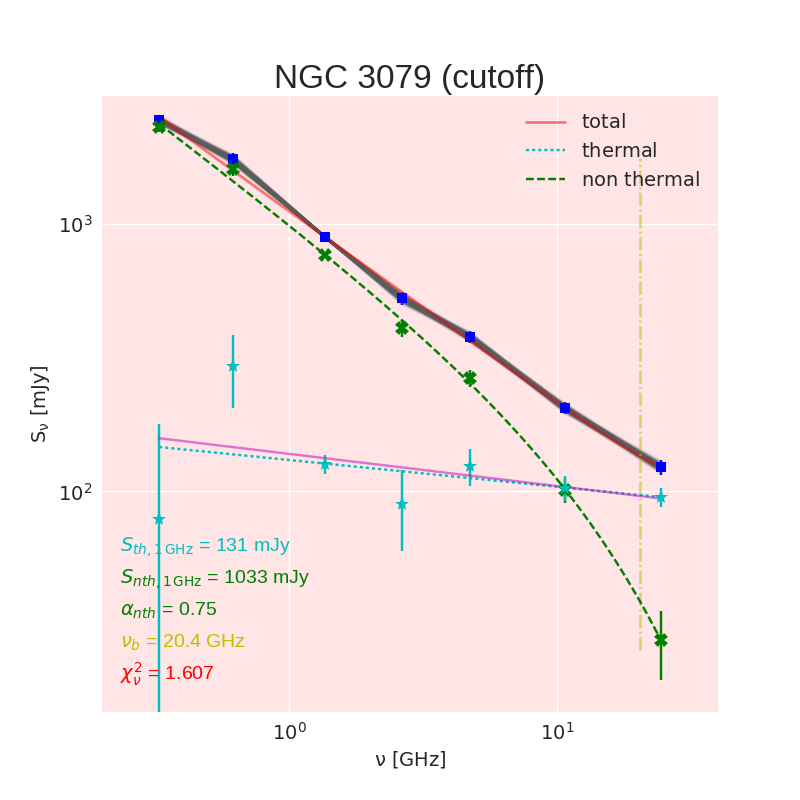}\\

\image{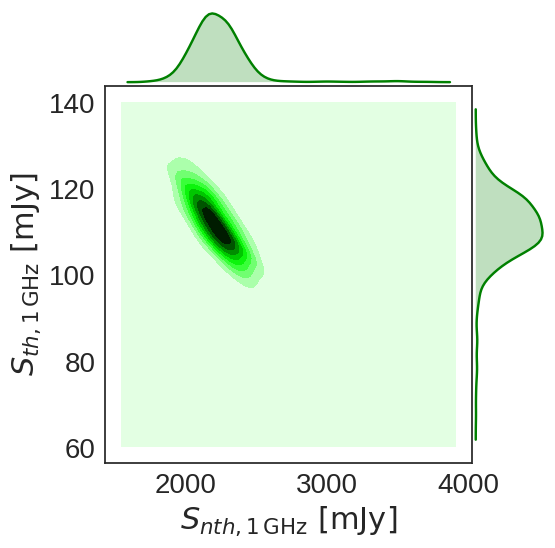}
\image{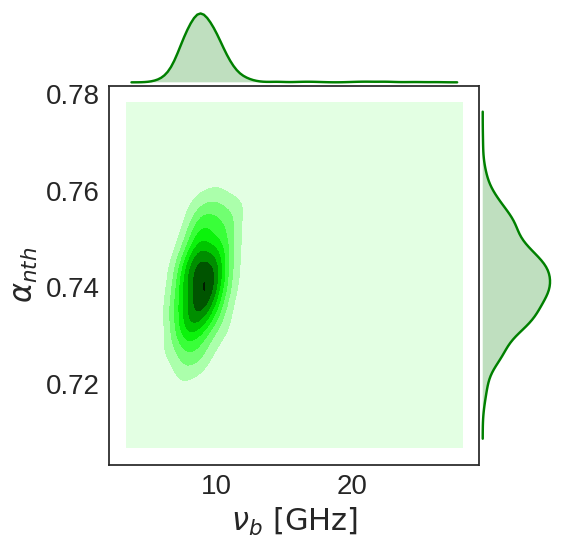} \hfill
\image{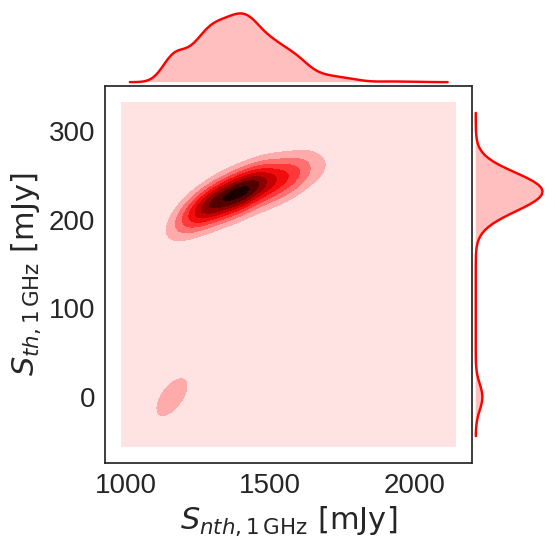}
\image{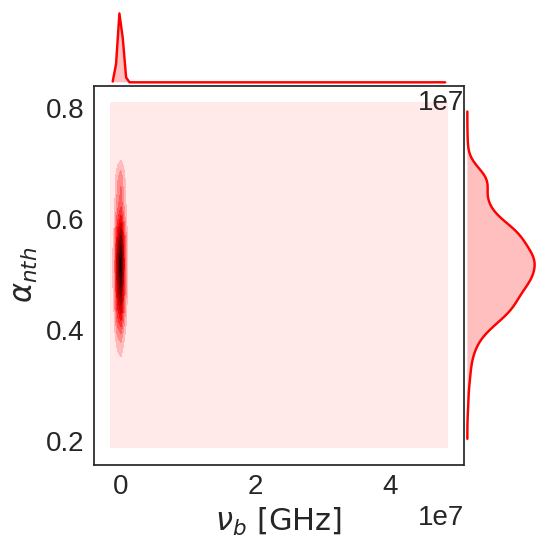}
\end{figure*}
\newpage
\clearpage

\begin{figure*}
\im{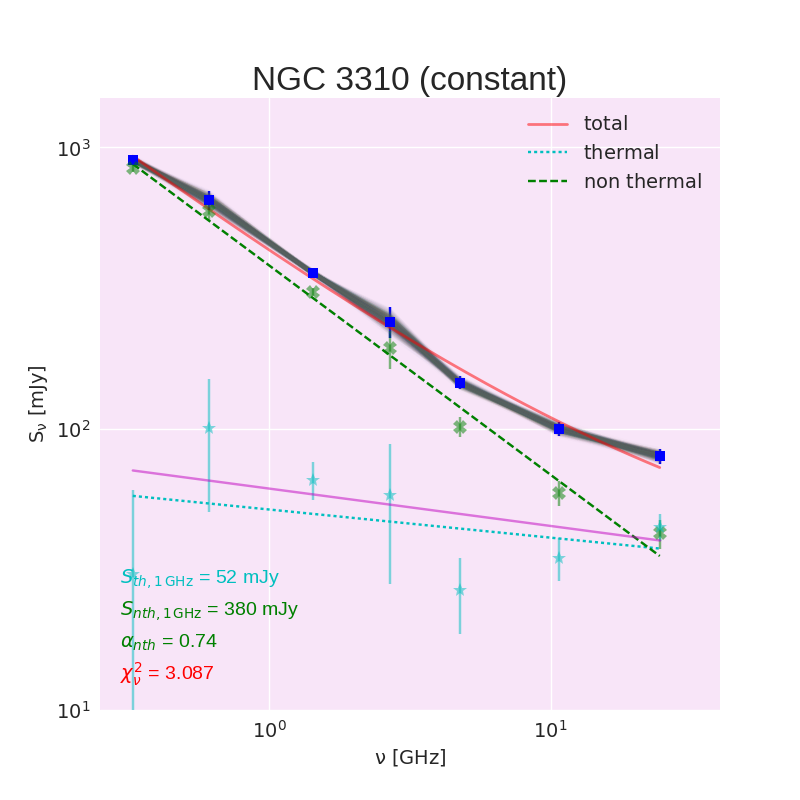} \hfill
\im{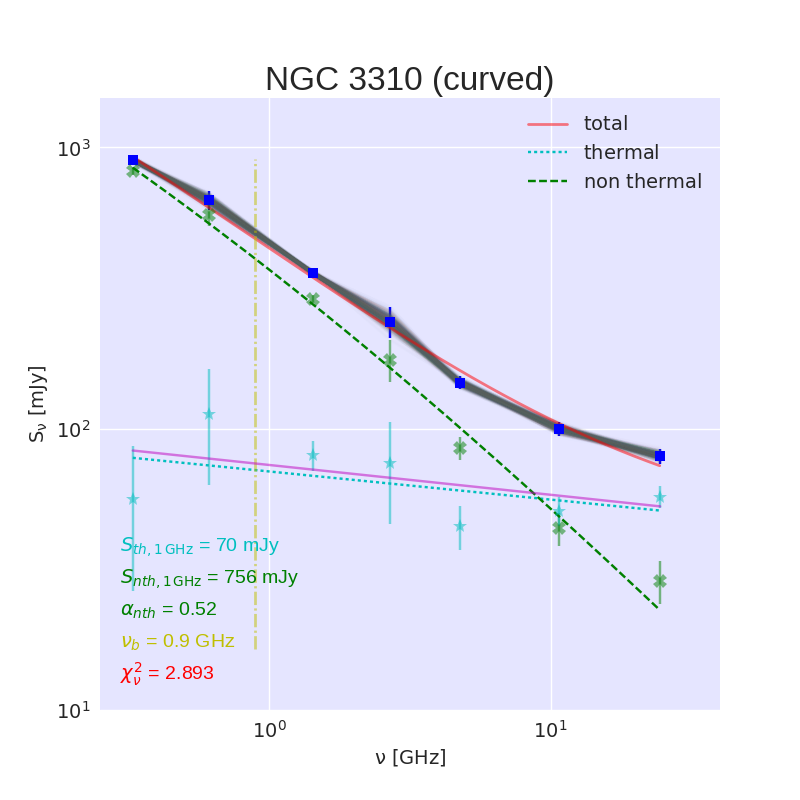}\\

\image{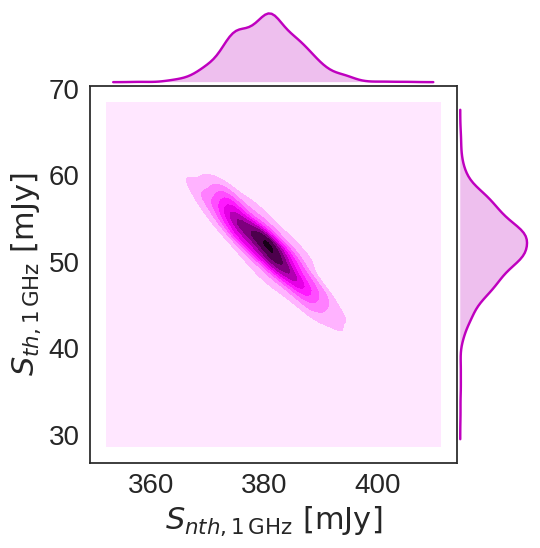}
\image{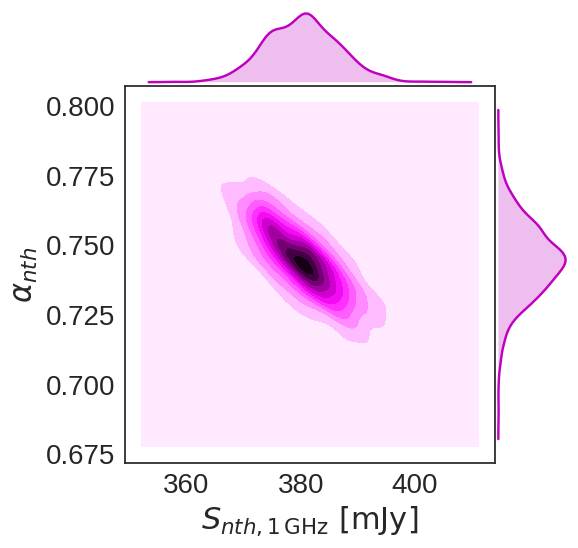} \hfill
\image{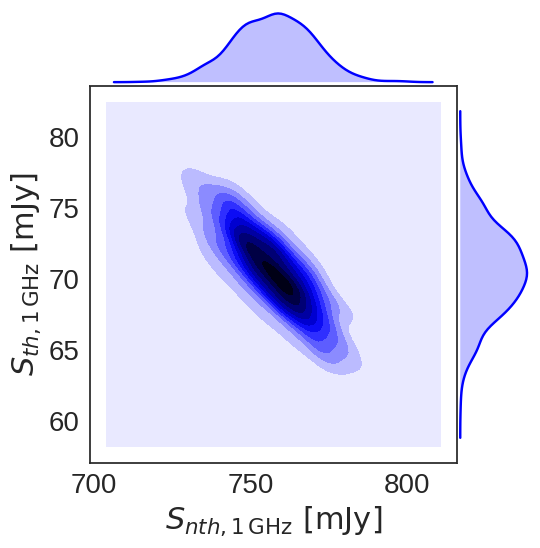}
\image{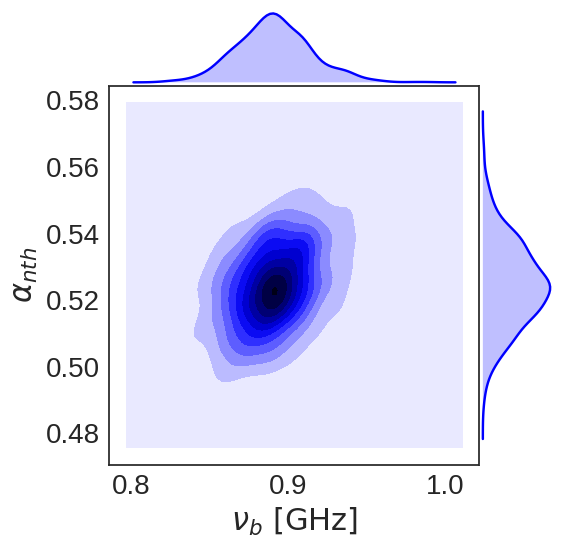}\\

\im{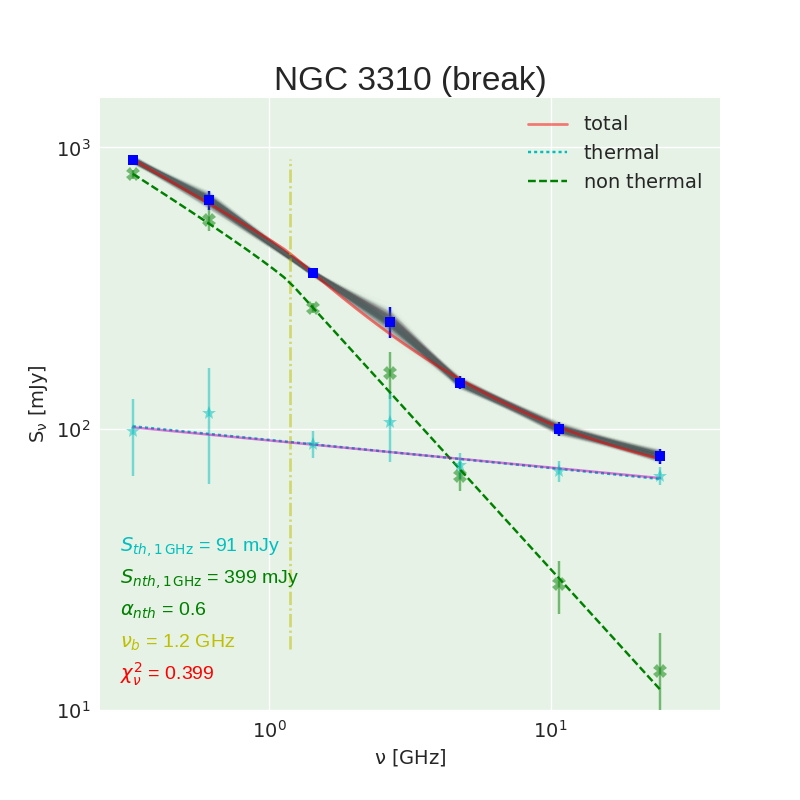} \hfill
\im{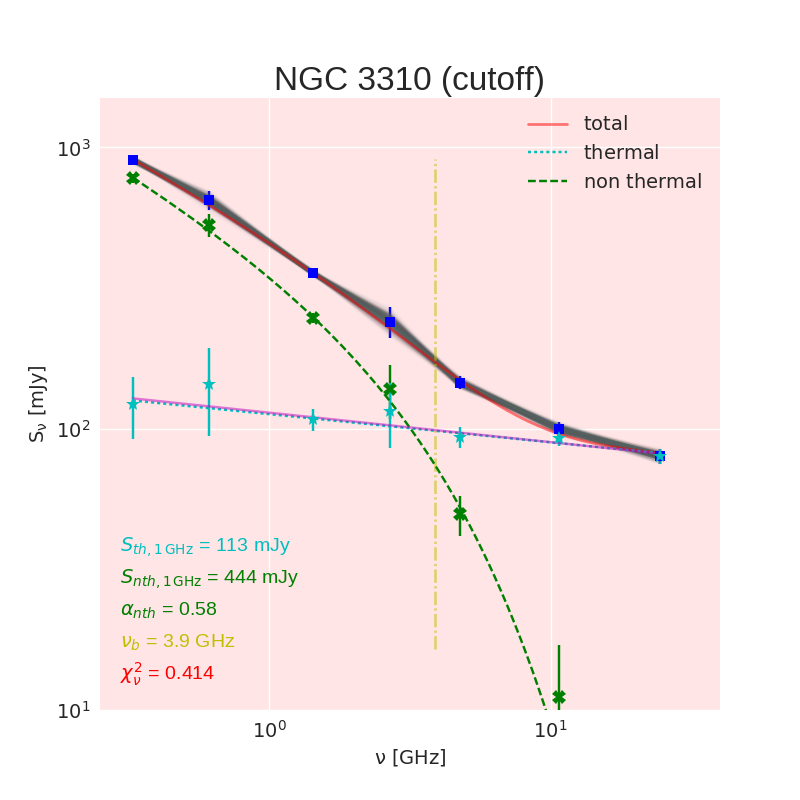}\\

\image{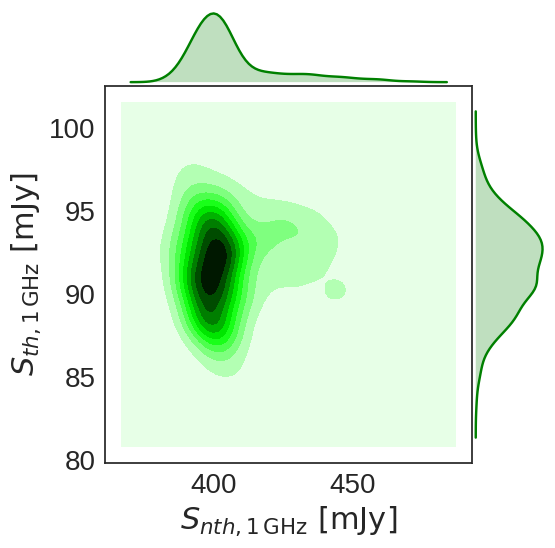}
\image{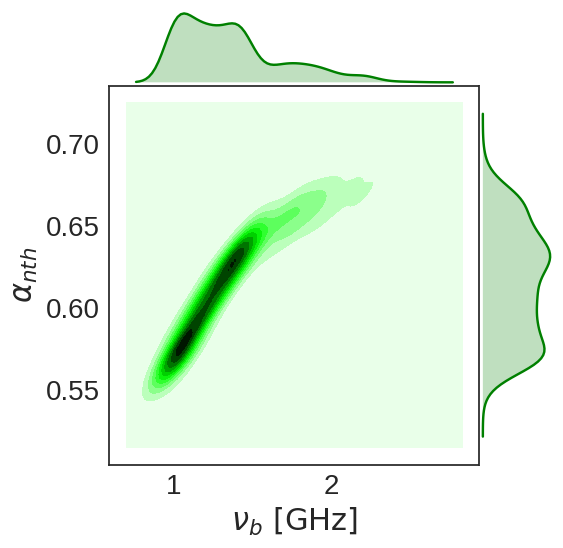} \hfill
\image{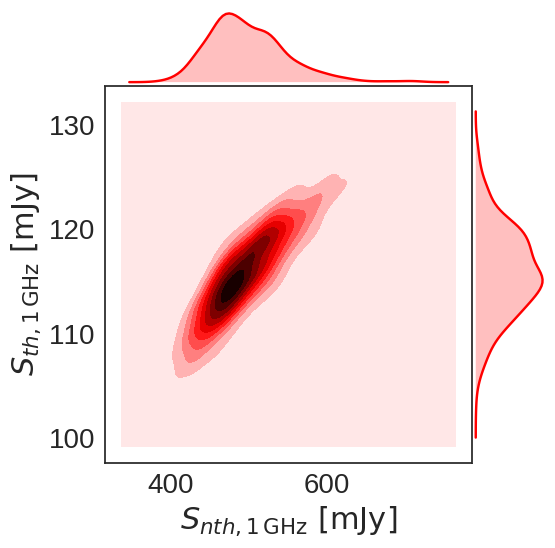}
\image{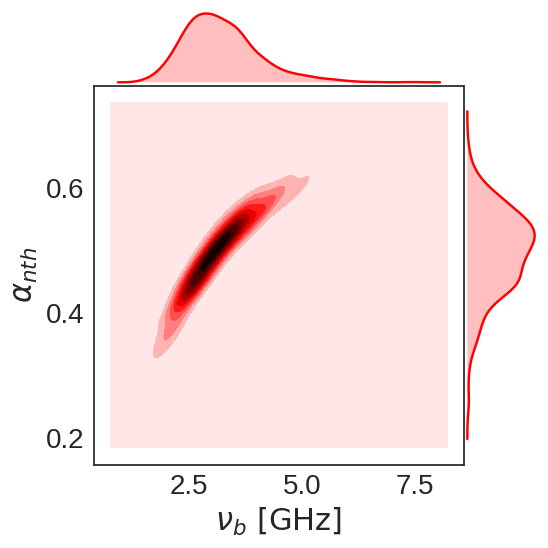}
\end{figure*}
\newpage
\clearpage

\end{document}